\documentclass{aa}  

\usepackage{graphicx}
\usepackage{subcaption}
\usepackage{soul}
\usepackage{ulem}
\usepackage{txfonts}
\usepackage{natbib}
\usepackage{caption}

\usepackage[maxfloats=256]{morefloats}
\usepackage[bookmarks=false, colorlinks=true, citecolor=blue, linkcolor=blue]{hyperref}
\def\kms{\,km\,s$^{-1}$}

\begin{document} 

\title
{Multi-frequency VLBI observations of maser lines during the 6.7~GHz maser flare in the high-mass young stellar object G24.33$+$0.14}

   \author{A. Kobak
         \inst{1}
         \href{https://orcid.org/0000-0002-1206-9887}{\includegraphics[scale=0.5]{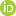}}
          \and
          A. Bartkiewicz
          \inst{1} \href{https://orcid.org/0000-0002-6466-117X}{\includegraphics[scale=0.5]{orcid.png}}
         \and
          M. Szymczak
          \inst{1} \href{https://orcid.org/0000-0002-1482-8189}{\includegraphics[scale=0.5]{orcid.png}}
          \and
          M. Olech
          \inst{2} \href{https://orcid.org/0000-0002-0324-7661}{\includegraphics[scale=0.5]{orcid.png}}
          \and
          M. Durjasz
          \inst{1} \href{https://orcid.org/0000-0001-7952-0305}{\includegraphics[scale=0.5]{orcid.png}}
          \and
          P. Wolak
         \inst{1} \href{https://orcid.org/0000-0002-5413-2573}{\includegraphics[scale=0.5]{orcid.png}}
          \and
          J.O. Chibueze
         \inst{3,4} \href{https://orcid.org/0000-0002-9875-7436}{\includegraphics[scale=0.5]{orcid.png}}
         \and
         T. Hirota
         \inst{5,6} \href{https://orcid.org/0000-0003-1659-095X}{\includegraphics[scale=0.5]{orcid.png}}
         \and
         J. Eisl\"offel
         \inst{7} \href{https://orcid.org/0000-0001-6496-0252}{\includegraphics[scale=0.5]{orcid.png}} 
         \and
         B. Stecklum
         \inst{7} \href{https://orcid.org/0000-0001-6091-163X}{\includegraphics[scale=0.5]{orcid.png}} 
        \and
         A. Sobolev
        \href{https://orcid.org/0000-0001-7575-5254}{\includegraphics[scale=0.5]{orcid.png}} 
        \and
         O. Bayandina
         \inst{8} \href{https://orcid.org/0000-0003-4116-4426}{\includegraphics[scale=0.5]{orcid.png}} 
        \and
         G. Orosz
         \inst{9} \href{https://orcid.org/0000-0002-5526-990X}{\includegraphics[scale=0.5]{orcid.png}} 
        \and
         R.~A. Burns
         \inst{5,10,11} \href{https://orcid.org/0000-0003-3302-1935}{\includegraphics[scale=0.5]{orcid.png}} 
        \and
         Kee-Tae Kim
         \inst{11,12} \href{https://orcid.org/0000-0003-2412-7092}{\includegraphics[scale=0.5]{orcid.png}}
         \and
         S.~P. van den Heever
         \inst{13}
         \href{https://orcid.org/???????}{\includegraphics[scale=0.5]{orcid.png}}
         }
         
 \institute{Institute of Astronomy, Faculty of Physics, Astronomy and Informatics, Nicolaus Copernicus University, Grudziadzka 5, 87-100 Torun, Poland
 \and Space Radio-Diagnostics Research Centre, University of Warmia and Mazury, ul.~Oczapowskiego 2, 10-719 Olsztyn, Poland
\and Centre for Space Research, North-West University, Potchefstroom 2520, South Africa
\and Department of Physics and Astronomy, Faculty of Physical Sciences,  University of Nigeria, Carver Building, 1 University Road, Nsukka, Nigeria
\and Mizusawa VLBI Observatory, National Astronomical Observatory of Japan, Osawa 2-21-1, Mitaka-shi, Tokyo 181-8588, Japan
\and Department of Astronomical Sciences, SOKENDAI (The Graduate University for Advanced Studies), Osawa 2-21-1, Mitaka-shi, Tokyo 181-8588, Japan
\and Th\"uringer Landessternwarte Tautenburg, Sternwarte 5, 07778 Tautenburg, Germany
\and INAF – Osservatorio Astrofisico di Arcetri, Largo E. Fermi 5, 50125 Firenze, Italy
\and Joint Institute for VLBI ERIC, Oude Hoogeveensedijk 4, 7991 PD, Dwingeloo, Netherlands
\and Department of Science, National Astronomical Observatory of Japan, 2-21-1 Osawa, Mitaka, Tokyo 181-8588, Japan
\and Korea Astronomy and Space Science Institute, 776 Daedeokdae-ro, Yuseong-gu, Daejeon 34055, Republic of Korea
\and  University of Science and Technology, Korea (UST), 217 Gajeong-ro, Yuseong-gu, Daejeon 34113, Republic of Korea
\and SARAO, Hartebeesthoek site, Krugersdorp, South Africa, 1739
}

  \date{Received 19 August 2022 / Accepted 22 December 2022 }

 
  \abstract
{Recent studies have shown that 6.7 GHz methanol maser flares can be a powerful tool for verifying the mechanisms of maser production and even the  specific signatures of accretion rate changes in the early stages of high-mass star formation.}
{We characterize the spatial structure and evolution of methanol and water masers during a flare of methanol maser emission at 6.7~GHz in the high-mass young stellar object (HMYSO) G24.33$+$0.14.} 
{Very Long Baseline Array (VLBA) was used to image the 6.7 and 12.2\,GHz methanol and 22.2\,GHz water vapor masers at three epochs guided by monitoring the methanol line with the Torun 32m\, telescope. The 6.7\,GHz maser maps were also obtained with the European VLBI Network (EVN) and Long Baseline Array (LBA) during the flare. The Wide-field Infrared Survey Explorer (WISE) data were used to find correlations between the 6.7\,GHz maser and infrared (IR) fluxes.}
{The 6.7\,GHz methanol maser cloudlets are distributed over $\sim$3500\,au, and the morphology of most of them is stable although their brightness varies following the course of the total flux density on a timescale of two months. The 12.2\,GHz methanol maser cloudlets cover an area an order of magnitude smaller than that of 6.7\,GHz emission, and both transitions emerge from the same masing gas. The 22.2\,GHz maser cloudlets lie in the central region and show a systematic increase in brightness and moderate changes in size and orientation, together with the velocity drift of the strongest cloudlet during two months of the Very Long Baseline Interferometry (VLBI) observing period. Time lag estimates imply the propagation of changes in the physical conditions of the masing region with a subluminal speed ($\sim$0.3c). A tight correlation of IR (4.6$\mu$m) and 6.7\,GHz flux densities is found, supporting the radiative pumping model. Proper motion analysis does not reveal any signs of expansion or inflow of the methanol cloudlets within $\sim$6\,mas over $\sim$10\,yr. Comparison with the 230\,GHz Atacama Large Millimeter Array (ALMA) data indicates that the methanol masers are distributed in the inner part of the rotating disk, whereas the 22.2\,GHz emission traces the compact inner component of the bipolar outflow or a jet structure.  
}
{The maser morphology in the target is remarkably stable over the course of the flare and is similar to the quiescent state, possibly due to less energetic accretion events that can repeat on a timescale of $\sim$8\,yr.}

   \keywords{masers -- stars: massive -- stars: formation --  stars: flare -- radio lines: ISM --  individual: G24.33+0.14}

\titlerunning{Multi-frequency observations of the maser flare in G24.33$+$0.14.}
\authorrunning{A. Kobak et al.}

   \maketitle

\section{Introduction}\label{sec:introd}
Spectral line observations of high-mass young stellar object (HMYSO) are valuable probes of the kinematics and dynamics of the star formation processes. Methanol maser lines are of special interest in this context as unambiguous signposts of star formation activity \citep[e.g.,][]{menten1991, caswell2003, breen2015}. 
Owing to their high brightness and compactness, they can probe neutral gas cloudlets of a few tens of au in size that reside in rotating structures, such as toroids and disks (\citealt{beltran2016} for review and references therein), or around powerful jets (e.g.,~\citealt{anglada2018}). Moreover,  brightening of the centimeter methanol maser lines is easily detectable in single-dish monitoring, and it has recently proven to be a signature of accretion events in HMYSO.

\citet{fujisawa2015} reported the outburst of the 6.7~GHz methanol maser line in S255IR-NIRS3, the first case since the discovery of this transition by \citet{menten1991}. This target has been presenting moderate variability on timescales of months to years over $\sim$23~yr \citep{szymczak2018b}. The follow-up near-infrared imaging showed the brightening of the central source and its outflow cavities, which provided evidence of an accretion burst onto the HMYSO \citep{caratti2017}.
A similar 6.7~GHz methanol maser flare was reported in the accretion burst source NGC6334I$-$MM1 by \citet{macleod2018}. They presented a contemporaneous flaring event of ten maser transitions in hydroxyl, methanol, and water. The Very Large Array (VLA) imaging by \citet{hunter2018} showed new methanol masers that appeared toward the MM1 protostellar system. \citet{brogan2018} reported flaring of water masers with a bow-shock pattern, and proper motions studies by \citet{chibueze2021} confirmed the radio jet influence.

Under the international collaboration M2O{\footnote{https://www.masermonitoring.com/}}, which is focused on maser flare detections and follow-ups by radio and infrared telescopes globally  \citep{burns2022}, interesting phenomena in the HMYSO G358.93$-$0.03 were reported (e.g., \citealt{burns2020}). Different methanol maser lines flared simultaneously, and Very Long Baseline Interferometry (VLBI) observations of methanol masers could outline the exact picture of a heat wave propagation in an accretion disk. 
Therefore, when the flare of the 6.7~GHz methanol maser line in the HMYSO G24.33$+$0.14 was reported \citep{wolak2019}, we started following up observations of the 6.7~GHz and 12.2~GHz methanol and 22.2~GHz water maser transitions with diverse interferometric networks within the M2O collaboration. The aim of the present paper is to study the changes in the maser morphology as an effect of possible accretion burst in the source. 

G24.33$+$0.14 (hereafter G24) is a massive young protostar, noted as an extended green object (i.e.,  an embedded high-mass star with molecular outflow) (\citealt{cyganowski2008}). The source is known to harbor methanol, water, and hydroxyl masers \citep{caswell2011}.
 The 6.7~GHz maser imaged using the European VLBI Network (EVN) in June 2009 consists of three emission groups separated by 0\farcs2$-$0\farcs4 \citep{bartkiewicz2016}. In mid-2011 the source experienced a strong flare of all the features at 6.7\,GHz, lasting 200-400 days \citep{szymczak2018}. Recent observations show that the source harbors lines of ammonia at 23.9\,GHz and formaldehyde at 4.8\,GHz \citep{McCarthy2021}.
The distance derived from H$_{{\rm I}}$ self-absorption is 9.5~kpc \citep{green2011} or 7.20$\pm$0.76~kpc (with 90\% probability) according to the Bayesian distance estimator based on accurate distances and proper motions of high-mass star-forming regions across the Milky Way \citep{reid2019}. This value is adopted in the present paper. \citet{hirota:2022} compared the 230~GHz Atacama Large Millimeter Array (ALMA) data at pre- and post-flare epochs of 2016 and 2019, respectively.
The 2019 dataset was obtained as a Director's Discretionary Time (DDT) project 20~days after the onset of the 6.7~GHz methanol maser flare. Three continuum sources were identified, indicating that the 6.7~GHz methanol maser coincides with the brightest one, named C1. The velocity structures of millimeter methanol lines at the C1 position show velocity gradients, implying the existence of a rotating ring-like structure of a disk--envelope system. The continuum and spectral observations suggest radiative heating in the central part of the disk, due to two flaring events involving a short-period episodicity on an 8 yr timescale traced by the methanol maser flares (Sect.~3.2), in addition to a 7000~yr timescale episodic outflow derived from multiple bow shocks traced by the 229.758~GHz methanol emission \citep{hirota:2022}. The detailed chemistry was studied by \citet{beak2022}; the results suggest that enhanced abundances of several molecules in the hot cores could be caused by the active accretion as well as different physical conditions of cores.

\section{Observations}\label{sect:obser}
\subsection{Torun 32m monitoring}
The target was observed in the 6668.519\,MHz methanol line about once a week as part of the monitoring project using the Torun 32m radio telescope. The beam full width at half maximum (FWHM) for this transition was 5\farcm8 with an rms pointing error of $\sim$25\arcsec\, before mid-2016 and $\sim$10\arcsec\, afterward. The data was taken in dual-polarization frequency-switching mode. The averaged spectra of 0.09\kms\, resolution (after Hanning smoothing) had a typical rms noise level of about 0.35\,Jy. The uncertainty of the flux density calibration was $\sim$10\% (for details on the method of observation, calibration, and measurement of the flux density, see  \citealt{szymczak2018}).

\subsection{VLBI}
The observations were conducted with diverse interferometers after detecting a flare of the 6.7\,GHz methanol line. Below we briefly describe them, and in Table~\ref{table1} we summarize the information on the maser transitions; dates of observations, day-of-the year (DOY) in 2019 is added for clarity; synthesized beams used in imaging; and the noise (1\,$\sigma_\mathrm{rms}$) in the emission-free channel map.

G24.33+0.14 was observed with the  Very Long Baseline Array\footnote{The National Radio Astronomy Observatory is a facility of the National Science Foundation operated under cooperative agreement by Associated Universities, Inc. Scientific results from data presented in this publication are derived from the following VLBA project code: BB416} (VLBA), European VLBI Network\footnote{The European VLBI Network is a joint facility of independent European, African, Asian, and North American radio astronomy institutes. Scientific results from data presented in this publication are derived from the following EVN Target of Opportunity project code: RB006B} (EVN), and Long Baseline Array\footnote{The Long Baseline Array is part of the Australia Telescope National Facility, which is funded by the Australian Government for operation as a National Facility managed by CSIRO. Scientific results from data presented in this publication are derived from the following LBA project code: VX026D} (LBA) interferometers. The methanol 6668.519\,MHz (hereafter 6.7\,GHz) line was observed by all these networks. Frequency switching was used in VLBA experiments to observe both the 12178.597\,MHz methanol and 22235.08\,MHz water (hereafter 12.2 and 22.2\,GHz) masers. 
We used a standard approach to reduce the data using the National Radio Astronomy Observatory (NRAO) Astronomical Image Processing System (AIPS) package. The quasar 3C345 was used as a delay and bandpass calibrator. The observations were conducted in phase-referencing mode with J1825$-$0737 as phase calibrator and J1835$-$1115 as the additional calibrator in LBA. The data were corrected for the effects of Earth's rotation and its motion within the Solar System and toward the Local standard of rest (LSR).  

The VLBA observed the target in three epochs. The average on-source time per band was 1.5\,h per epoch and the cycle time for phase-referencing was 1~min on phase calibrator and 2 min on the maser. We configured the baseband converter to a bandwidth of 32~MHz in dual polarization, averaged to increase the signal-to-noise ratio in the continuum data. Using the DiFX software correlator \citep{Deller11}, channel spacings were set to 0.176 (6.7~GHz methanol), 0.096 (12.2~GHz methanol), and 0.053~km~s$^{-1}$ (22.2~GHz water). We note that for the  22.2~GHz data at the first epoch neither the phase-referencing technique nor the inverse phase-referencing was successful; therefore, we fringe-fitted the data on the brightest maser channel and estimated the absolute positions by identifying the same structures of maser spots in the second epoch.  

For LBA the DiFX correlator was configured with a bandwidth of 4~MHz divided into 4092 channels and with an integration time of 2\,s. The on-source time for the target was 4.48\,h with a cycle time of 135\,s both on the source and phase calibrators. Due to the nature of LBA observations, the initial gain amplitude calibration for the individual antennas was carried out by using the ACFIT task in AIPS, with Parkes as the reference antenna. Inspection of the autocorrelation spectra has shown that the target source flux density is $\sim$30$\%$ higher than for the Torun observations in the same period. Therefore, these data are considered unreliable for the recovered flux density estimation, and only the measured emission structure is analyzed in this publication as a general overview of the structure during the flare. 

In EVN we used a 2~min cycle time between source and calibrator,  with a total time on the source of 2.5\,h. The bandwidth of 16~MHz was divided into 8192 spectral channels, yielding the velocity resolution of 0.087~km~s$^{-1}$.

In order to analyze the parameters of milliarcsecond maser structures, we determined the position of emission peaks of maser spots that are detected in a velocity channel of an image cube and we defined a maser cloudlet as a group of maser spots that appear in at least three contiguous channels and coincide in positions within half the synthesized beam (e.g., \citealt{sanna2017}).

For the amplification along the same line-of-sight path (i.e., co-propagation of 6.7 and 12.2\,GHz maser transitions observed with the VLBA), we estimated the astrometric accuracy of single spot measurements on the phase-referenced images to be 0.07~ms in rectascension (RA) and 1\,mas in declination (Dec) and 0.06~ms in RA and 0.8~mas in Dec for the 6.7 and 12.2\,GHz data, respectively. These values were obtained considering the following factors, as recommended in \cite{Richards2022} (Sect.~3): (1) inaccuracy of the phase-reference source position of 0.3\,mas \citep{charlot2020}; (2) the position uncertainty due to noise, given by the ratio (beam size)/(signal-to-noise), which is $<$0.3\,mas; (3) the phase interpolation errors due to the phase calibrator--target time difference (0.03~mas); (4) the uncertainties introduced by the separation between the target and phase reference sources (2.4\degr), which are 0.7 and 0.6\,mas at 6.7 and 12.2 \,GHz, respectively;
and (5) the antenna position errors (with the antenna position error of 2~cm) of 0.5 and 0.3\,mas at  6.7 and 12.2 \,GHz, respectively. Similarly, the positional accuracy of the 22~GHz water maser line observed using the VLBA is 0.08~ms in RA and 1.2~mas in Dec.

\begin{table}
\centering
\caption{Observing parameters}
\label{table1}
\begin{tabular}{llcc}
\hline
 Transition &  Date  & Synthesized beam & 1$\sigma_{rms}$\\ 
& 2019&  (a$\times$b; PA)\\
(MHz) & (day month) & (mas$\times$mas; \degr)&  (mJy~beam$^{-1}$)\\
\hline
\multicolumn{4}{l}{\bf VLBA} \\
6668.519   & 27 Sep & 2.7$\times$1.3; $+$2 & 5.5 \\
12178.597  & (DOY: 270)  & 1.9$\times$0.7; $-$18 & 9.0 \\
22235.08  &             & 0.9$\times$0.4; $-$9 & 16  \\
6668.519   & 27 Oct & 6.5$\times$2.1; $-$13 & 4.8\\
12178.597  & (DOY: 300)  & 2.1$\times$0.6; $-$20 & 7.4\\
22235.08  &              & 2.4$\times$0.9; $-$14 & 7.1\\
6668.519   & 2 Dec & 5.2$\times$1.3; $-$19 & 3.9 \\
12178.597 &  (DOY: 336)  & 2.4$\times$0.8; $-$21 & 9.8 \\
22235.08    &            & 3.8$\times$1.3; $+$23 & 7.2 \\
\multicolumn{4}{l}{\bf LBA}\\
6668.519 & 28 Sep & 8.0$\times$6.6; $-$5 & 12 \\
 & (DOY: 271) & \\
\multicolumn{4}{l}{\bf EVN}   \\  
6668.519 & 7 Oct  & 16.3$\times$6.2; $+$7 & 3.4\\
 & (DOY: 280) & \\
\hline
\end{tabular}
\end{table}

\section{Results}\label{sec:results}
Table~\ref{table2} lists the absolute coordinates of the brightest maser spots for each transition, along with the peak velocity ($V_{\mathrm p}$), the peak intensity ($S_{\mathrm p}$), and the corresponding LSR velocity range of emission ($\Delta V_{\mathrm{LSR}}$), for all the experiments. In Figure~\ref{fig:multiobs} we present a combined map resulting from the multi-frequency maser observations at epoch DOY:270. 

The 6.7 and 12.2~GHz methanol lines are within the LSR velocity of 107.5 to 120.2\,\kms\, and from 110.0 to 112.1\,\kms, with the peaks at 115.3 and 110.4\,\kms, respectively. The water maser emission ranges from 122.1 to 127.1\,\kms\, with the peak at 125.1\,\kms. The 6.7~GHz methanol maser spots are spread over and area of $\sim$0\farcs5$\times$0\farcs5,  corresponding to $\sim$3500~au$\times$3500~au (Fig.~\ref{fig:sixepochs}). 
The 12.2~GHz methanol maser emission appears in a much smaller region of $\sim$0\farcs02$\times$0\farcs06 ($\sim$150\,au$\times$450\,au), and its velocity range is a factor of 6 narrower than that of the 6.7~GHz emission. The two southern 12.2~GHz cloudlets disappeared in the third epoch (Fig.~\ref{fig:vlba3meth12}). 
The 22.2~GHz water masers are located in the central part of the 6.7~GHz maser region (Fig.~\ref{fig:multiobs}). They form two tiny clusters of 4$-$12\,au in length, elongated in the SE-NW direction with a clear velocity gradient (Fig.~\ref{fig:vlba3water}). The gray dashed-line ellipse represents the position of the 44 GHz radio jet reported by \cite{purser2021}. There is a $\sim$100\,mas ($\sim$720\,au) offset between the peak position of the radio jet and the 22\,GHz water masers, perhaps caused by the masers tracing the entrained gas around the jet. A small fraction of the offset could be attributed to the 7.5\,mas astrometric accuracy of the VLA observations.

\subsection{Emission structure during the flare and variability}
\subsubsection{6.7\,GHz}

\begin{table*}[ht!]
\centering
\caption{Results of the VLBA, LBA, and EVN observations.}
\label{table2}
\begin{tabular}{lcllccc}
\hline
Interferometer & Transition & \multicolumn{4}{c}{Results for the brightest maser spot} &  \\
&  & RA (J2000) & Dec (J2000) & $V_{\mathrm p}$ & $S_{\mathrm p}$ & $\Delta V_{\mathrm {LSR}}$\\
Day-of-the-year & (GHz) & (h m s) & (\degr~\arcmin~\arcsec) & (km s$^{-1}$)  & (Jy beam$^{-1}$) & (km s$^{-1}$) \\
\hline
VLBA DOY: 270& 6.7 CH$_3$OH & 18 35 08.14243$\pm$0.00007 & $-$07 35 03.9242$\pm$0.0010& 115.3 & 2.34 & 107.5--120.0 \\ 
 & 12.2 CH$_3$OH & 18 35 08.13433$\pm$0.00006 & $-$07 35 04.2806$\pm$0.0008  & 110.3 & 0.97 & 110.1--112.1 \\
& 22.2 H$_2$O$^a$ &  18 35 08.13231$\pm$0.00008 & $-$07 35 04.1437$\pm$0.0012  & 125.1 & 7.72  & 122.1--127.1\\
VLBA DOY: 300  &6.7 CH$_3$OH& 18 35 08.14235$\pm$0.00007 & $-$07 35 03.9269$\pm$0.0010 & 115.4 & 5.80 & 107.5--120.1\\ 
&12.2 CH$_3$OH& 18 35 08.13428$\pm$0.00006 & $-$07 35 04.2815$\pm$0.0008 & 110.3 & 1.03 & 110.0--112.1\\
&22.2 H$_2$O& 18 35 08.13234$\pm$0.00008 & $-$07 35 04.1443$\pm$0.0012  & 125.0 & 8.17 & 122.2--127.0\\
VLBA DOY: 336&6.7 CH$_3$OH& 18 35 08.14234$\pm$0.00007 & $-$07 35 03.9266$\pm$0.0010 & 115.4 & 2.59 & 109.8--120.2\\
 &12.2 CH$_3$OH & 18 35 08.13430$\pm$0.00006 & $-$07 35 04.2821$\pm$0.0008 & 110.3 & 0.56 & 110.2--110.6\\
&22.2 H$_2$O& 18 35 08.13233$\pm$0.00008 & $-$07 35 04.1453$\pm$0.0012  & 124.8 & 9.39 & 122.1--127.1\\
LBA ~~~DOY: 271& 6.7 CH$_3$OH& 18 35 08.14238$\pm$0.00009 & $-$07 35 03.9259$\pm$0.0014 & 115.3 & 8.97  & 107.6--120.1\\
EVN ~~DOY: 280 & 6.7 CH$_3$OH& 18 35 08.14231$\pm$0.00010 & $-$07 35 03.9247$\pm$0.0010  & 115.2 & 7.34 & 108.1--120.2\\
\hline
\multicolumn{7}{l}{$^a$ - the coordinates taken from VLBA DOY: 300 (see Sect.~2.1 for details).}
\end{tabular}
\end{table*}

\begin{figure*}
\centering
\includegraphics[width=\textwidth]{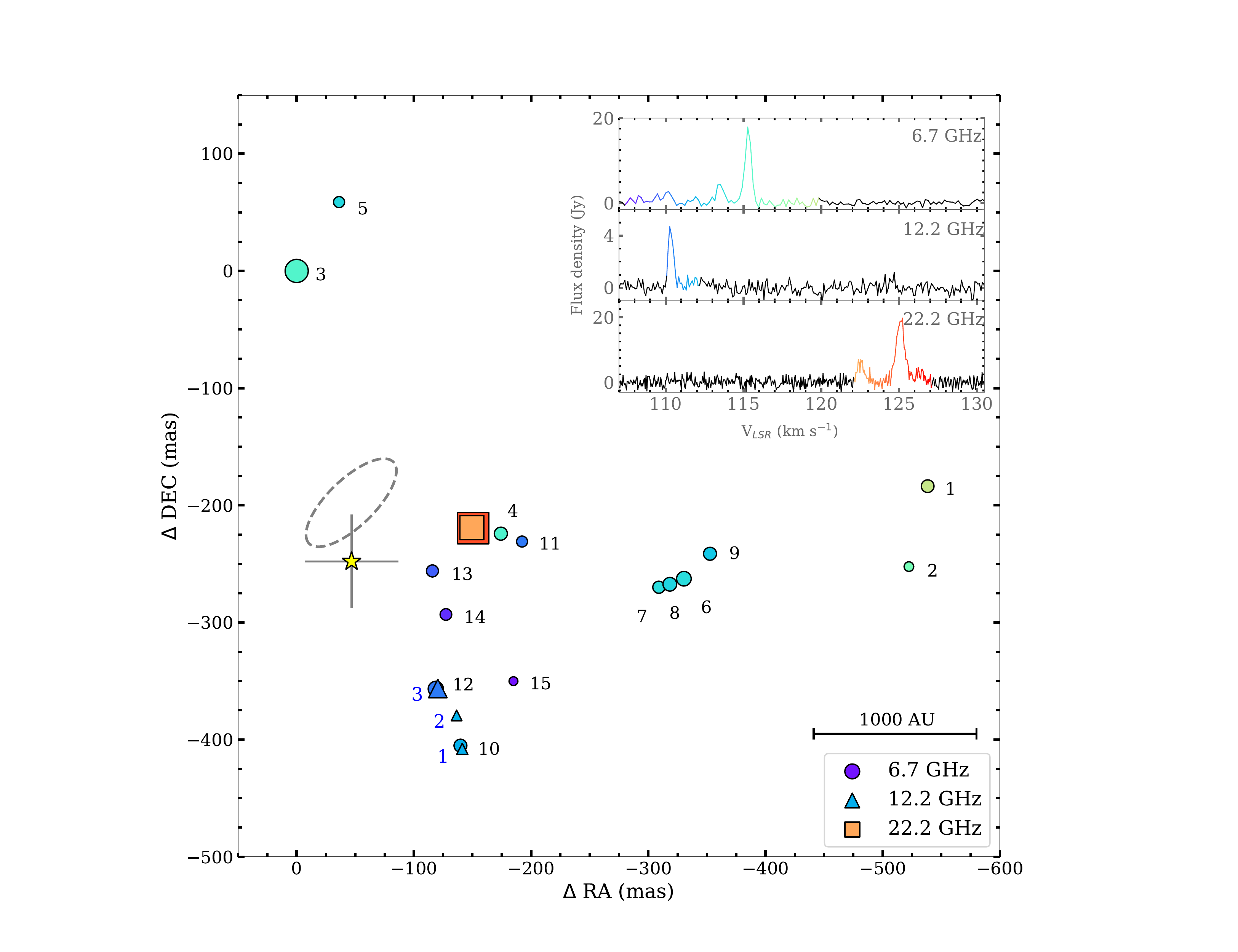}
\caption{Distribution of methanol and water maser cloudlets in G24 obtained with VLBA at DOY: 270. The inset shows the corresponding spectra. The circles, triangles, and squares correspond to the 6.7~GHz, 12.2~GHz methanol, and 22.2~GHz water transitions, respectively. The symbol sizes are proportional to the square root of the maser intensities. The symbol color corresponds to the LSR velocity, as indicated in the inserted spectra. The black numbers refer to groups of the 6.7~GHz methanol maser spots identified as cloudlets (Table~\ref{tablegaussmeth6}), while the blue numbers to the 12.2~GHz methanol maser cloudlets (Table~\ref{tablegaussmeth12}). The yellow star represents the brightest 230\,GHz continuum source at the position of RA(J2000)=18$^{\mathrm{h}}$35$^{\mathrm{m}}$08\fs13928 and Dec(J2000)=$-$07\degr35\arcmin04\farcs1721, while the  gray cross gives its  astrometric uncertainty \citep{hirota:2022}. The (0,0) point corresponds to {RA(J2000)=18$^{\mathrm{h}}$35$^{{\rm m}}$08\fs14243  and Dec(J2000)=$-$07\degr35\arcmin03\farcs9242, which is the position of the brightness maser spot at 6.7~GHz at DOY: 270. The gray dotted ellipse indicates the position of the Q-band jet reported in \cite{purser2021}.}}
\label{fig:multiobs}
\end{figure*}

\begin{figure*}[ht!]
\centering
\includegraphics[width=\textwidth]{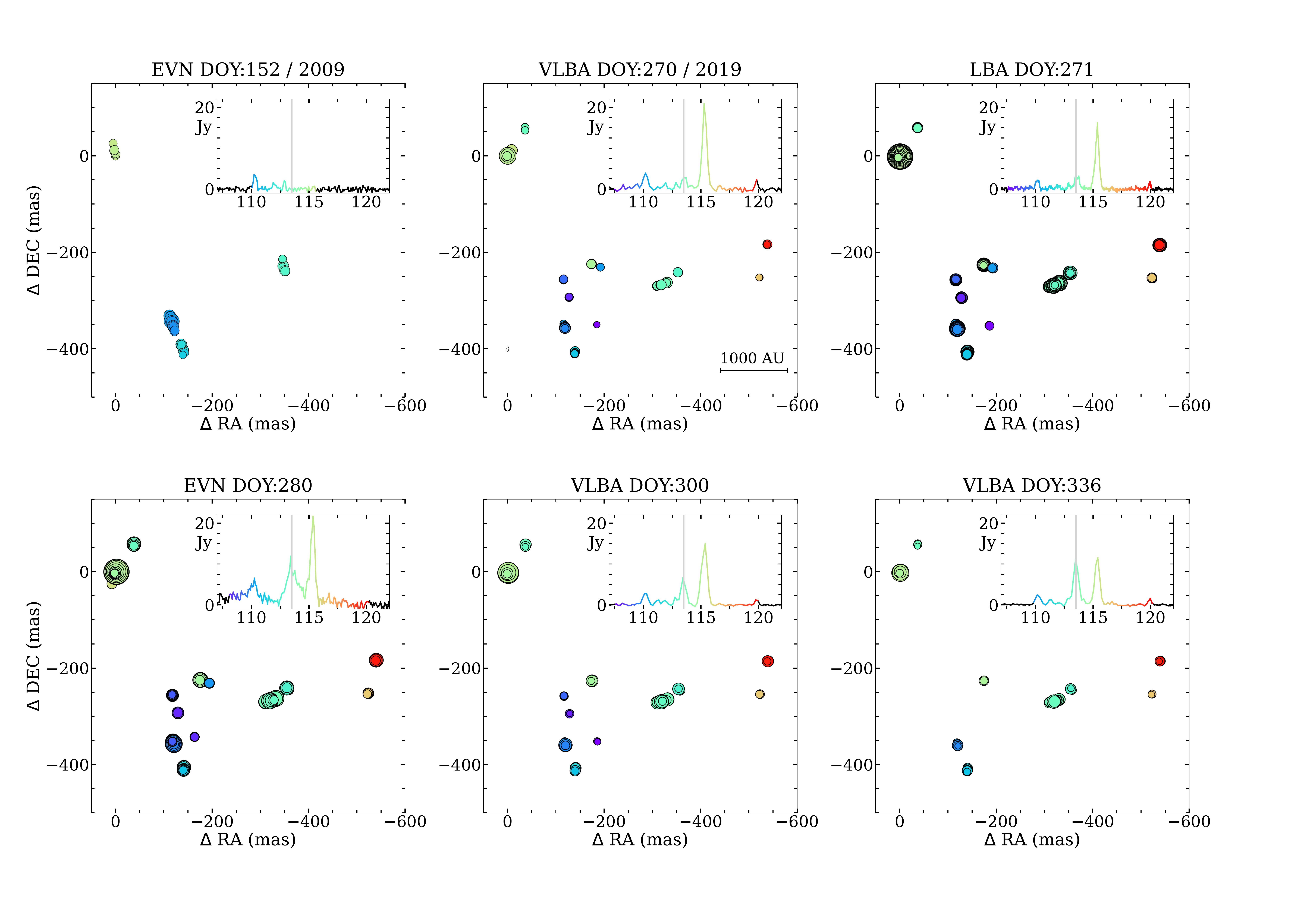}
\caption{Distribution of the 6.7~GHz methanol maser spots, as derived from the EVN \citep[including archival data from][]{bartkiewicz2016}, VLBA, and LBA observations at all epochs. The size of the circles is proportional to the square root of the intensity of a given spot. The symbol color corresponds to the LSR velocity, as indicated in the inserted spectra. The vertical gray lines at the spectra indicate the systemic velocity \citep{hirota:2022}. The (0,0) point in 2019 corresponds to the coordinates given in Fig.~\ref{fig:multiobs} and in 2009 to the coordinates RA(J2000)=18$^{{\rm h}}$35$^{{\rm m}}$08\fs12683 and    Dec(J2000)=$-$7\degr35\arcmin03\farcs7868 (from the BeSSeL survey, EVLA,  by \citealt{hu2016}).}
\label{fig:sixepochs}
\end{figure*}

\begin{figure*}[ht!]
\centering
\includegraphics[width=\textwidth]{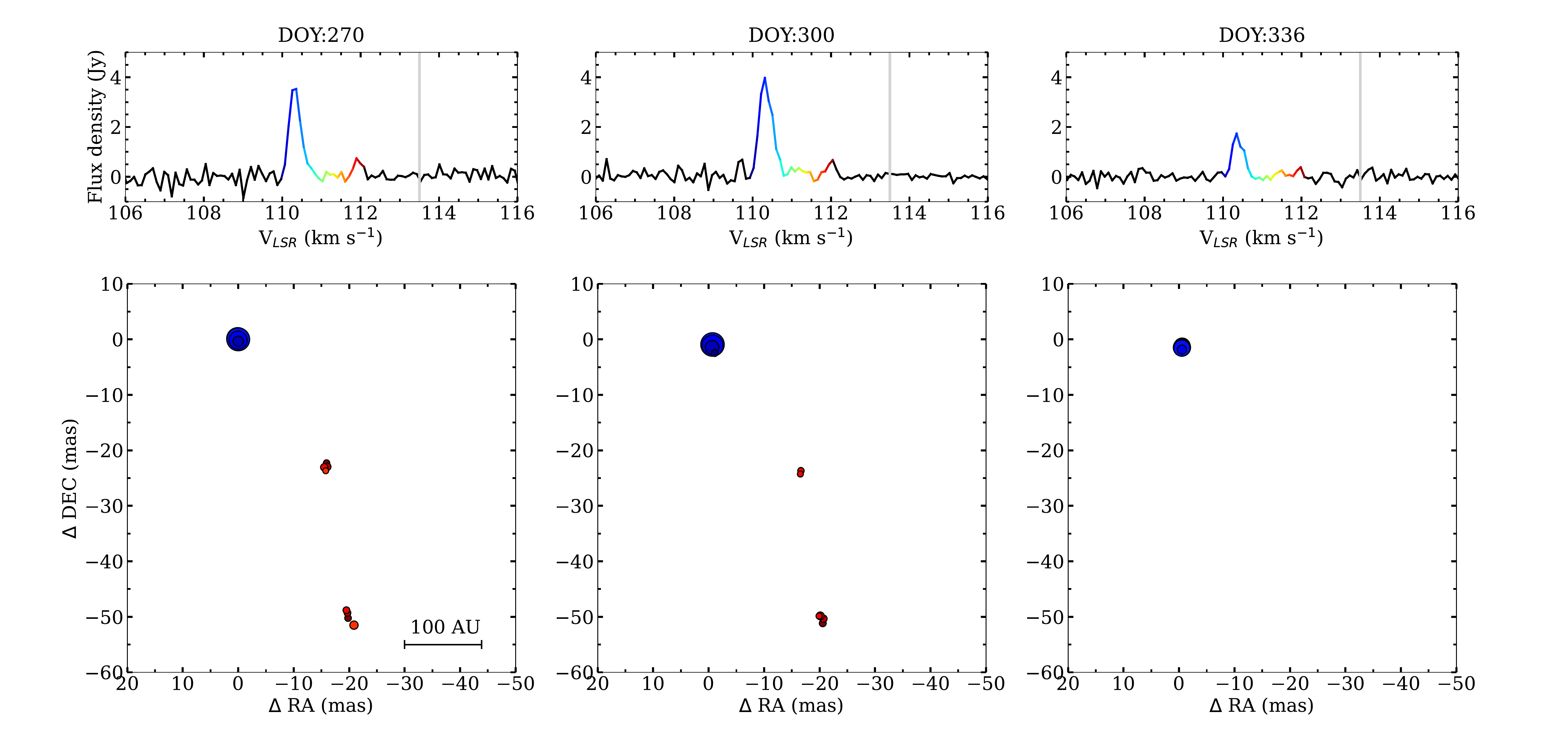}
\caption{VLBA maps of the 12.2~GHz methanol maser spots in G24 at three epochs. The size of the circles is proportional to the square root of the intensity of a given spot. The symbol color corresponds to the LSR velocity as indicated in the inserted spectra. The vertical gray lines at the spectra indicate systemic velocity \citep{hirota:2022}. The point (0,0) corresponds to RA(J2000)=18$^{{\rm h}}$35$^{{\rm m}}$08\fs13433 and Dec(J2000)=$-$07\degr35\arcmin04\farcs2806,  the position of the brightest spot at DOY: 270.} 
\label{fig:vlba3meth12}
\end{figure*}

\begin{figure*}[ht!]
\centering
\includegraphics[width=\textwidth]{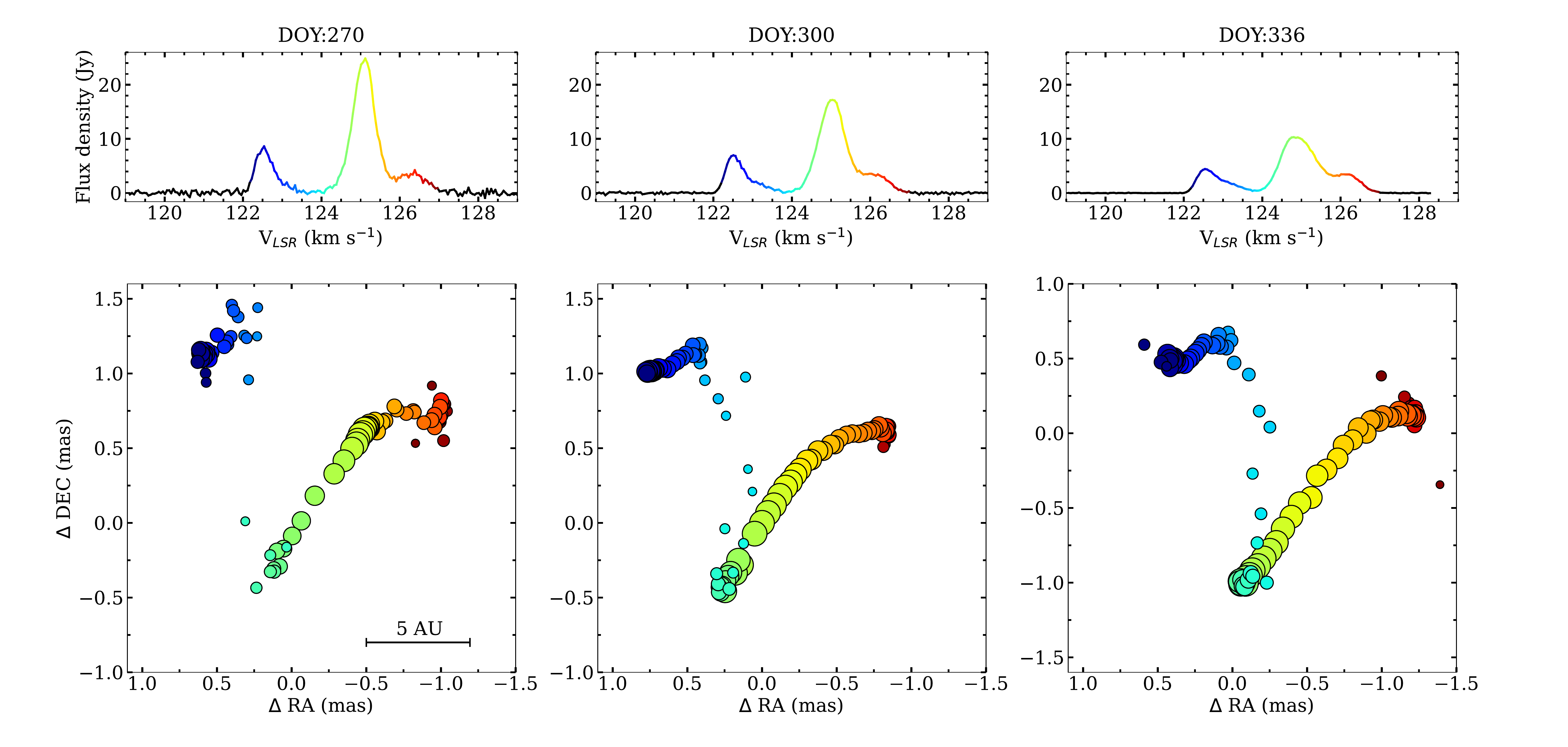}
\caption{Same as Fig.~\ref{fig:vlba3meth12}, but for 22.2~GHz water masers. The point (0,0) corresponds to RA(J2000)=18$^{{\rm h}}$35$^{{\rm m}}$08\fs13231 and Dec(J2000)=$-$07\degr35\arcmin04\farcs1447, the position of the brightest spot at DOY: 300. Coordinates for VLBA DOY: 270 are from VLBA DOY: 300 (see Sect. 2.2.).}
\label{fig:vlba3water}
\end{figure*}


Figure~\ref{fig:sixepochs} presents the 6.7~GHz methanol emission derived from all our EVN, VLBA, and LBA observations, including the archival EVN data from 2009 \citep{bartkiewicz2016}. We note that the three masing regions seen in 2009 have persisted for more than 10~yr and that the overall distribution of all 6.7\,GHz clusters was stable during the 2019 flare, at least for the 66 days of our interferometric observations. The position angle of the whole structure has also been constant over the last 10 years with the values from 70$^{\circ}$ to 77$^{\circ}$. The following analysis of the emission at the mas scale indicates that the brightness and size of individual cloudlets varied in phase with the flare profile.

In Table \ref{tablegaussmeth6} we list the following parameters of the 15 maser cloudlets detected in 2019: the LSR velocity of the brightest spot ($V_{\mathrm p}$) and the projected area (i.e., the smallest rectangle that covers the whole emission, $\Delta$RA$\times\Delta$Dec); if the Gaussian fit of the emission spectrum is successful, the fitted peak velocity $(V_{\mathrm {fit}}$),  FWHM,  and amplitude ($S_{\mathrm {fit}}$). The Gaussian fits to velocity profiles of single maser cloudlets are presented in Fig.~\ref{fig:gaussmeth6}.
 In addition, we calculated the values of the correlation coefficients of the linear fits to the spot positions on the sky plane ($r_{\mathrm s}$) and to the LSR velocity change with position along the major axis of the spot distribution ($r_{\mathrm v}$), as defined in \cite{moscadelli2011}. We used spots with a brightness above 15$\times\sigma_{rms}$ at 6.7\,GHz and 20$\times\sigma_{rms}$ at 12.2\,GHz and 22.2\,GHz to calculate the above  coefficients. In Table~\ref{tablegaussmeth6} we present values when $r_s \ge 0.5$ and $r_v \ge 0.7$, meaning that the correlations are reliable. In such cases, we calculated the position angle (PA) of the major axis  of the spot distribution using the least-squares fit and defined positive from N to E, and the velocity gradient ($V_{grad}$) measured as the maximum difference in velocity between spots divided by their distance. Similar to \citet{moscadelli2011}, we neglect the sign of the gradient for analysis.
 
\begin{table}
\centering
\caption{Number of cloudlets and FWHM mean value for each epoch of VLBA observations.}
\label{table3}
\begin{tabular}{lcc}
\hline
DOY & $N/N_\mathrm{0}$ & mean FWHM  \\
   &    & (\kms) \\
\hline
{\bf 6.7~GHz}\\
270 & 7/15  &  0.24$\pm$0.04 \\
300 & 13/14 &  0.18$\pm$0.02 \\
336 & 10/11 &  0.20$\pm$0.02 \\
{\bf 12.2~GHz}\\
270 & 2/3 &  0.18$\pm$0.01 \\
300 & 3/3 &  0.18$\pm$0.01 \\
336 & 1/1 &  0.17          \\
{\bf 22.2~GHz}\\
270 & 2/2 &  0.35$\pm$0.01 \\
300 & 2/2 &  0.40$\pm$0.01 \\
336 & 2/2 &  0.43$\pm$0.01 \\
\hline
\end{tabular}
\label{table:par_cloudlets_vlba}
\end{table}

\begin{figure}[ht!]
\begin{center}
\includegraphics[width=0.80\columnwidth]{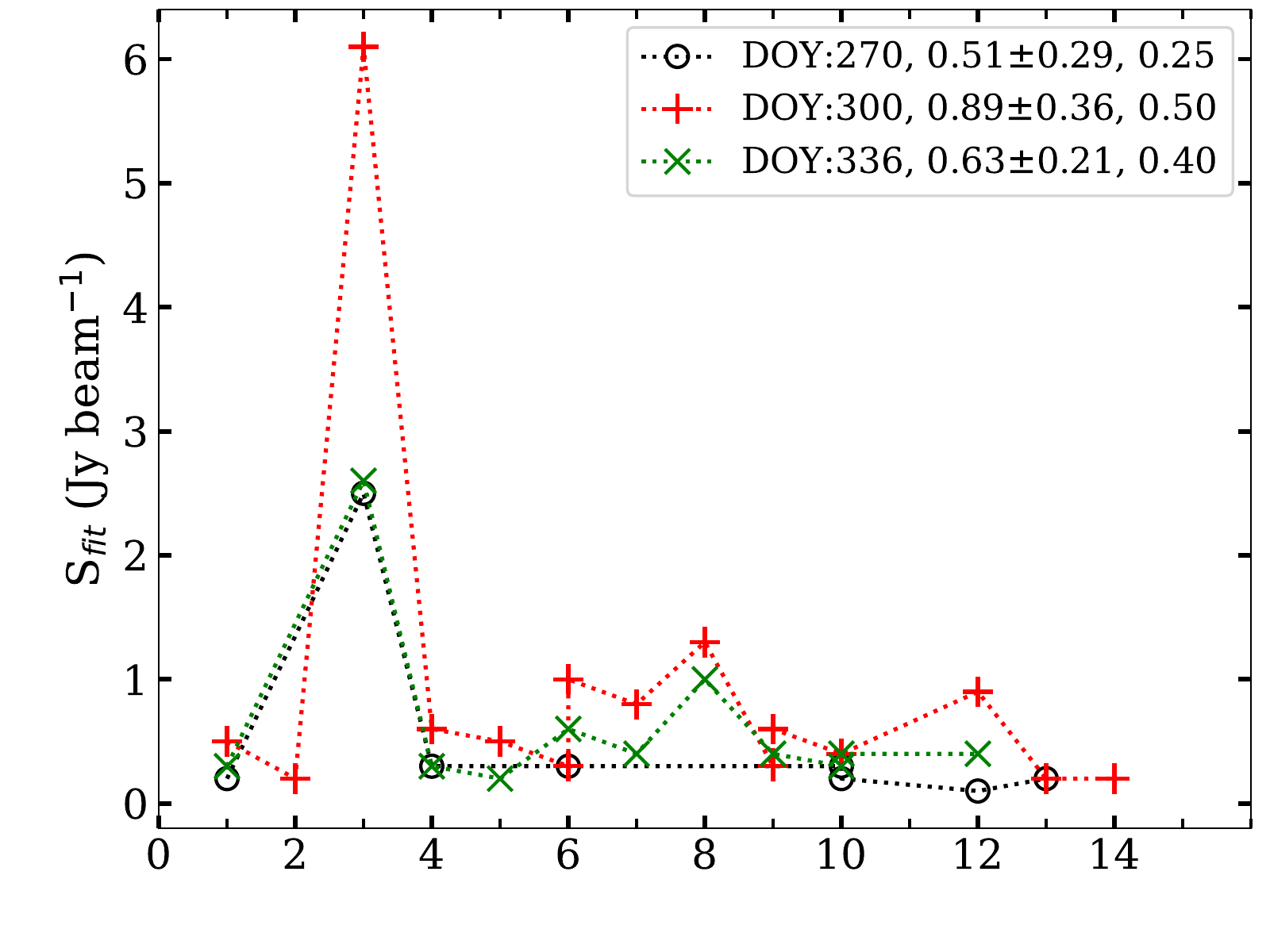}
\includegraphics[width=0.80\columnwidth]{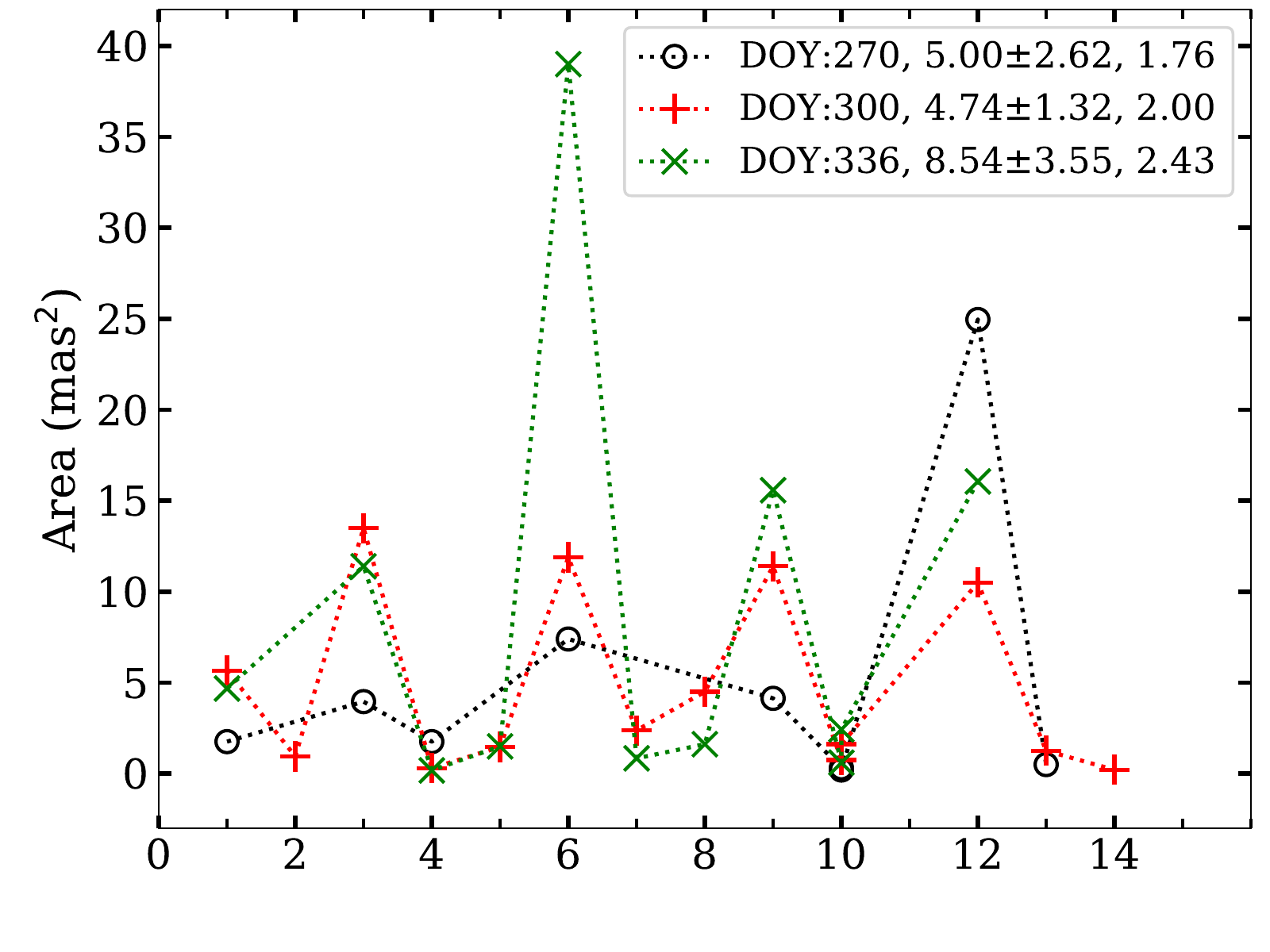}
\includegraphics[width=0.80\columnwidth]{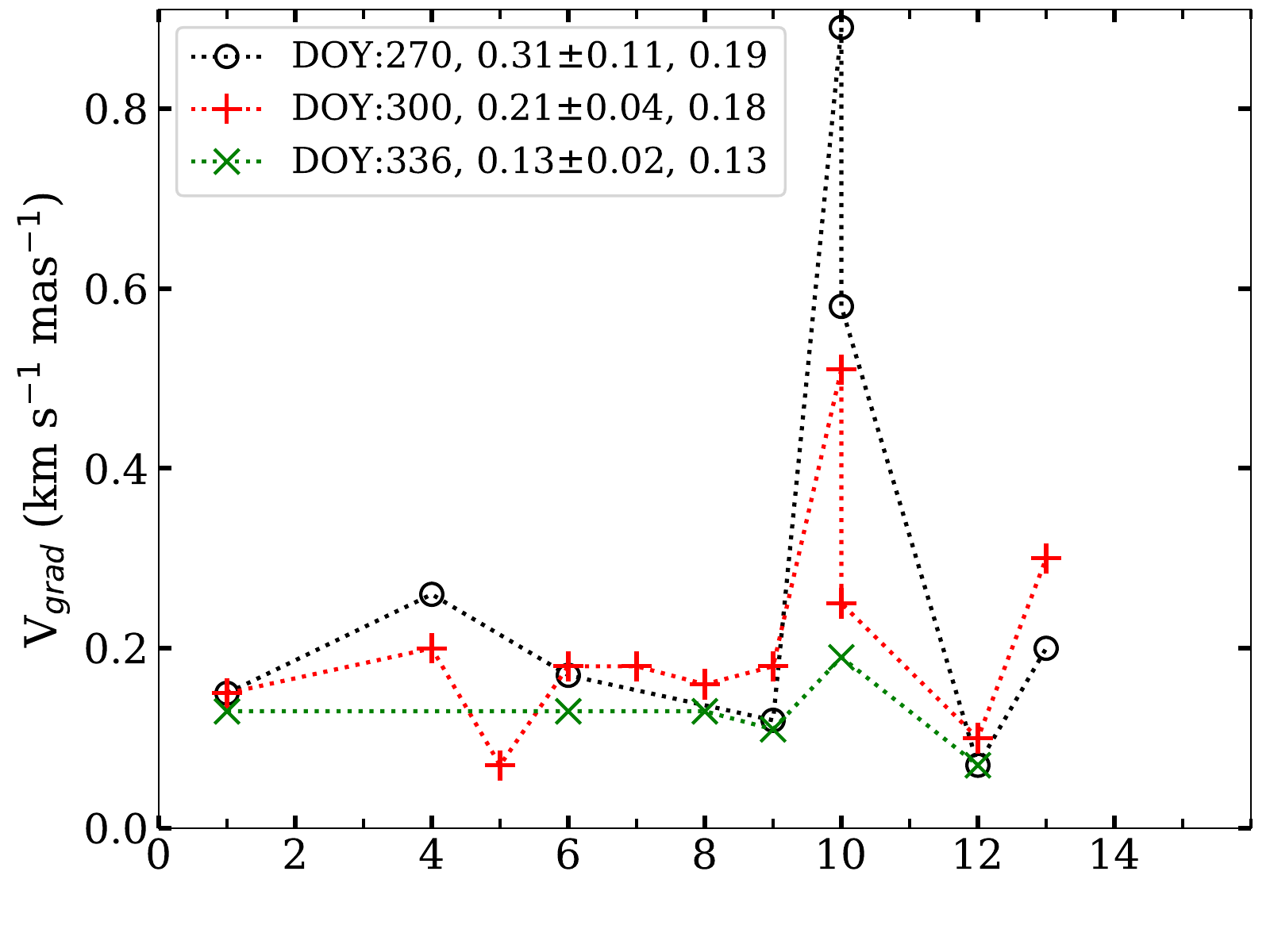}
\includegraphics[width=0.80\columnwidth]{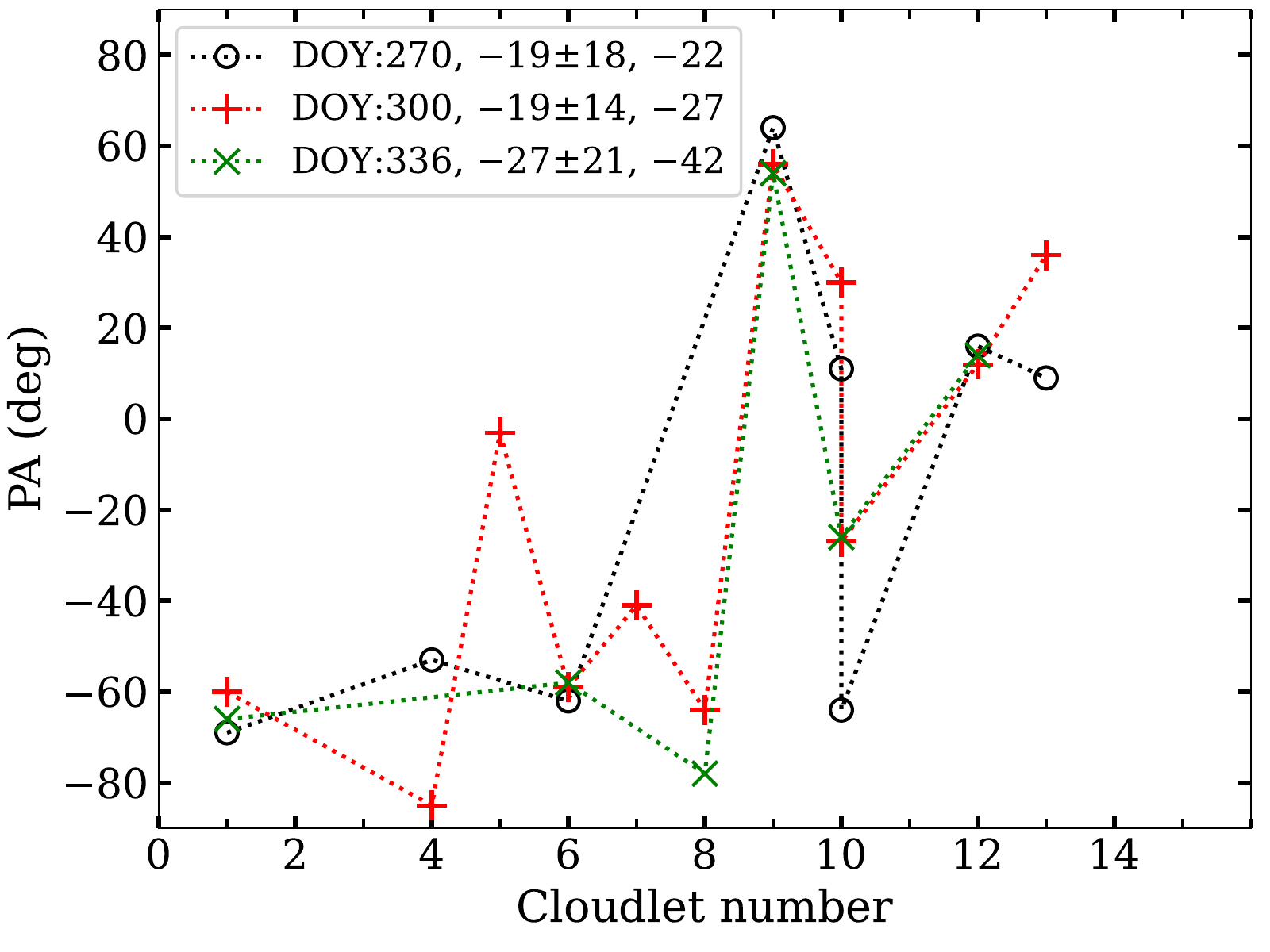}
\caption{Evolution of the VLBA 6.7~GHz methanol maser cloudlets listed in Table~\ref{tablegaussmeth6}. The mean of each parameter together with the standard error and the median value are given for each epoch.}
\label{fig:histovlba}
\end{center}
\end{figure}

We used the VLBA data to analyze the 6.7~GHz methanol cloudlet behavior during the flare, and the results are summarized in Table~\ref{table:par_cloudlets_vlba} and Figure~\ref{fig:histovlba}. Table~\ref{table:par_cloudlets_vlba} lists the total number of cloudlets ($N_\mathrm{0}$), the number of cloudlets with single- or multiple-Gaussian velocity profiles ($N$), and the mean value of FWHM.

From the above analysis, we note that on DOY: 270, six cloudlets show single-Gaussian velocity profiles, and one shows a double-Gaussian velocity profile (Cloudlet~10). 
On DOY: 300, ten cloudlets show a single-Gaussian. Cloudlets 6, 9, and 10 show double-Gaussian profiles, while Cloudlet~11 is not detected. 
On DOY: 336, nine cloudlets show single and Cloudlet 10 shows double-Gaussian profiles. Cloudlets~13--15 are no longer detected (Table~\ref{tablegaussmeth6}). The  FWHM values are constant on a timescale of $\sim$2 months during the flare. The VLBA data indicate that cloudlets 11, 13, and 14 are among the most variable, and they are located closest to the central object (Fig.\ref{fig:multiobs}), whereas the corresponding single-dish features showed a double profile during the 2011 flare (Fig.\ref{fig:light_curve_g24}).

The typical projected size of an individual cloudlet (i.e., the diagonal of the area) is 3--4~mas (20--30\,au) (Table\,\ref{tablegaussmeth6}), but there are more extended structures; for example, Cloudlet~12 is $\sim$10~mas (70\,au) long. The cloudlets change their brightness and sometimes their size, while their velocity gradients are roughly constant (Fig.~\ref{fig:histovlba}). In general, the cloudlets are the brightest at DOY: 300; this means that their $S_\mathrm{fit}$ were 60\%\ and 40\% greater than those at DOY: 270 and 336, respectively (Fig.~\ref{fig:histovlba}). The sizes of the majority of the cloudlets are constant within the measurement accuracy, but Cloudlets 1, 6, 9, and 10 exhibit a gradual increase during the flare of $\sim$42\%, $\sim$55\%, $\sim$48\%, and $\sim$78\%, respectively, while Cloudlet 12 shrinks by $\sim$23\% (Fig.~\ref{fig:histovlba}). The swelling cloudlets do not change their brightness and tend to lie farther from the central object. 

The correlation coefficients of the linear fit to the spot positions on the sky plane and to the LSR velocity variation with position ($r_\mathrm{s}, r_\mathrm{v}$) are listed in Table~\ref{tablegaussmeth6}. The cloudlets exhibit commonly linear or elongated structures with high correlation coefficients in all the maps with the three instruments and are remarkably stable over two months during the flares. The maps of sample cloudlets are shown in Figure~\ref{fig:fitmeth67}. Cloudlet\,3 represents a case of arched morphology{\bf{,}} the best resolved with VLBA at DOY:300. Cloudlet\,4 has a barely resolved structure resulting in uncertain estimates of correlation coefficients. For all the cloudlets we note the effect of probing more extending emission by the EVN and LBA compared to the VLBA. We conclude that the morphology of 6.7\,GHz maser cloudlets is generally preserved on a two-month timescale during the flare.  

\subsubsection{12.2~GHz}
A similar analysis of cloudlet parameters was done for the 12.2~GHz methanol transition during the 6.7~GHz flare ({Fig.\,\ref{fig:gaussmeth12}}, Table\,\ref{tablegaussmeth12}). We identified three 12.2\,GHz cloudlets, which are numbered from south to north (Fig.\,\ref{fig:multiobs}). A Gaussian profile is successfully fitted to each cloudlet at all the epochs, except for Cloudlet\,1 at DOY: 270. The mean and median FWHMs are similar to the FWHMs of the 6.7\,GHz profiles (Table~\ref{table:par_cloudlets_vlba}). The projected size of the cloudlet is typically 1.5\,mas (10\,au), which is comparable with the beam size, and a linear morphology is usually  visible. The cloudlets display considerable variability; Cloudlets 1 and 2 are detected only at DOY: 270 and DOY: 300, whereas Cloudlet 3 is present at all three epochs (Table~\ref{tablegaussmeth12}, Fig.\,\ref{fig:fit12group}). The intensity and size of Cloudlet 3 decreased at the last epoch (DOY: 336) by a factor of two compared to the values measured at the first two epochs.

\subsubsection{22.2~GHz}
The parameters of the 22.2~GHz water maser cloudlets are presented in Figs.\,\ref{fig:gausswater} and \ref{fig:fit22group} and in Table~\ref{tablegausswater}. Two cloudlets identified in the 22.2~GHz water maser line occurred at all three epochs and always showed  double-Gaussian velocity profiles. The profiles are broader than the methanol line profiles (Table~\ref{table:par_cloudlets_vlba}). We note a systematic increase in the brightness of all the profiles by a factor of 1.3--3.2 during the VLBA observations. The structure of Cloudlet\,1 (southern) elongated at PA of about $-$45\degr\, is stable, and  shows a clear velocity gradient of $\sim$2~km~s$^{-1}$~mas$^{-1}$ ($\sim$0.3~km~s$^{-1}$~au$^{-1}$). Furthermore, it exhibits a velocity drift of $-$0.17\kms\,month$^{-1}$. Cloudlet\,2 (northern) is complex, and its size increased by a factor of 2 over the 66~days (Fig.~\ref{fig:fit22group}). 

\subsection{Long-term variability}
\begin{figure*}[h!]
\centering
\includegraphics[width=0.92\textwidth]{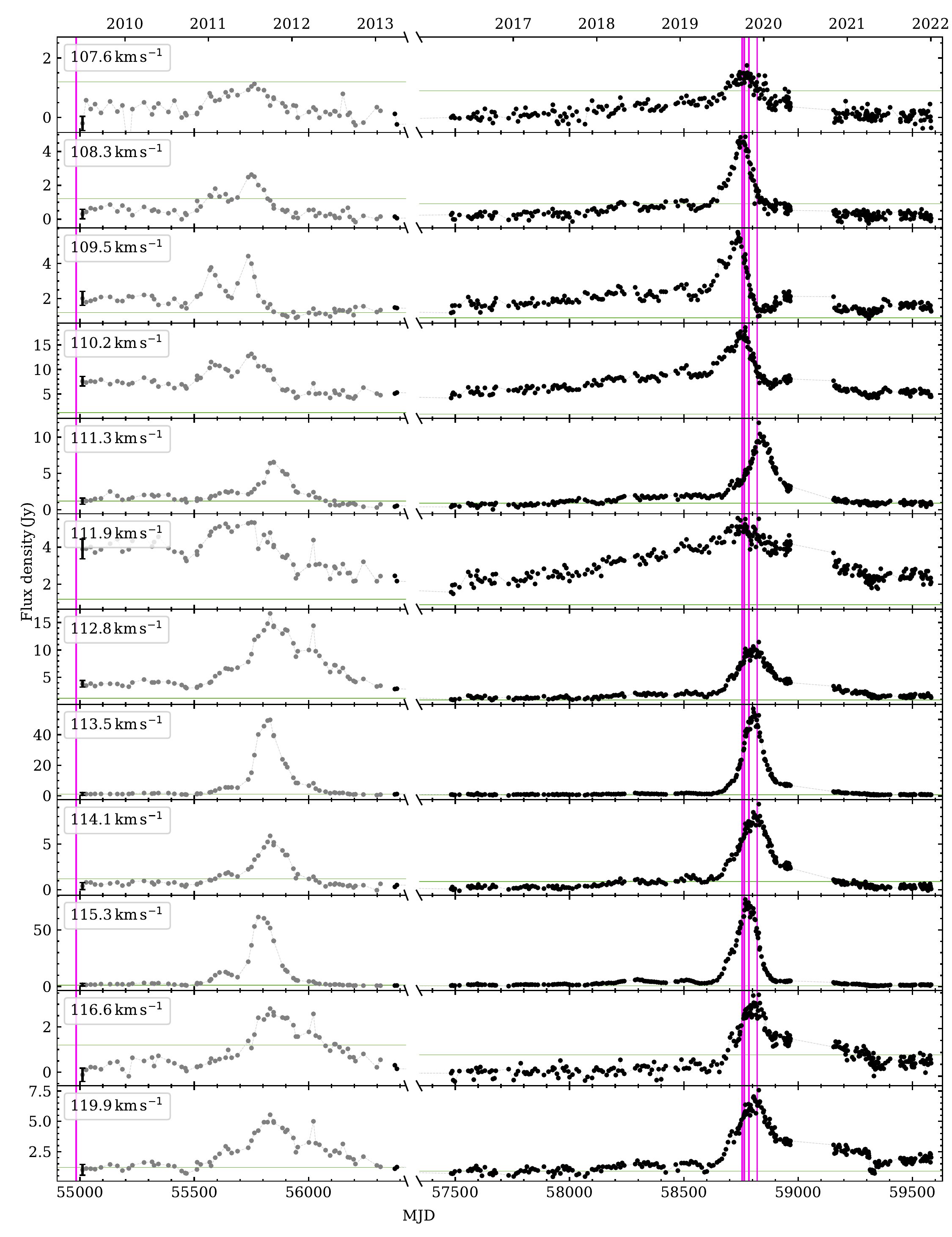}
\caption{Light curves of the 6.7\,GHz methanol maser features. The typical 3$\sigma$ noise level is shown as the green horizontal line. The observational data from before March 2013 (gray dots) were published in \citet{szymczak2018}. The typical measurement uncertainty is shown by the bar for the first data point of each curve. No observations were made in the period of 56387$<$MJD$<$57482. The vertical magenta lines indicate the epochs of the VLBI projects presented in this paper.
 \label{fig:light_curve_g24}}
\end{figure*}

\begin{figure}[h!]
\centering
\includegraphics[width=0.5\textwidth]{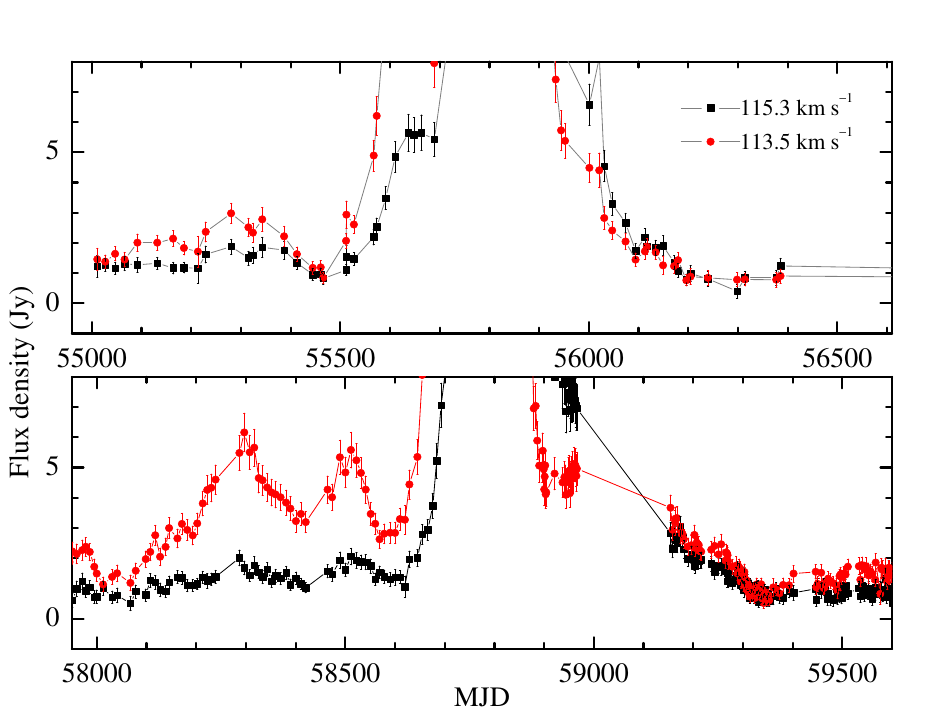}
\caption{Magnifed images of selected parts of light curves of the two strongest features highlighting an oscillating flux variation before the flares.}
 \label{fig:g24pre-flare-var}
\end{figure}

In the 6.7\,GHz maser single-dish data, there are four intermittent velocity features (107.6, 108.3, 114.1, and 116.6\kms) whose intensity was above the 3$\sigma$ detection limit for specific time intervals. The light curves of these four intermittent and eight persistent spectral features of the 6.7~GHz methanol maser are presented in Fig.~\ref{fig:light_curve_g24}. The overall picture of variability is shown in Fig.~\ref{fig:g24_dynamic_spectrum_by_md}. 
The major flares of the strongest features at 113.5 and 115.3\kms\, are preceded by small variations in amplitude (Fig.\,\ref{fig:g24pre-flare-var}). The periodogram of the light curve of 115.3\kms\, calculated for the time interval of MJD 57482$-$58620 implies a periodicity of six\,months. The pre-flare variability is synchronized for all features, but differs significantly between the 2011 and 2019 events. This may be related to pulsation of powering HMYSO postulated by \cite{inayoshi2013}. The general flare profiles of strong features appear to be similar in the two flare events, which peaked in August 2011 and November 2019, in contrast to most of the weak features. Considering the brightest feature at the LSR velocity of 115.3\kms, we estimate that the 2019 flare lasted 280$\pm$3~days. 

\begin{figure}
\includegraphics[width=0.5\textwidth]{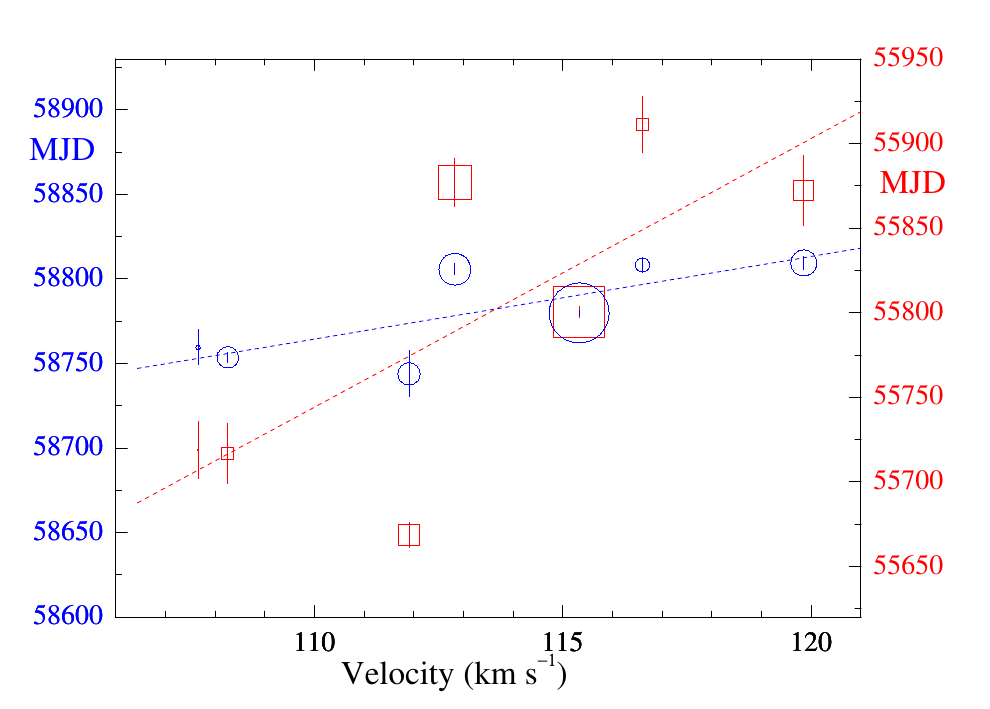}
\caption{Dates of the flux density maxima of the maser spectral features at 6.7\,GHz at the first (red) and second (blue) flare events in 2011 and 2019, respectively.  The symbol sizes are proportional to the logarithm of the flux density and the vertical bars indicate the fivefold standard error of the peak flare time maxima. The dashed lines denote the best fits to the data. \label{fig:g24p33delay}}
\end{figure}

The time of the flux maximum was calculated by fitting a skewed Gaussian function using the method \href{https://docs.scipy.org/doc/scipy/reference/generated/scipy.optimize.curve_fit.html?highlight=curve_fit}{CURVE\_FIT} from the \href{https://docs.scipy.org/doc/scipy/reference/optimize.html?highlight=optimize#module-scipy.optimize}{SCIPY.OPTIMIZE} package \citep{scipyPaper}, for the features shown in Fig.~\ref{fig:light_curve_g24}, except for those with double peaks (108.3, 109.5, and 110.2\,\kms). The average flux density-weighted interval between the two flares is 2982$\pm12$\,d. We   checked whether each feature in the single-dish spectrum has a counterpart in the cross-correlation spectrum and a unique cloudlet in the VLBI map. It appeared that the emission of eight features (119.9, 116.6, 115.3, 112.8, 111.9, 109.5, 108.3, and 107.6\,\kms) in the autocorrelation spectrum unequivocally emerge from Cloudlets 1--3, 9, 10, 13, 14, and 15 (Fig.\,\ref{fig:multiobs}). 
The remaining single-dish features are the effect of emission blending from spatially different cloudlets. There are two features at 111.23 and 114.1\,\kms\, that were not detected with VLBI, indicating that their emission comes from extended structures resolved by the VLBI beam. The results of time lag estimates are summarized in Figure\,\ref{fig:g24p33delay}. For the 2019 flare, the time lag of the flare peaks of the features at two extreme velocities of 107.6 and 119.9\kms\, is 59.5$\pm$4.2\,d. For the 2011 flare, the time lag estimates are highly scattered and the average lag is almost a factor of three higher. The first flare was probed with low cadence and is discarded in further analysis because the errors of the time lags are higher than those for the 2019 epoch (Sect.\,\ref{sec:cloudlet-3months}).

\subsection{The Wide-field Infrared Survey Explorer (WISE) observations}
\begin{figure}
\begin{center}
\includegraphics[width=\columnwidth]{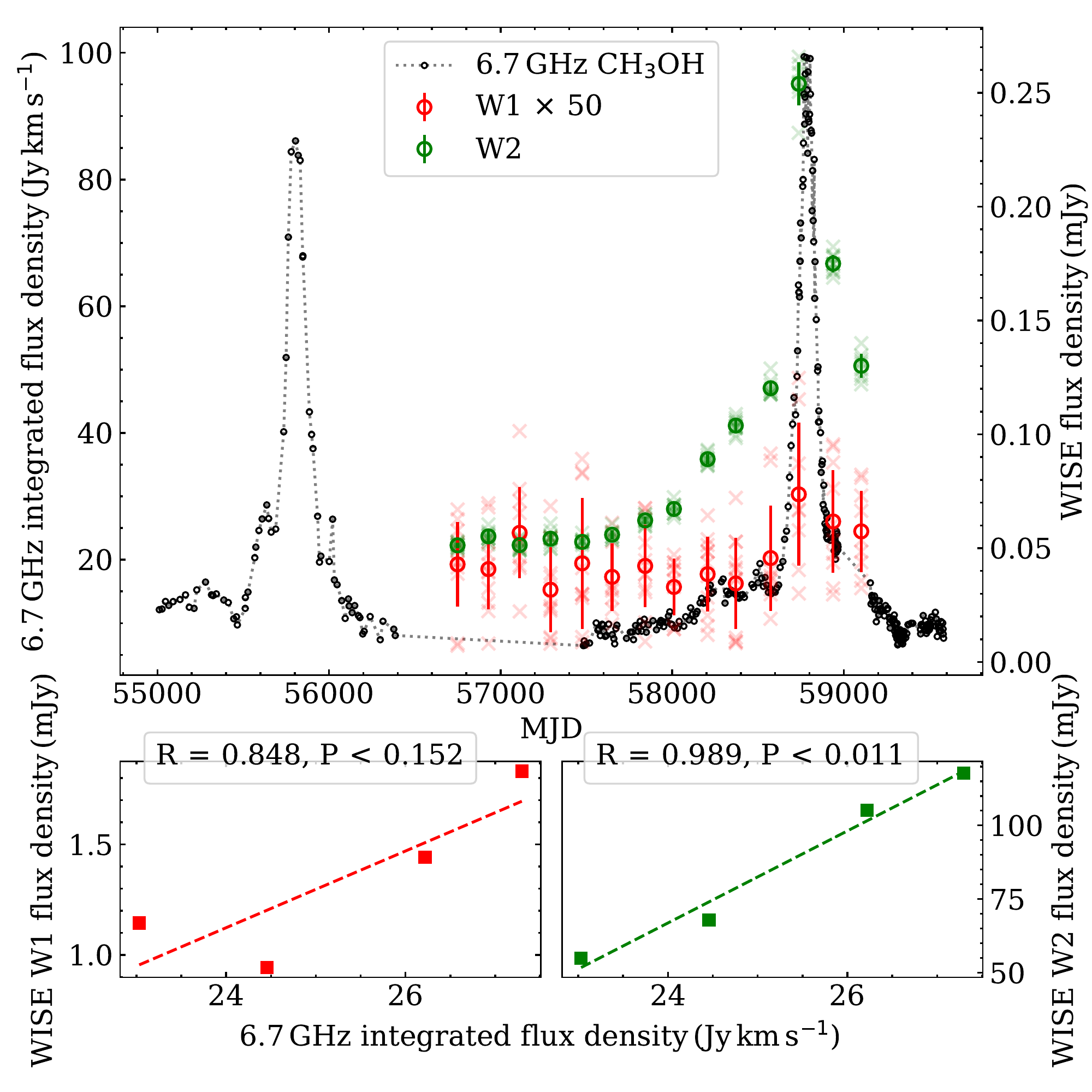}
\captionof{figure}{WISE W1 and W2 photometric measurements superimposed with 6.7\,GHz CH$_3$OH maser integrated flux density. Circles indicate the values of the WISE measurements for every epoch; transparent crosses indicate the singular WISE photometric measurements. The error bars correspond to standard deviation. Only measurements with cc\_flags of 0 (unaffected by known artifacts) or h (possible contamination by scattered light from a nearby bright source), and qual\_frame values no lower than 5 (high image quality) are shown. Two bottom panels show 6.7\,GHz integrated flux density vs WISE singular measurements. Here a stricter quality criterion was also used: only ph\_qual of A (source detected in this band with  S/N>10). Values in the boxes are Pearson correlation coefficients.
\label{fig:W2_correlation}}
\end{center}
\end{figure}

Photometry in the W1 (3.4\,$\mu$m) and W2 (4.6\,$\mu$m) bands, as well as astrometry for G24 from the WISE \citep{2010AJ....140.1868W} and subsequent NEOWISE \citep{2014ApJ...792...30M} missions, were retrieved from Infrared Science Archive{\footnote{\url{https://irsa.ipac.caltech.edu}}} (IRSA) covering observations until the end of 2020. A saturation correction was applied\footnote{See\,\,\url{http://wise2.ipac.caltech.edu/docs/release/neowise/expsup/sec2\_1civa.html}} to account for a photometric bias due to the detector warm-up. The (NEO)WISE infrared (IR) fluxes rose along with the methanol flare, as   pointed out by \citet{hirota:2022}. NEOWISE W1 and W2 photometric measurements superimposed with 6.7\,GHz CH$_3$OH maser integrated fluxes are shown in Fig.\,\ref{fig:W2_correlation}, accompanied by plots revealing the clear correlation between the maser and the IR flux for both bands. This is expected for the radiative excitation of this maser transition.

\begin{figure}
\begin{center}
\includegraphics[width=\columnwidth]{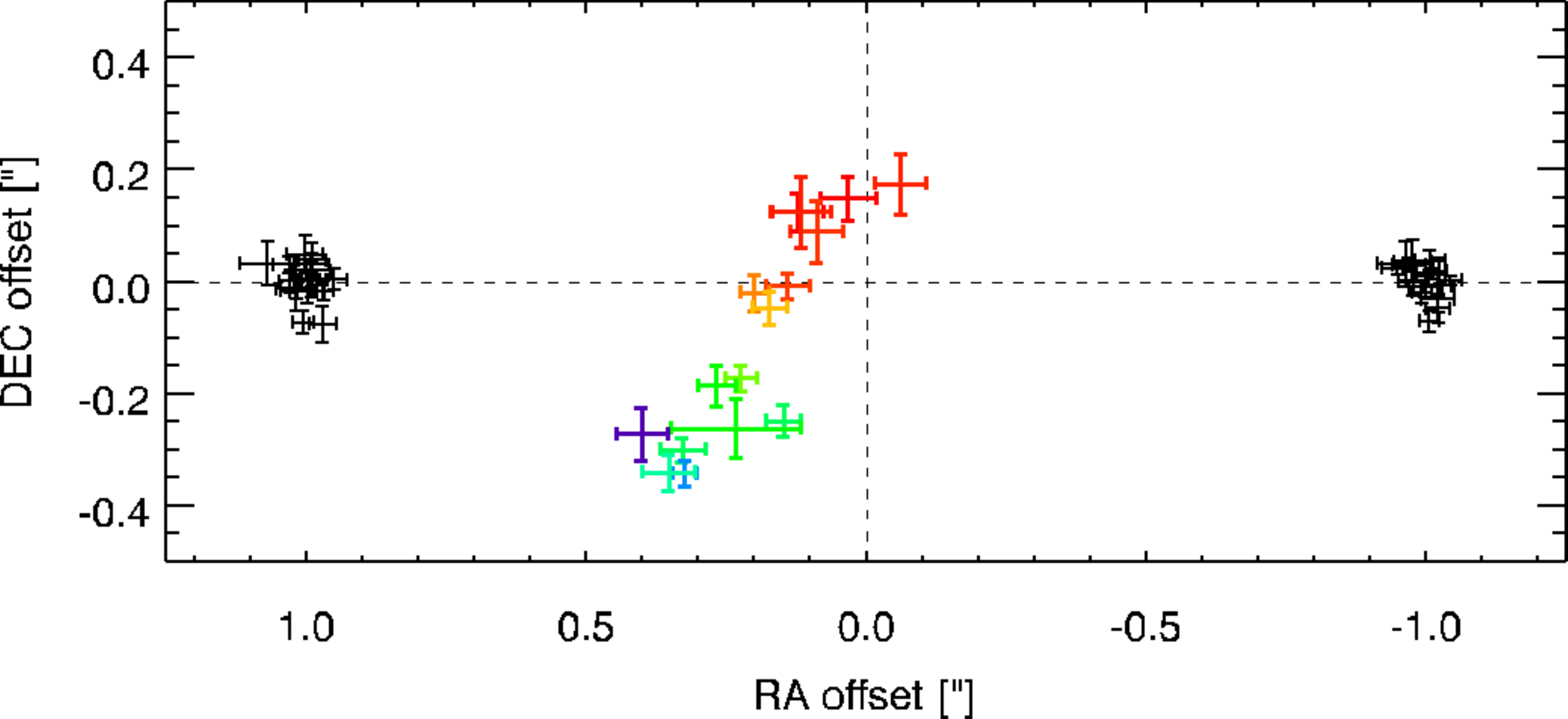}
\captionof{figure}{
Offsets from the overall mean position for each epoch for G24 (colored symbols) and two adjacent objects of similar brightness (black, shifted by $\pm$1\arcsec{} in right ascension for clarity). Colors indicate the W2 brightness (dark blue for the brightest state and red for the dimmest). The size of the symbol corresponds to the 1$\sigma$ uncertainty. The intersection of the dashed lines gives the ALMA reference position of \cite{hirota:2022}.
}
\label{fig:W2_centroid}
\end{center}
\end{figure}

In Fig.~\ref{fig:W2_centroid} we present the mean centroid positions of the 4.6\,$\mu$m emission from G24 and two adjacent objects for each epoch, spanning approximately 10.5\,yr. The centroid distribution of G24 is clearly different from that of the comparison objects. It is resolved and aligned with the outflow orientation (see Fig.~\ref{fig:g24_33_alma_vlbi}). During the dim state, the 4.6\,$\mu$m centroid is closer to the ALMA reference position, while it moves by $\sim$0\farcs5 to the southeast along the outflow direction during the brightening of the source.

\section{Discussion}
\subsection{Evidence of a disk-jet system}
\begin{figure*}
\centering
\includegraphics[width=0.99\textwidth]{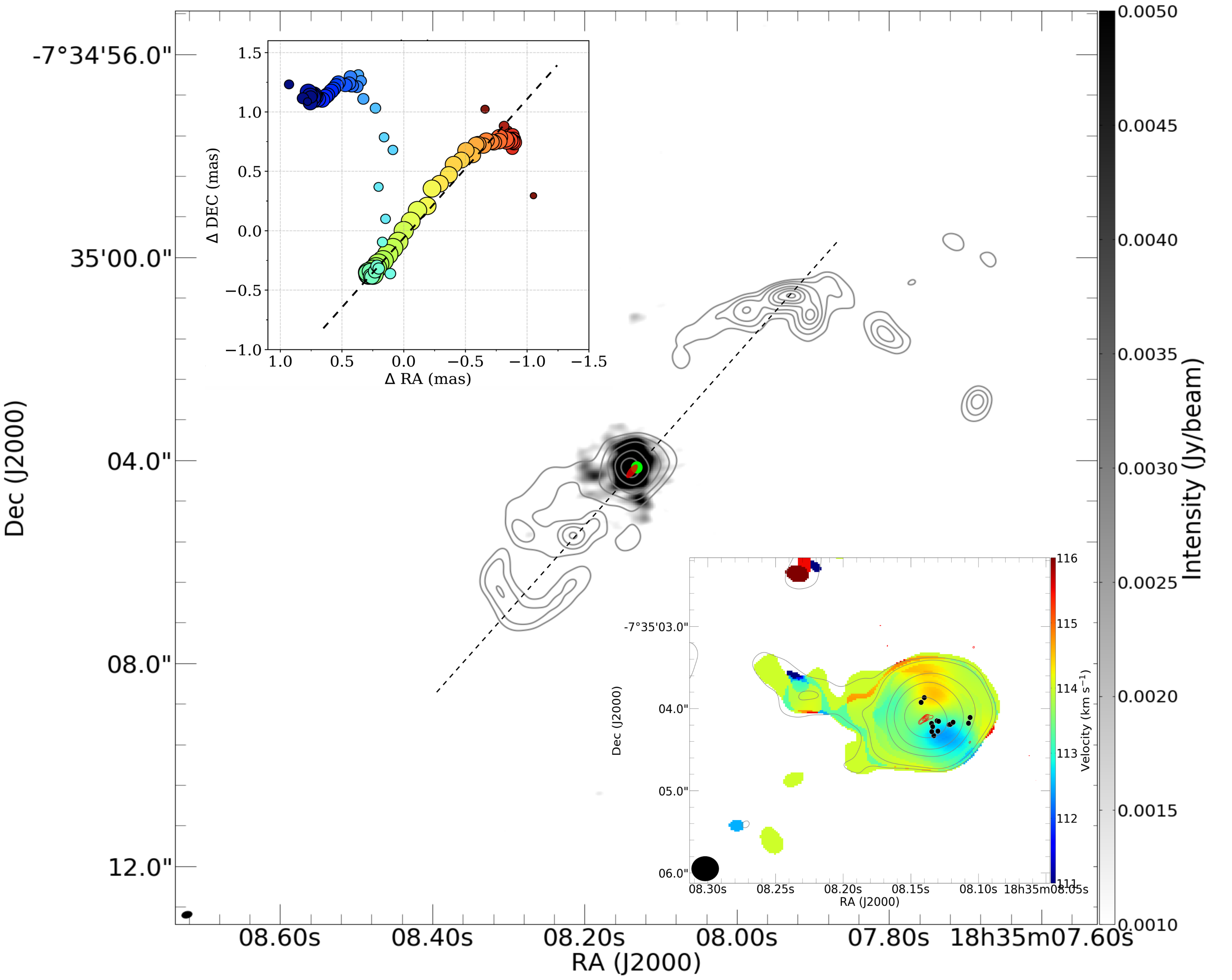}
\caption{Aligned large-scale and compact bipolar outflow. The grayscale background represents ALMA 1.3~mm dust continuum emission of G24, while the gray contours (levels: [0.2, 0.5, 1, 2, 3, 4, 5, 6] Jy\,beam$^{-1}$~km\,s$^{-1}$) represent the moment 0 (integrated intensity) map of the 229.758\,GHz class I CH$_3$OH maser line. The red ellipse (same as red contours in the bottom right inset) indicates the position of the 7\,mm~(Q-band) ionized jet reported in \citet{purser2021} and elongated along the axis of the ALMA CH$_3$OH line and the VLBA 22.2\,GHz water maser linear structure (top left inset). The green filled circle represents the position of the water masers in G24. The top left inset is the VLBA DOY: 336 epoch of the water maser (see Fig. \ref{fig:vlba3water}). The dashed lines show the axes of the large-scale (in the ALMA image) and compact (in the VLBA water maser map) outflows driven by G24. The position angle derived from the ALMA millimeter data is $\sim$136$^\circ$. The bottom right inset shows the  ALMA moment 1 map of the 229.589\,GHz CH$_3$OH line with its moment 0 maps overlaid as contours (levels: [0.04, 0.08, 0.5, 1, 2] Jy\,beam$^{-1}$~km\,s$^{-1}$). The VLBI DOY: 270 positions of the 6.7\,GHz methanol masers are represented by the black filled circles.
} 
\label{fig:g24_33_alma_vlbi}
\end{figure*}

Linear structures in the 22.2\,GHz water maser distribution around massive protostellar objects can be indicative of the presence of a bipolar outflow or jet \citep{2012ApJ...748..146C, Burns2016}. Figure~\ref{fig:g24_33_alma_vlbi} shows an overlay of the ALMA 229.758\,GHz class I methanol maser line (gray contours) on the 1.3\,mm dust continuum in grayscale (see also \citealt{hirota:2022}). The green filled circle represents the position of the 22.2~GHz water maser emission (to the west of the continuum source), while the red ellipse indicates the position of the Q-band ionized jet reported by \citet{purser2021}. The dashed line connects the prominent northwest and southeast knots of the 229.758\,GHz class I methanol maser line through the center of the dust core cutting through the ionized jet axis and the water maser position. The northwest--southeast morphology of the ALMA 229.758~GHz CH$_3$OH emission is indicative of the presence of a bipolar outflow in G24 along the axis marked by the dashed line. The inset in Fig.~\ref{fig:g24_33_alma_vlbi} shows that the 22.2~GHz water masers are aligned along the same axis (black dashed line) of the ALMA class I methanol emission. The water maser could trace either the compact inner component of the bipolar outflow or a jet structure within G24. The PA of the axis of the jet--outflow, as indicated by the dashed lines in Fig.~\ref{fig:g24_33_alma_vlbi}, is about $-$44$^{\circ}$. The PA of the axis of the jet is similar to the PAs of water maser Cloudlets~1 (see Fig.~\ref{fig:vlba3water} and Table~\ref{tablegausswater}). This could be an indication that this cloudlet is excited by the ionized jet \citep{2012ApJ...748..146C}. 
We note that some of the 6.7\,GHz methanol maser features used in the Keplerian disk model presented in Sect. 4.3 do not fit well, which could be attributed to the influence of the jet. Neglecting the weaker maser spots, cloudlets 1 and 2 have similar PAs, and the difference in the V$_{LSR}$ of the two cloudlets could be attributed to jet rotation. Precession of the jet or rotation of the outflow could be responsible for the water maser excitation and the observed differences in the V$_{LSR}$. The Q-band jet ($\sim$\,100\,mas in size) \citep{purser2021} has a knot at the northwest end. This knot is slightly south of the jet axis, and could be an indication of precession and/or rotation. 

In agreement with the above, the ALMA 229.758\,GHz CH$_3$OH emission is a maser line which traces the outflow cavity wall, while the  229.589\,GHz CH$_3$OH emission is thermal emission tracing the rotation disk structures around the protostar. The eastern elongation of the 229.758\,GHz CH$_3$OH emission could be an indication of the fragmentation of the rotating disk structure and/or spiral arm system. The position angle of the Keplerian disk is $15^{\circ}\pm3^{\circ}$, as traced by the 229.589\,GHz CH$_3$OH emission (see Fig. \ref{fig:g24_33_alma_vlbi} \citep{hirota:2022}, this is roughly perpendicular to the average of the position angles of the water masers, indicating a disk-jet--outflow system in the protostar.




\subsection{Diversity of kinematics traced by the water and methanol masers}
The water maser Cloudlet\,1 lies on the sky within 7\,mas ($\sim$50\,au) of the 6.7\,GHz methanol maser Cloudlet\,4, but the separation in the velocity is $\sim$3\,\kms. This is different from the case in G31.581$+$00.077 presented by \cite{bartkiewicz2012}, where both masers appeared at the same LSR velocity (within 0.1~\kms), but they were separated by 133~au, {\bf{(}}considering the distance to the source of 5.5~kpc; \citealt{reid2019}). Similarly \cite{darwish2020},  toward the region IRAS 19410+2336, found methanol and water masers separated by $\sim$205\,au and only by $\sim$0.17\,\kms\,in the velocity. In G24 the velocity difference implies the physical separation of the two maser species and confirms that these masers probe different parts of the environment of HMYSO, which is strongly consistent with theoretical models showing that these two maser transitions require different physical conditions for the pumping (\citealt{cragg2005,gray2022}).

Two 12.2~GHz methanol maser cloudlets coincide with the positions of 6.7~GHz methanol masers. The 12.2~GHz Cloudlet~3 and the 6.7~GHz Cloudlet~12 with the peak at the LSR velocity of $\sim$110.2~km~s$^{-1}$ coincide within 1.8~mas in RA and 1~mas in Dec in the VLBA images taken simultaneously in both transitions at three epochs. That is well within the position uncertainties of a single maser spot at each transition (Sect.~2.1.1). Moreover, they both become weaker with time (Figs.\,\ref{fig:vlba3meth12} and \ref{fig:sixepochs}). This indicates that the 6.7 and 12.2\,GHz lines originate in the same gas volume. Similarly, the 12.2~GHz Cloudlet~1 and the 6.7~GHz Cloudlet~10 with the peak at the LSR velocity of $\sim$111.9~km~s$^{-1}$ coincide within 0.5~mas in RA and in Dec in the two first VLBA epochs DOY: 270 and 300. On DOY: 336 the 12.2~GHz emission in Cloudlet 1 was not visible, with an upper limit of 30\,mJy (Fig.~\ref{fig:vlba3meth12}). The corresponding 6.7~GHz Cloudlet 10 was very weak on this date. 

\begin{figure*}
\centering 
\includegraphics[scale=0.3,trim=100 100 0   200]{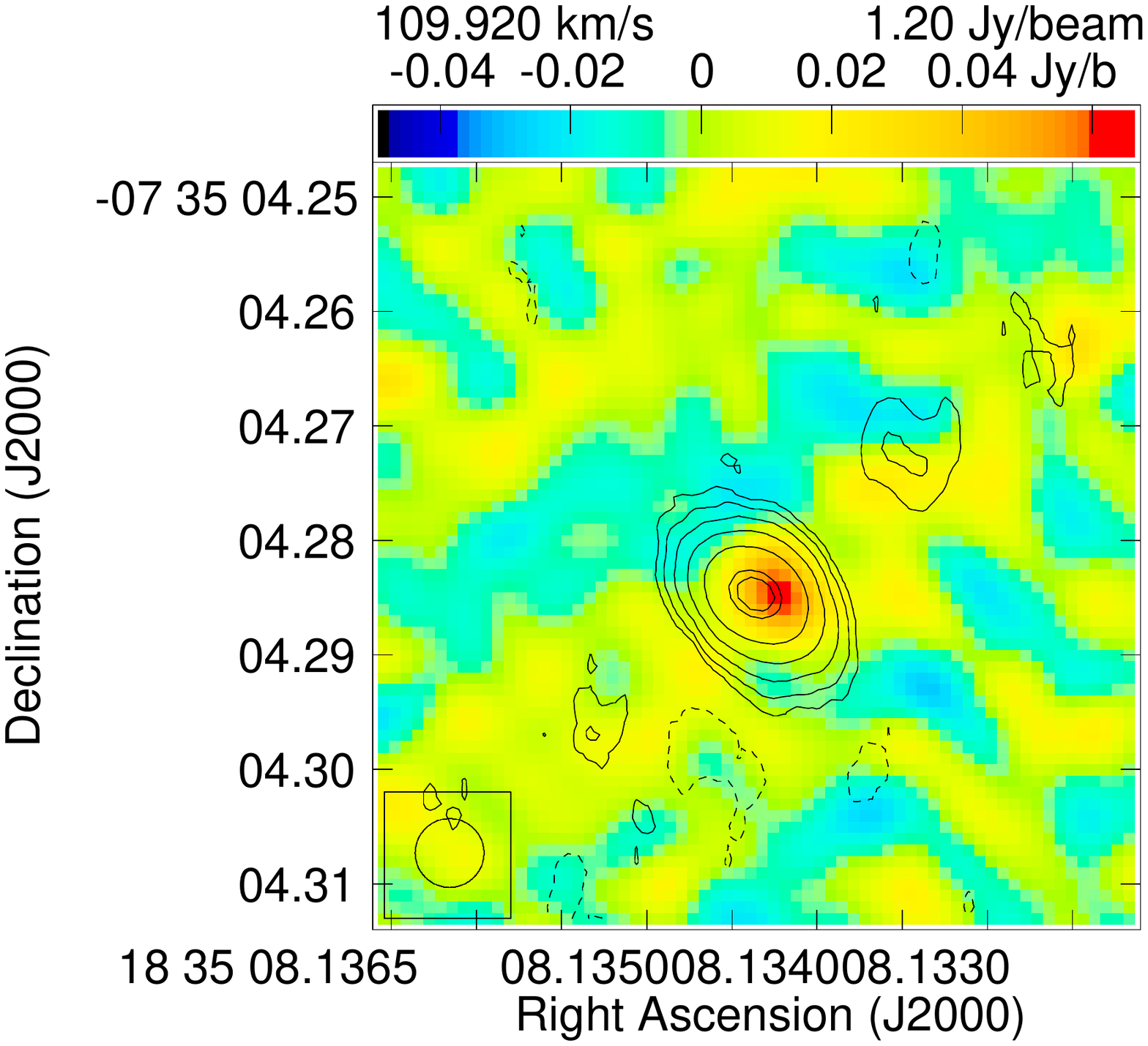}
\includegraphics[scale=0.3,trim=120 100 0 200]{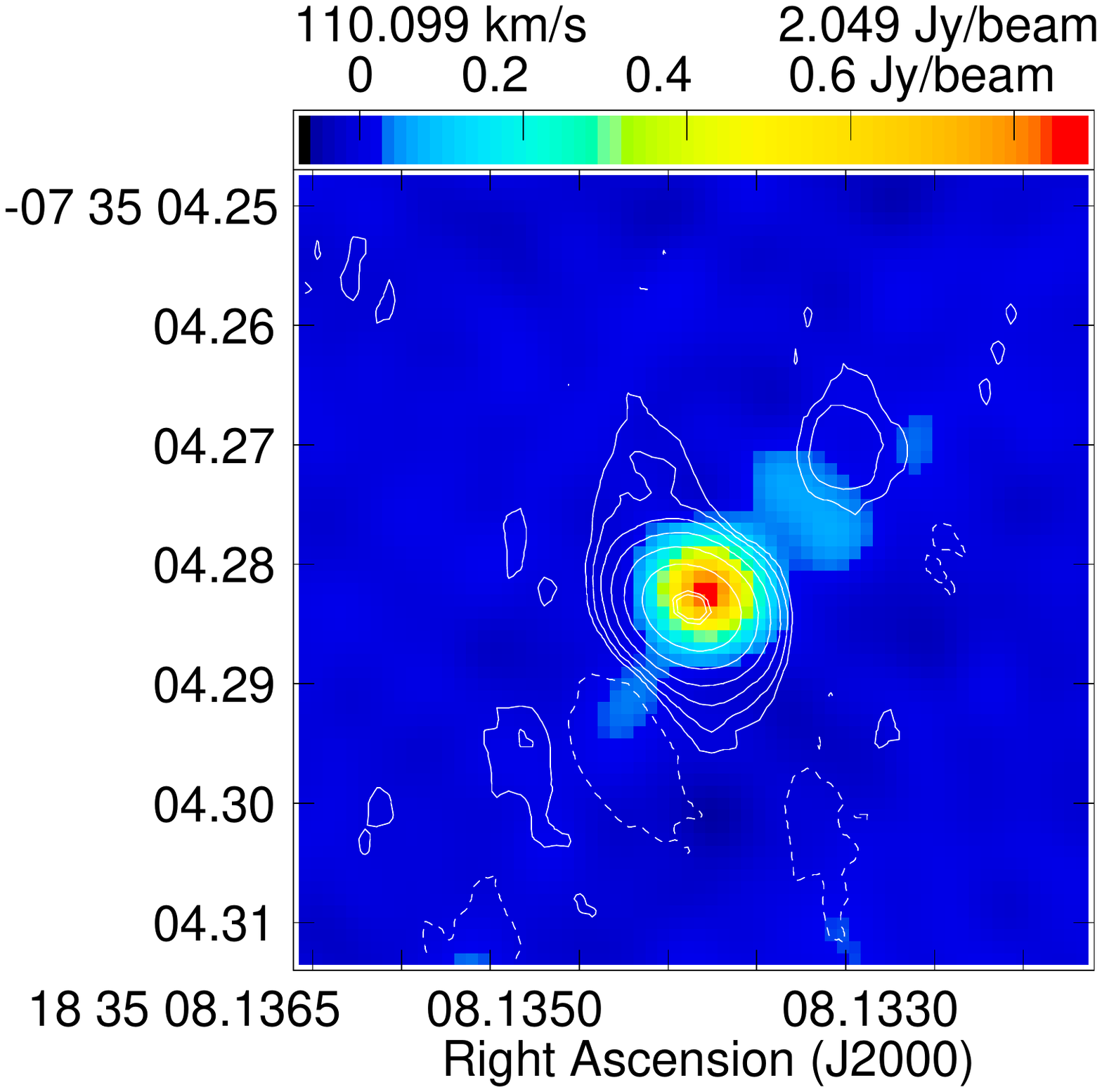}
\includegraphics[scale=0.3,trim=120 100 0 200]{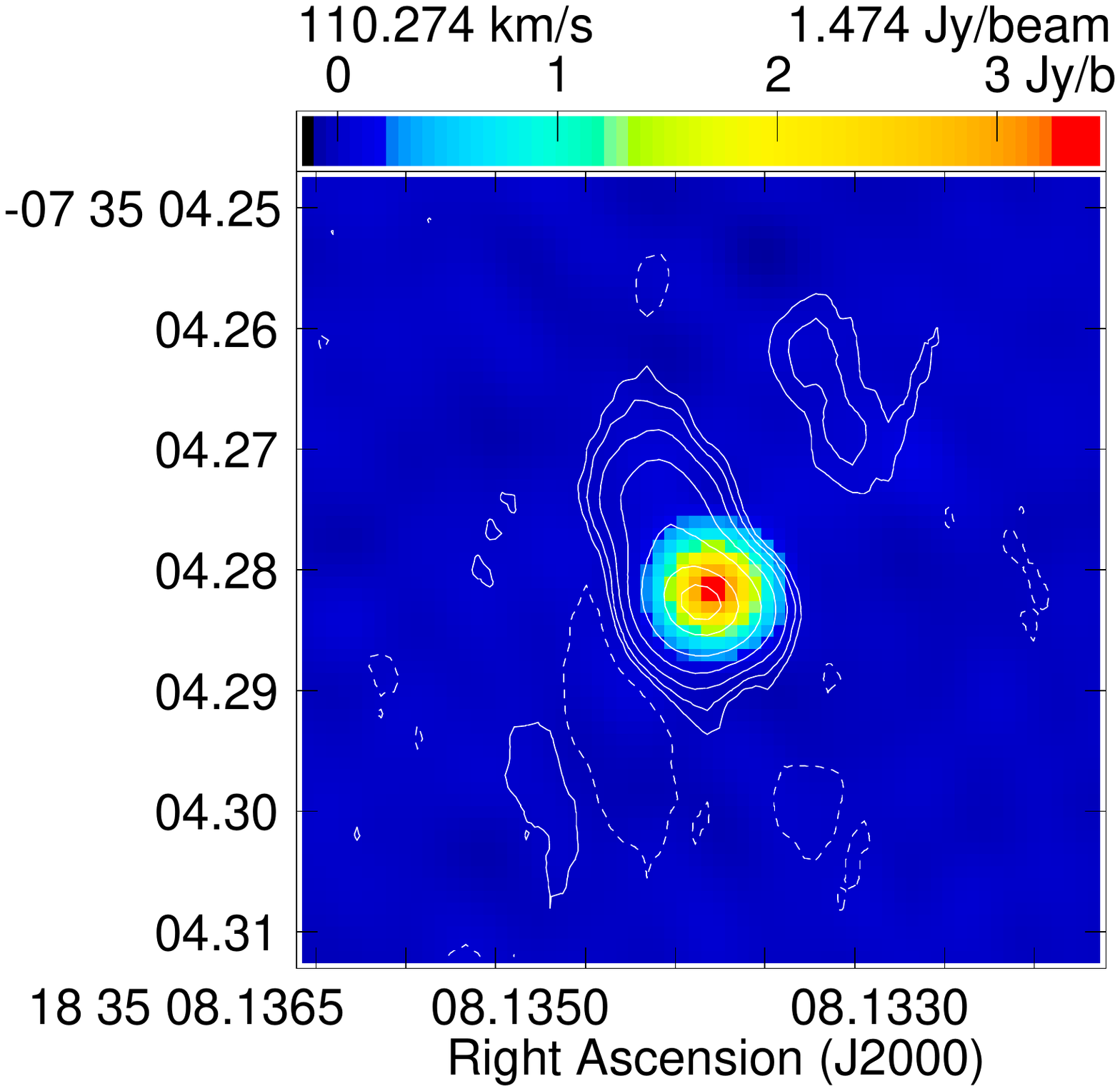}
\includegraphics[scale=0.3,trim=120 100 0 200]{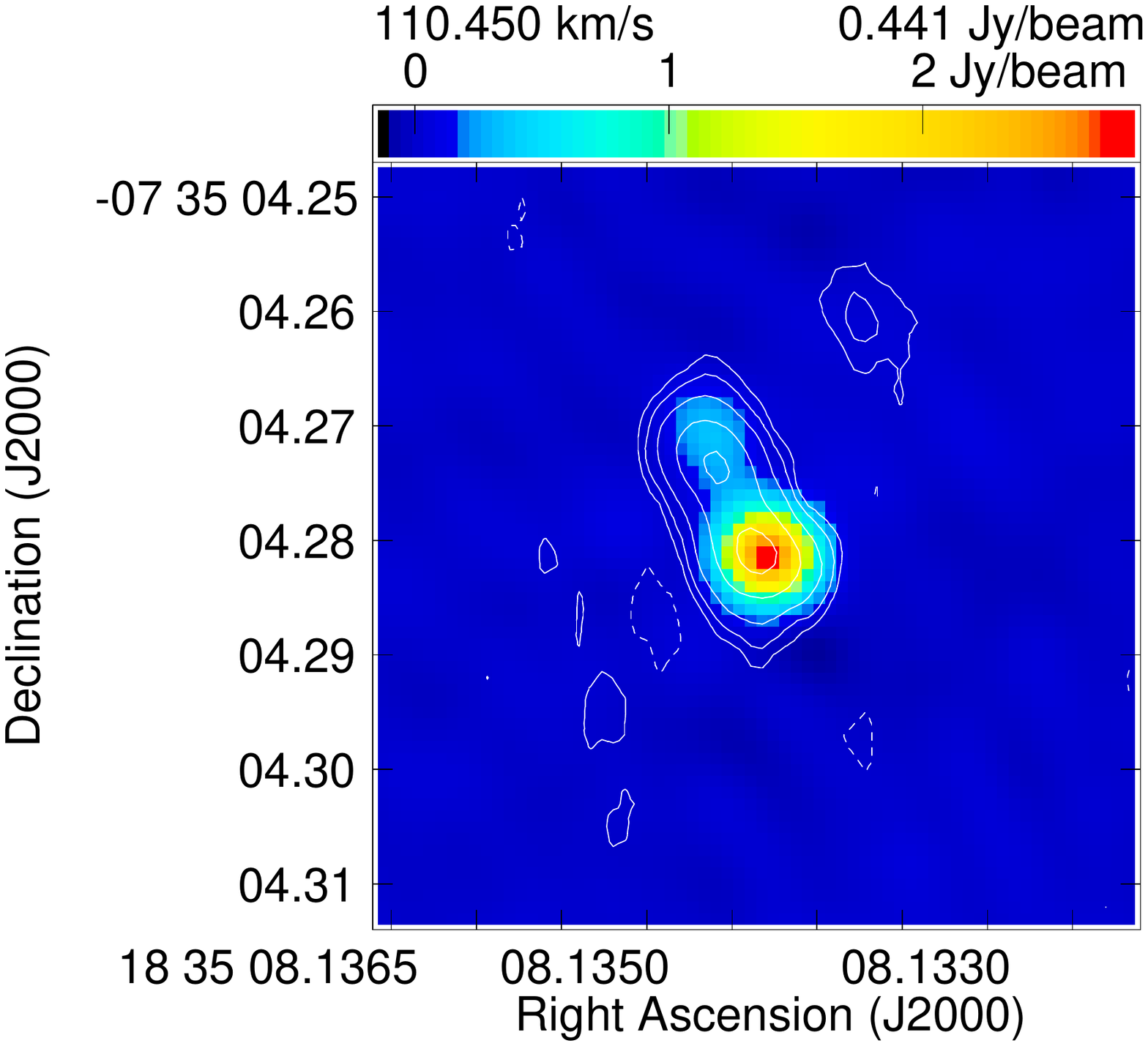}
\includegraphics[scale=0.3,trim=120 100 0 200]{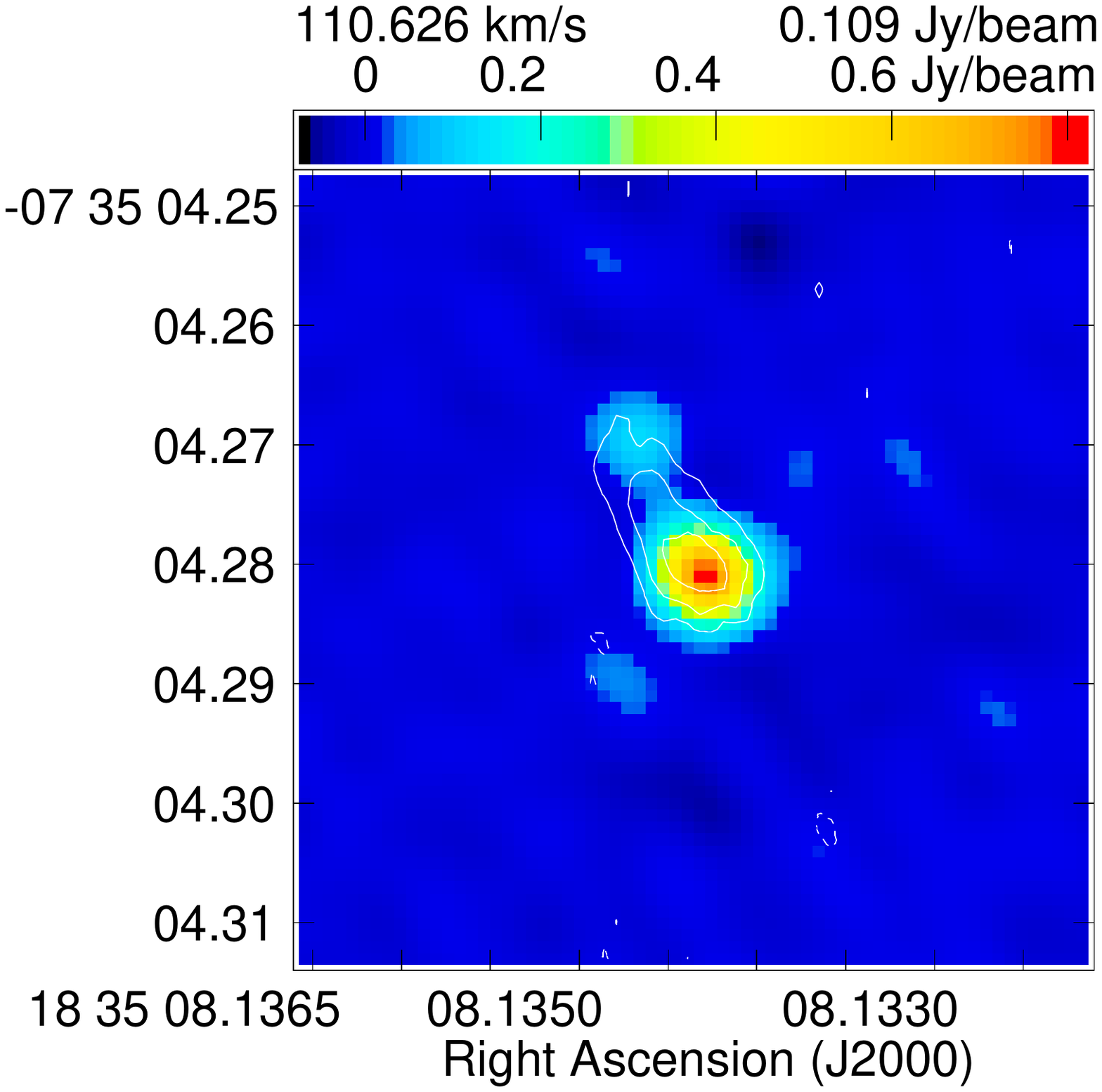}
\includegraphics[scale=0.3,trim=120 100 0 200]{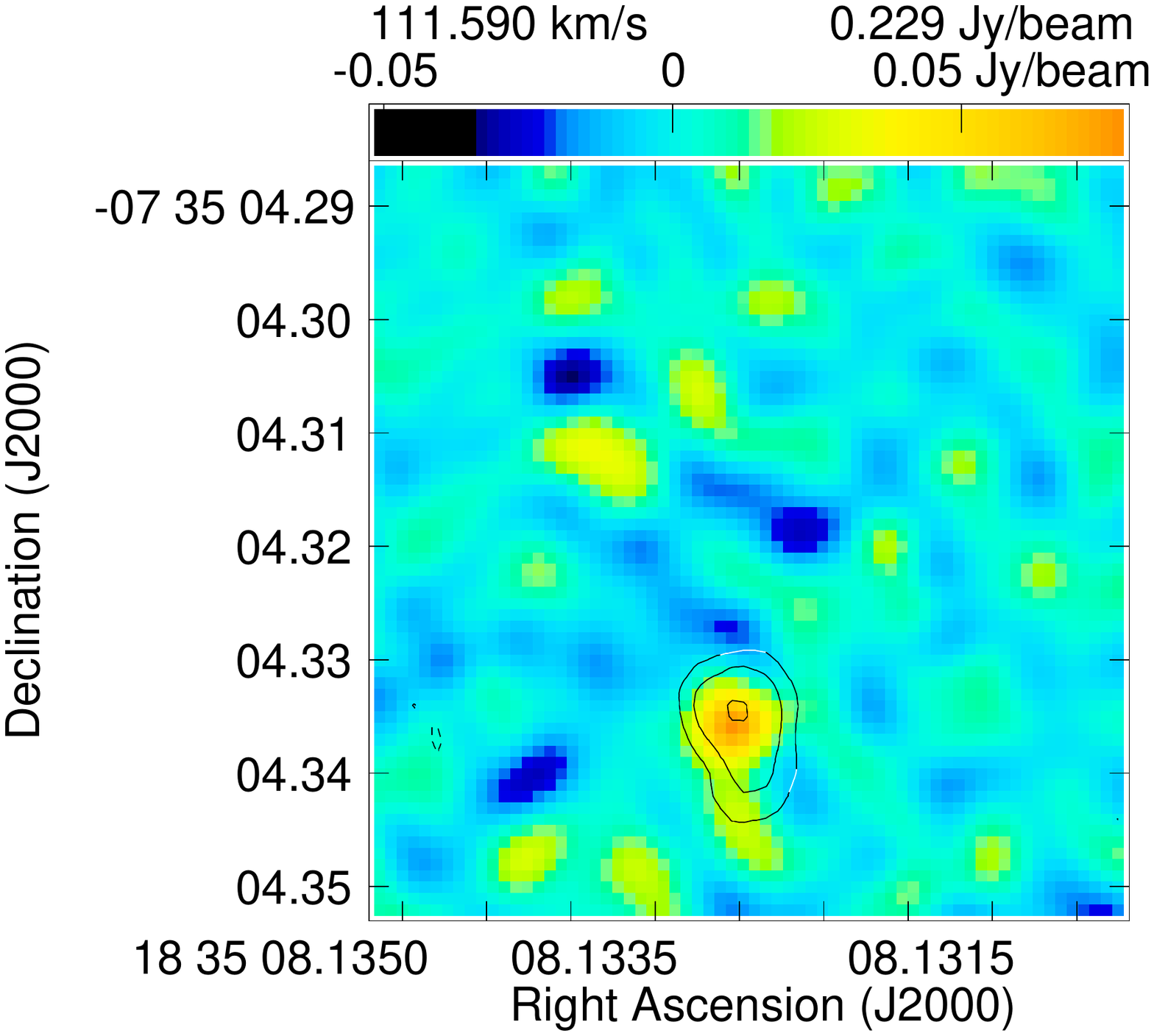}
\includegraphics[scale=0.3,trim=120 100 0 200]{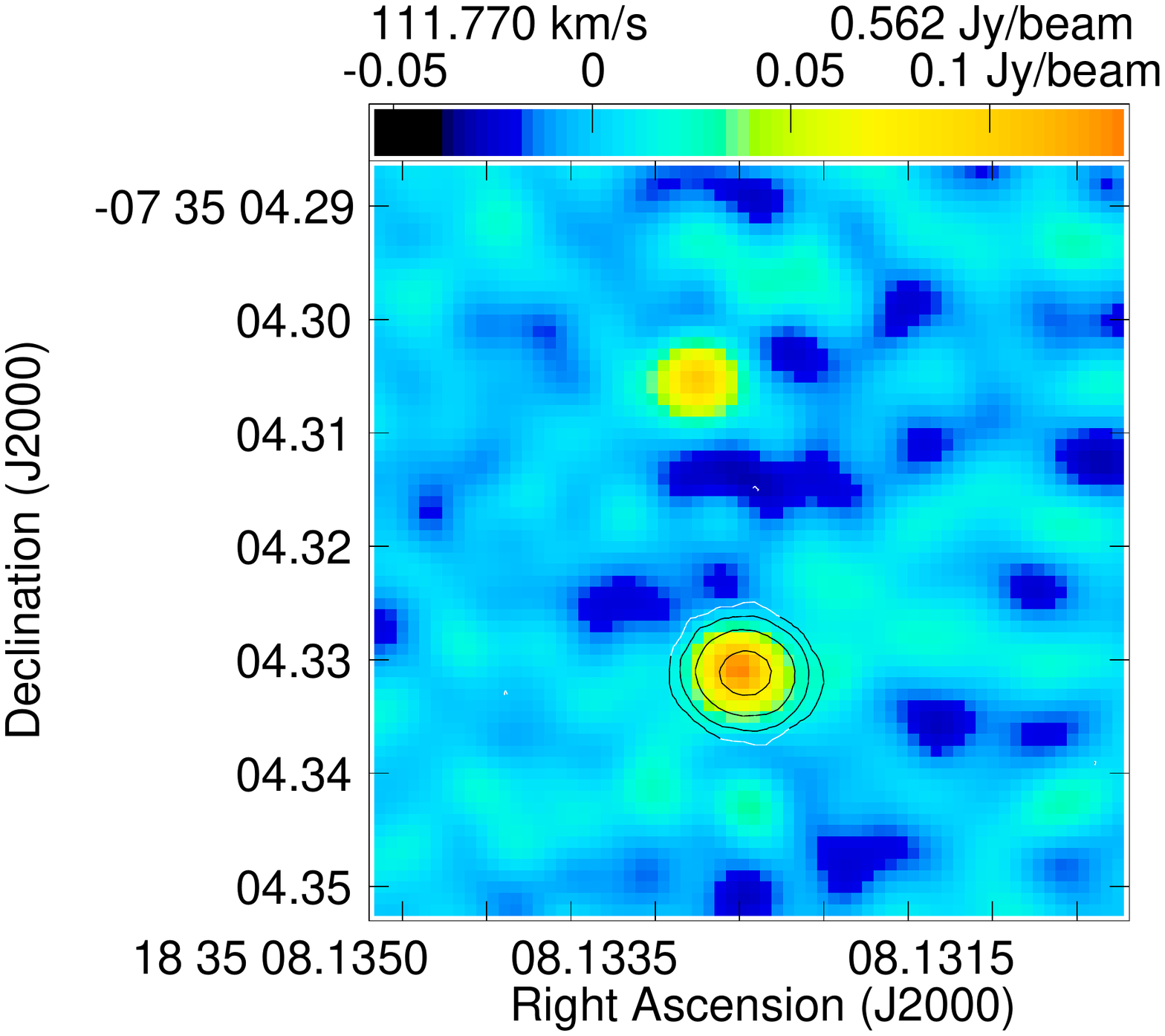}
\includegraphics[scale=0.3,trim=120 100 0 200]{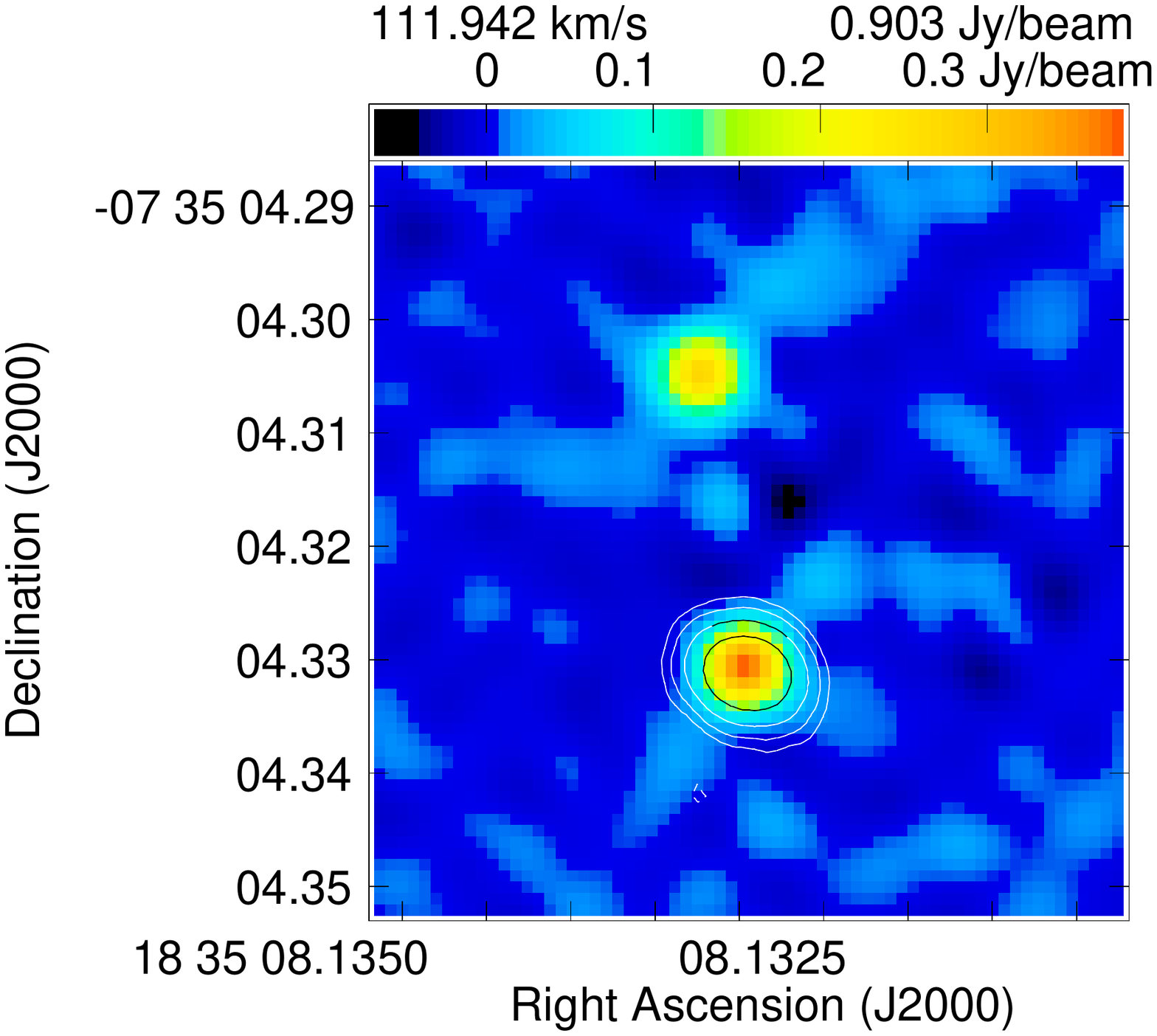}
\includegraphics[scale=0.3,trim=120 100 0 200]{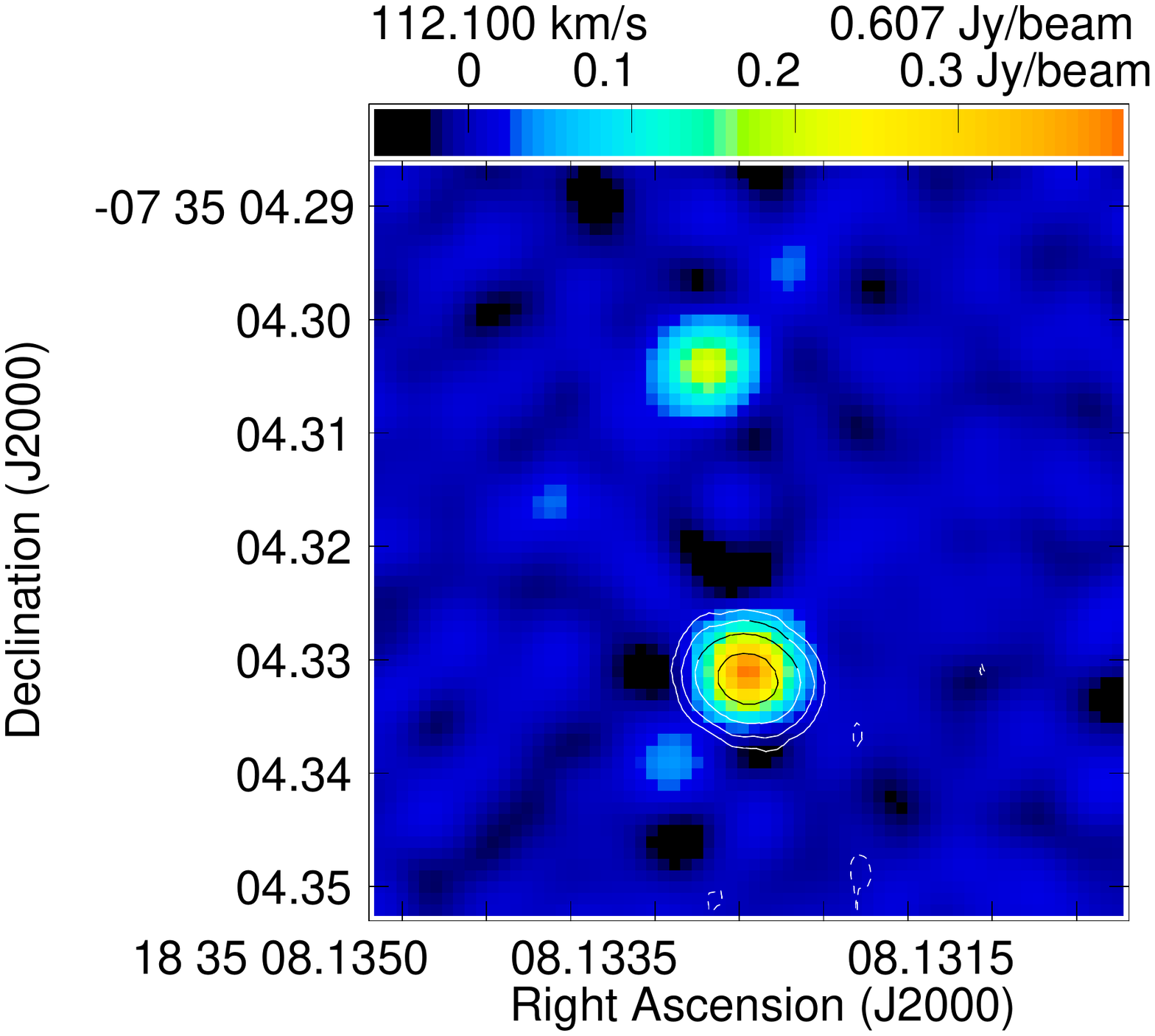}
\caption{Channel images showing co-propagation of 6.7 and 12.2~GHz methanol masers obtained at DOY:300. The first five images at velocities of 109.92 to 110.63\,\kms\, represent 6.7\,GHz Cloudlet 12 and 12.2\,GHz Cloudlet 3, while the next four images show Cloudlets 10 (6.7\,GHz) and 1 (12.2\,GHz) for the velocity range of 111.59$-$112.10\,\kms\, (Tables\,\ref{tablegaussmeth6} and  \ref{tablegaussmeth12}). Contours represent the 6.7~GHz methanol maser emission; the first contour corresponds to 30~mJy~beam$^{-1}$ ($\sim3\sigma_\mathrm{rms}$). The next contours are at 2, 4, 8, 16, 32, 64, and 128 times 3$\sigma_\mathrm{rms}$. The first negative contour is also shown, and the contour at the 90\% of the peak emission. The numbers at the top of each panel correspond to the LSR velocities of each spectral channel and the peak intensity of the 6.7~GHz methanol maser emission. The color scale presents the 12.2~GHz methanol emission with ranges as indicated at the top of each map in Jy~beam$^{-1}$.
}
\label{fig:coop1} 
\end{figure*}

\begin{figure*}
\centering
\includegraphics[width=\textwidth]{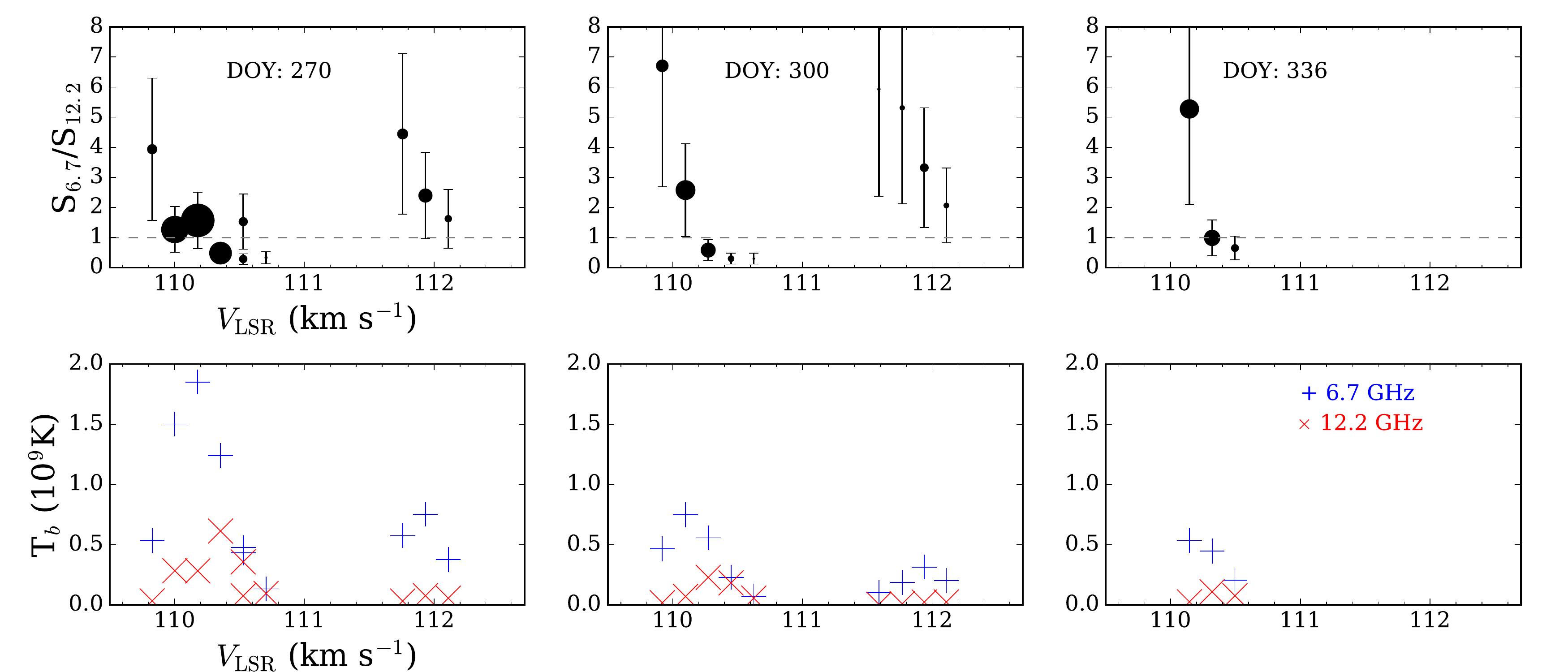}
\caption{Details of co-propagating 6.7 and 12.2~GHz methanol masers through the same gas volume. {\bf Top:} Ratios of flux densities. The size of the circles is proportional to the flux density of the 6.7~GHz emission. {\bf Bottom:} Brightness temperatures of co-propagating spots.}
\label{fig:coop2}
\end{figure*}

To compare the intensities of the two transitions in the two cloudlet sets, we created images  with a similar spectral resolution by averaging two channels at the 12.2~GHz maser lines, and  with the same circular synthesized beams of 6~mas$\times$6~mas. Single-channel images for DOY: 300 are presented in Fig.~\ref{fig:coop1} for both cloudlet groups. The ratios of the flux densities to the lower limits of brightness temperatures are presented in Fig.~\ref{fig:coop2}. The maximum uncertainties of these ratios were calculated using the total differential method, assuming   50\% and 10\% uncertainties in the flux densities at 12.2 and 6.7~GHz, respectively. The brightness temperature ($T_\mathrm{b}$) is calculated according to Eq.~9$-$27 in \cite{wrobel99}. Co-propagation of both methanol maser transitions is predicted in a broad range of physical conditions considering gas temperature and density by \cite{cragg2002} and \cite{cragg2005}. 
These conditions are characterized by a relatively large difference between the temperatures of the pumping dust emission and the masering gas, which may arise due to the flare nature of accretion onto a young star formed in G24. This difference appears because the heating of the dust occurs due to the emission of the accretion flare while the gas in this case is heated by collisions with the dust and has a considerable time lag. In our data, we do not find cases where the brightness temperature of the 12.2~GHz transition exceeds that of the 6.7~GHz, which is entirely consistent with the modeling results of \cite{cragg2005}. The variability of the two transitions is synchronized, as predicted by the model. 
It is interesting that the 12.2~GHz Cloudlet~2 that is visible in DOY: 270 and DOY: 300 appeared without 6.7~GHz counterpart detected within an upper limit of 210~mJy~beam$^{-1}$ (3$\sigma$). This value is higher than the flux values of the detected 12.2~GHz Cloudlet~2. So, this   does not describe a contradiction with the model of \cite{cragg2005}.

\subsection{Keplerian model - Derivation of disk inclination}

We fitted the 6.7 GHz maser positions projected along the disk major axis (PA = 15$^{\circ}$), perpendicular to the axis of the radio jet:  the slice along the line connecting RA(J2000)=18$^{\mathrm{h}}$35$^{{\rm m}}$08\fs125, Dec(J2000)=$-$07\degr35\arcmin04\farcs5 and RA(J2000)=18$^{\mathrm{h}}$35$^{{\rm m}}$08\fs145, Dec(J2000)=$-$07\degr35\arcmin03\farcs5). Using the peak position of the dust continuum as the YSO position (P$_0$), we fitted the maser positions (P$_j$), along the slice, and V$_{LSR}$, (V$_j$) to a Keplerian rotation profile around a central mass (M$_{core}$) (8.8\,M$_\odot$ from \citealt{hirota:2022}) and line-of-sight inclination ($i$) using
\begin{equation}
     V_j = V_{sys} \pm \sqrt{\frac{G\,M_{core}}{P_j-P_0}}\,sin\,i
,\end{equation}
where G is the gravitational constant, P$_j$\,$-$\,P$_0$ is the radius from the center of the disk (inner radius of 10\,au and outer radius of 120\,au), V$_{sys}$=114.5\,km\,s$^{-1}$ is the systemic velocity of the core, and  $i$ varies from 0$^{\circ}$ to 90$^{\circ}$ in steps of 5$^{\circ}$. The range of inclination is chosen to cover edge-on and face-on cases \citep{Seifried2016}.
Figure~\ref{fig:g24_33_pv} shows the result of the Keplerian rotation fit, corresponding to the best fit inclination angle of 85$^{\circ}\pm$3$^{\circ}$. 
Our best fit at the inclination angle of 85$^{\circ}$ falls within the range of angles where a recognizable Keplerian disk rotation profile is observable for edge-on disks \citep{Seifried2016}.
The outcome of our Keplerian disk rotation fit indicates that the 6.7~GHz methanol masers are tracing the inner part of the disk enclosing a significant fraction of the mass. However, we note that the result depends on the sensitivity of the interferometric observations used in making the maser map. Non-detection of outward red- or blueshifted maser features may have impacted our findings. \citet{2004MNRAS.351..779D} observed that masers can be several au to 10s of au in size. Very long baselines could resolve out slightly extended maser features. Therefore, more sensitive maser mapping observations, especially sensitive to extended maser emission, will be required to improve our fit results.

\subsection{Maser internal structure with respect to  source kinematics} \label{sec:cloudlet-3months}
Considering the pre-flare data obtained in 2009, we  attempted to identify cloudlets that persisted over 10~yr. As we mention above, the 2009 data show three masing regions, to the NE, W, and SE. Assuming the same persistent LSR velocities, as we did not notice any velocity drifts, we note that the NE masers correspond to the 2019 epoch Cloudlet~3, the W masers have a  position similar to that of Cloudlet 9, while the SE masers correspond to Cloudlets~10 and 12 (Fig.~\ref{fig:sixepochs}). For proper motion analysis, we determined the barycenters, weighted by the spot brightness, of individual cloudlets, as in \cite{bartkiewicz2020}. Relative motions with respect to the barycenters do not show any obvious and regular motions indicating expansion, infall, or rotation with an accuracy of 6~mas over 10~yr, corresponding to an upper limit of velocity of 19.1 \kms. The main difference between the 2009 and 2019 epochs is the appearance of new maser features, in particular to the S and W where the redshifted spots are located. We found Gaussian characteristics in eight velocity profiles of ten cloudlets seen in 2009 (Table~\ref{tablegaussmeth2009}). The average and median FWHM values in 2009 were 0.14$\pm$0.01~\kms~and 0.14~\kms, respectively, and they were smaller than in 2019 (0.18-0.24 \kms and 0.17-0.21 km/s). These values are two times smaller than the typical values of FWHMs in the sample presented in \cite{bartkiewicz2016}. In 2019 VLBA DOY: 270, the spectral feature at the LSR velocity of 115.3~\kms~  dominated the 6.7~GHz spectrum with the flux density of 20.9~Jy and it decreased by a factor of two after 66 days (Fig.~\ref{fig:sixepochs}). We note that it was the weakest feature in 2009 in the interferometric and single-dish observations. 

A detailed analysis of the 6.7~GHz methanol maser cloudlets distribution shows a diverse behavior of these mas structures on a three-month timescale. Cloudlet~3 shows a remarkable arched and stable emission structure with the same velocity gradient in all five epochs. Cloudlets 1, 6, and 9 show a linear morphology of increasing length and preserved PAs over two months (see Cloudlet~6 in Fig.~\ref{fig:fitmeth67} for example). It suggests that the flare causes an increase in the gas volume where the maser amplification takes place. In Cloudlet~12 the PA was stable, but the structure shrank, suggesting a reduction in the gas volume sustaining the maser amplification. There is a hint that some cloudlets may have changed their orientation during the observing period. For instance, monotonic changes in the PA of Cloudlet\,4 might indicate the occurrence of ordered motion (Table~\ref{tablegaussmeth6}, Fig.~\ref{fig:fitmeth67}). Similar variations may have occurred in Cloudlets\,2, 13, and 14, but the emission is too weak to draw a firm conclusion. 
These cloudlets are close to the central object, while those preserving PAs lie at farther distances (Fig.~\ref{fig:multiobs}). We note a general trend that the PA of the W cloudlets is roughly $-$70\degr\,; the exception is Cloudlet~9, which persisted ten~years and whose PA is close to the PA of the whole structure. The remaining cloudlets in W (Cloudlets 1, 4, 6, 7, and 8) appear during the flare, and their PAs are perpendicular to the PA of the whole structure. The PAs of the SE cloudlets are typically $\sim$17\degr, which may indicate a large-scale order.

\begin{figure}[h!]
\centering
\includegraphics[width=\columnwidth]{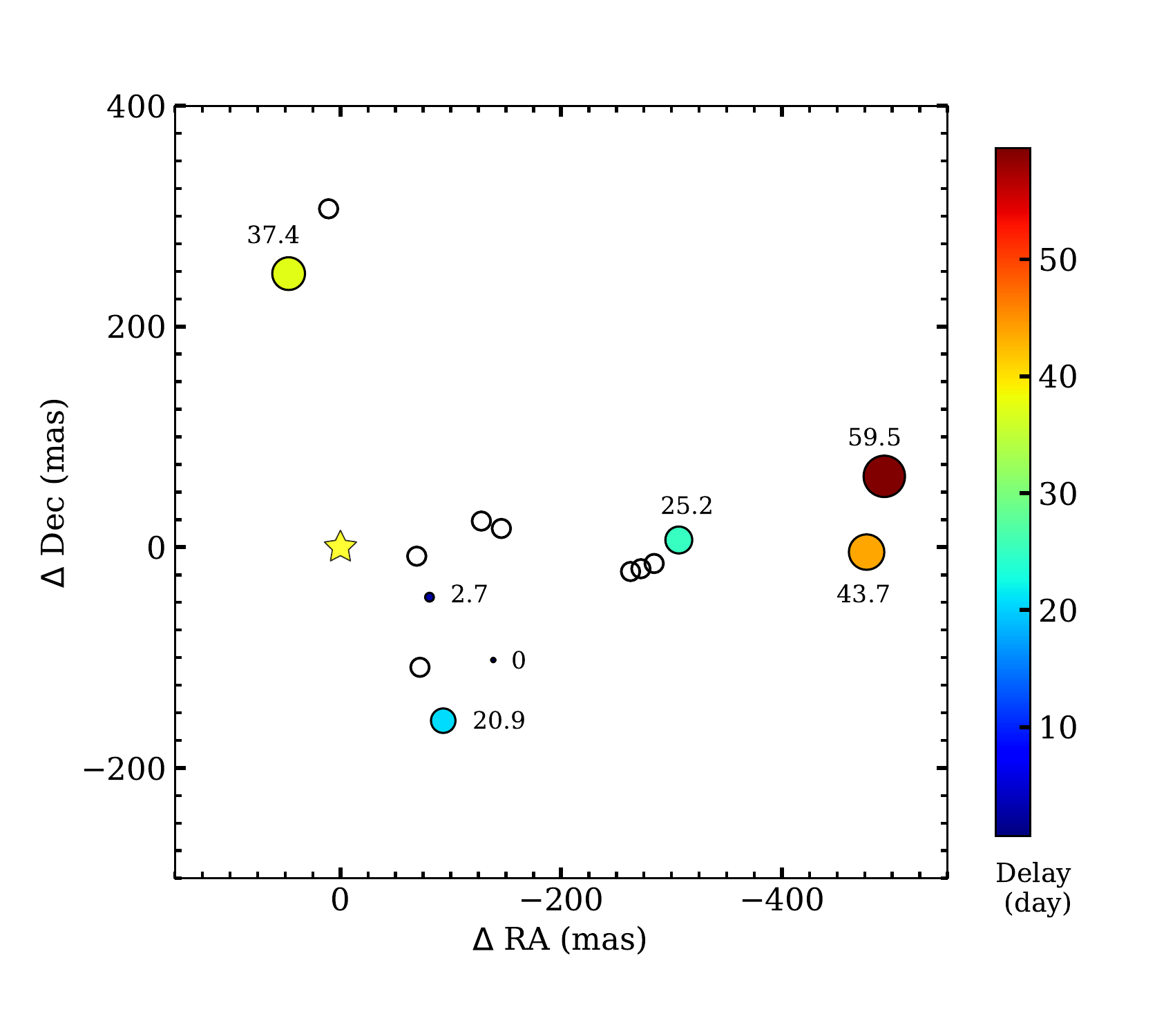}
\caption{Map of the 6.7\,GHz methanol maser cloudlets detected with VLBA at DOY: 270. The time lags as fitted to the 32\,m telescope measurements (Fig.\,\ref{fig:g24p33delay}) are marked by color and circle size, as indicated in the wedge. The size of symbols is proportional to the square root of delay. Cloudlets without meaningful measurements are marked as open circles. The yellow star is as in Fig.\,\ref{fig:multiobs}.} 
\label{fig:map_delays}
\end{figure}

Figure~\ref{fig:map_delays} shows the distribution of 6.7\,GHz cloudlets where the symbol size is proportional to the fitted time lags shown in Fig.\,\ref{fig:g24p33delay}. One of the cloudlets nearest to the central object is chosen as the reference. There is a clear trend of increase in the time lag with  projected distance from the central star, implying, for the assumed distance of 7.2\,kpc, propagation of excitation conditions at 0.3--0.5 the speed of light. For a simple disk--envelope model with a diameter comparable to the measured size of the  6.7\,GHz maser structure of 3500\,au (Fig.\,\ref{fig:multiobs}) the maximum time lag of 59.5\,d indicates that thermal heat inducing the maser flare propagates on average at 0.33 the speed of light. Our VLBI maps prove that this process does not affect the morphological structure of the 6.7\,GHz maser on a two-month timescale. It is in contrast to the extreme transformation of 6.7\,GHz maser morphology on a  26-day timescale reported in accreting HMYSO G358-MM1 that has been interpreted as an effect of propagation of thermal infrared radiation at 0.04--0.08 the speed of light \citep{burns2020}. We suggest that the flare in G24 is induced by an episode of enhanced production of infrared photons, but much less energetic than those observed in G358-MM1 and S255-NIRS3, likely triggered by accretion bursts \citep{burns2020,moscadelli2017}.

Considering the 12.2~GHz methanol maser emission, we note that only Cloudlet~3 was visible during all three epochs of VLBA observations. Its structure and velocity gradient persisted, but the flux density decreased (Fig.~\ref{fig:fit12group}). The remaining cloudlets at this transition gradually dimmed and disappeared at DOY: 336.

The 22.2~GHz water masers showed stable structure with velocity gradient over two months with emission between Cloudlets\,1 and 2 at the intermediate LSR velocities (Fig.~\ref{fig:fit22group}). Therefore, we state that during the flare, the majority of 6.7 and 12.2~GHz masers are not destroyed, but they brighten or dim. Furthermore, the 6.7~GHz methanol masers showed the peak of the emission on DOY: 330, which is different from the 22.2\,GHz water cloudlets that systematically increased in their intensities during the flare reaching the maximum at DOY: 336. We performed a proper motion analysis similar to that for the 6.7~GHz methanol maser emission, and we state that the  two elongated structures remained at a distance from each other at the level of 0.08~mas during 66 days, corresponding to an upper limit in velocity of 0.26\,\kms.

\subsection{Time-dependent IR position shift}
The temporal brightening of a YSO due to an accretion burst may lead to an echo seen in scattered light, for example, as in the case of S255IR-NIRS3 \citep{caratti2017}, or thermal emission. Because of the lower optical depth in the outflow cavities, photons will preferentially propagate along them. Only for an outflow in the plane of the sky does the centroid location remain unchanged during light propagation. In general, the projected distance of the light travel path is larger for light scattered from the lobe facing toward the observer. For the receding lobe, the light will travel backward first, and only then will it be scattered into the line of sight. Thus, the projected distance will be smaller. Moreover, since back scattering off dust grains is less efficient than forward scattering, the echo from the receding lobe will be fainter. Both effects lead to a displacement of the light centroid along the blueshifted cavity lobe. A temporal centroid shift caused by a light echo has been observed in the case of G107.298$+$5.639 \citep{stecklum:2018}.
 
For G24 the centroid shift of about 0\farcs5 amounts to 3600~au at the assumed distance of 7.2~kpc, which corresponds to a light travel time of approximately 20 days. So, if the shift during 10.5 years was due to a light echo, it must be highly subluminal. Such a slowdown of the light propagation would require extremely high optical depths over this distance, which is unrealistic.
Furthermore, the offset of the 4.6\,$\mu$m position from that of the dust continuum measured with ALMA \citep{hirota:2022} is smaller during the dim state between the flares and greater during the brightness peak. This is opposite to what has been observed for the light echo of G107.298$+$5.639 \citep{stecklum:2018}. Since the comparison of the G24 centroid positions to those of neighboring objects rules out an instrumental origin, the reason for the observed shift remains open.

\section{Conclusions}
We imaged the 6.7 and 12.2\,GHz methanol masers and the 22.2\,GHz water maser in HMYSO G24 during the 6.7\,GHz maser flare in 2019 and monitored the target in the 6.7\,GHz transition with the 32m telescope.

The 6.7\,GHz masers are spread over an area of  $\sim$3500\,au, whereas the 12.2\,GHz maser region is an order of magnitude smaller. The 12.2\,GHz maser cloudlets coincide with those at the  6.7\,GHz transition, supporting the standard models with a broad range of gas temperatures and densities of maser environments. The 22.2\,GHz masers located in the central region do not co-propagate with the methanol masers and could trace the base of bipolar outflow or jet observed with ALMA. The 6.7 and 12.2\,GHz methanol masers lie in the inner part of the rotating disk revealed by ALMA.

The maser cloudlets in G24 commonly show a clear velocity gradient and linear or elongated morphology that, for most of them, is preserved over  approximately two  months  of VLBI observations during the 6.7\,GHz flare. The brightness of the methanol cloudlets varies synchronously with the total 6.7\,GHz maser flux density monitored with the single dish. In contrast, the water cloudlets gradually increase with moderate changes in size and orientation and a velocity drift of $-$0.17\kms\,month$^{-1}$ was observed for the strongest 22.2\,GHz cloudlet. The sizes of specific 6.7\,GHz cloudlets increase at the flare maximum, which can be caused by changes in gas volume that favor maser emission. The observed variations in brightness and size of the methanol cloudlets are fully consistent with the radiative scheme of excitation of the methanol maser transitions indicated by a high correlation of the 6.7\,GHz and 4.6$\mu$m flux densities found in G24. 

The estimates of time lags of the flux density maxima of 6.7\,GHz features, together with the map of counterpart cloudlets, reveal that a change in physical conditions causing the maser flare propagates at one-third the speed of light. It can be a thermal heat wave postulated to explain dramatic changes in the 6.7\,GHz maser structure in accretion bursting HMYSOs (\citealt{burns2020}). We state that the overall maser structure in G24 is stable on a two-month timescale. Furthermore, comparison with the archival map taken in a quiescent state (i.e., before the discovery of the first flare), and proper motion analysis does not reveal any signs of expansion or infall within $\sim$6\,mas on a 10\,yr timescale.   

The high similarity of the flare profile in 2011 and 2019 suggests that G24 may be a long-period ($\sim$8.2\,yr)  maser source and can be a good objective for further testing of the causes of variability, and more generally the causes of the accretion bursts.

\begin{acknowledgements}
We acknowledge the useful discussions about astrometry with Dr. Anita Richards from the Jodrell Bank Centre for Astronomy, University of Manchester. The 32m radio telescope is operated by the Institute of Astronomy, Nicolaus Copernicus University and supported by the Polish Ministry of Science and Higher Education SpUB grant. AB, AK, MS, MD, PW acknowledge support from the National Science Centre, Poland through grant 2021/43/B/ST9/02008. 
TH is financially supported by the MEXT/JSPS KAKENHI Grant Numbers 17K05398, 18H05222, and 20H05845. MO thanks the Ministry of Education and Science of the Republic of Poland for support and granting funds for the Polish contribution to the International LOFAR Telescope (arrangement no 2021/WK/02) and for maintenance of the LOFAR PL-612 Baldy (MSHE decision no. 28/530020/SPUB/SP/2022). O.B. acknowledges financial support from the Italian Ministry of University and Research - Project Proposal CIR01$\_$00010. This work was supported by resources provided by the Pawsey Supercomputing Centre with funding from the Australian Government and the Government of Western Australia.
This publication makes use of data products from the Near-Earth Object Wide-field Infrared Survey Explorer ((NEO)WISE), which is a joint project of the Jet Propulsion Laboratory/California Institute of Technology and the University of Arizona. (NEO)WISE is funded by the National Aeronautics and Space Administration.
\end{acknowledgements}

\bibliography{librarian}{}
\bibliographystyle{aa}
 
\Online
\clearpage
\begin{appendix}
\section{Figures and tables}



\begin{table*}
\centering
\caption{Parameters of the 6.7~GHz methanol maser cloudlets with Gaussian velocity profiles in five epochs in 2019.}
\label{tablegaussmeth6}
\addtocounter{table}{-1}
\begin{tabular}{lccccrrccc}
\hline
No. of cloudlet & $V_{\rm p}$ & $V_{\rm fit}$ & FWHM & $S_{\rm fit}$ & $r_{\rm s}$ & $r_{\rm v}$ & $\Delta$RA$\times\Delta$Dec & $V_{\rm grad}$ & PA\\
Epoch & (km s$^{-1}$) & (km s$^{-1}$) & (km s$^{-1}$) & (Jy beam$^{-1}$) &  &  &  (mas$\times$mas) & (\kms~mas$^{-1}$) & (\degr) \\
\hline
\textbf{Cloudlet 1} \\ 
DOY: 270 (VLBA) & 119.84 & 119.85 & 0.17 & 0.2 & 1.00 & 1.00 & 2.2 $\times$ 0.8 & 0.15 & $-$69\\
DOY: 271 (LBA) & 119.86 & 119.89 & 0.12 &  1.0 & - & - & 2.9 $\times$ 1.6 & - & - \\
DOY: 280 (EVN) & 119.83 & 119.87 & 0.16 & 1.1 & 0.76 & 1.00 & 2.8 $\times$ 3.3 & 0.13 & $-$53 \\
DOY: 300 (VLBA)& 119.76 & 119.82 & 0.16 & 0.5 & 0.77 & 1.00 & 2.7 $\times$ 2.1 & 0.15 & $-$60\\
DOY: 336 (VLBA) & 119.98 & 119.90 & 0.18 & 0.3 & 0.99 & 1.00 & 3.9 $\times$ 1.2 & 0.13 & $-$66 \\
\textbf{Cloudlet 2} \\ 
DOY: 270  & 116.68 &- & - & - &- & - & - & - & -\\
DOY: 271  & 116.61 & 116.64 & 0.14 & 0.3 & 0.83 &  0.89 & 2.1 $\times$ 2.0 &  0.07 & $+$54 \\
DOY: 280  & 116.58 & 116.61 & 0.17 & 0.4 & 0.86 & 0.95 & 3.2 $\times$ 1.7 & 0.12 & $-$65\\
DOY: 300  & 116.60 & 116.56 & 0.17 & 0.2 & - & - & 1.9 $\times$ 0.5 & - & - \\
DOY: 336  & 116.64 &- & - & - &- & - & - & - & -\\
\textbf{Cloudlet 3} \\ 
DOY: 270 & 115.27 & 115.33 & 0.17 & 2.5 & - & - & 1.2 $\times$ 3.3 & -& -\\
DOY: 271 & 115.38 & 115.37 & 0.17 & 12.6 & - & - &  4.6 $\times$ 7.7 & -&- \\
DOY: 280 & 115.35 & 115.34 & 0.20 & 12.1 & - & - & 4.0 $\times$ 6.7 & -&- \\
DOY: 300 & 115.37 & 115.29 & 0.21 & 6.1 & - & - & 2.7 $\times$ 5.0 & - & -\\
DOY: 336 & 115.41 & 115.37 & 0.20 & 2.6 & - & - & 2.0 $\times$ 5.7 &- & -\\
\textbf{Cloudlet 4} \\ 
DOY: 270 & 115.10 & 115.02 & 0.20 & 0.3 & 0.93 & 0.98 & 1.7 $\times$ 1.6 & 0.26 & $-$53\\
DOY: 271 & 115.03 & 115.04 & 0.15 & 1.0 & 0.99 & 0.91 & 3.4 $\times$ 2.0 & 0.12 & $-$67\\
DOY: 280 & 115.00 & 115.02 & 0.17 & 1.6 & 0.52 & 1.00 & 2.1 $\times$ 0.5 & 0.25 & $-$69\\
DOY: 300 & 115.02 & 114.97 & 0.17 & 0.6 & 0.60 & 1.00 & 1.5 $\times$ 0.2 & 0.20 & $-$85\\
DOY: 336 & 115.06 & 115.04 & 0.15 & 0.3 & - & - & 0.2 $\times$ 0.9 & - & - \\
\textbf{Cloudlet 5} \\ 
DOY: 270 & 113.51 & - & - & - &- & - & - & - & -\\
DOY: 271 & 113.50 & 113.52 & 0.11 & 0.3 & - &- & 0.7 $\times$ 1.3 & - & -\\
DOY: 280 & 113.51 & 113.46 & 0.20 & 1.1 & - & - & 1.8 $\times$ 4.7 & - & - \\
DOY: 300 & 113.44 & 113.43 & 0.19 & 0.5 & 0.98 & 0.90 & 0.3 $\times$ 4.9 & 0.07 & $-$3\\
DOY: 336 & 113.48 & 113.55 & 0.18 & 0.2 & - & - &  0.3 $\times$ 5.0 & - & -\\
\textbf{Cloudlet 6} \\  
DOY: 270 & 113.69 & 113.76 & 0.30 & 0.3 & 1.00 & 0.98 & 3.7 $\times$ 2.0 & 0.17 & $-$62 \\
DOY: 271 & 113.76 & 114.17 & 0.12 & 0.5 & 0.93 & 0.97 & 5.7 $\times$ 2.0 & 0.14 & $-$70\\
        &        & 113.76 & 0.15 & 1.7 & & & & &\\
DOY: 280 & 113.77 & 114.14 & 0.14 & 0.7 & 0.99 & 0.98 & 3.6 $\times$ 2.2 & 0.17 & $-$60\\
        &        & 113.73 & 0.16 & 2.1 & & & & & \\
DOY: 300 & 113.61 & 114.11 & 0.15 & 0.3 & 0.92 & 0.97 & 4.1 $\times$ 2.9 & 0.18 & $-$59\\
   &        & 113.69 & 0.16 &  1.0& & & &\\
DOY: 336 & 113.66 & 113.66 & 0.33 & 0.6 & 0.94 & 0.98 & 7.8 $\times$ 5.0 & 0.13 & $-$58\\  
\textbf{Cloudlet 7} \\ 
DOY: 270 & 113.69 & - & - & - & - & - & - & - & -\\
DOY: 271 & 113.67 & 113.64 & 0.12 & 0.6 & 0.97 & 0.96 & 2.3 $\times$ 0.7 & 0.09 & $-$72\\
DOY: 280 & 113.60 & 113.61 & 0.17 & 1.3 & 0.98 & 0.97 & 1.8 $\times$ 1.6 & 0.11 & $-$48\\
DOY: 300 & 113.61 & 113.58 & 0.15 & 0.8 & 0.99 & 1.00 & 1.4 $\times$ 1.7 & 0.18 & $-$41\\
DOY: 336 & 113.66 & 113.65 & 0.18 & 0.4 & - & - & 0.6 $\times$ 1.4 & - & -\\
\textbf{Cloudlet 8} \\ 
DOY: 270 & 113.34 & - & - & - &- & - & - & - &- \\
DOY: 271 & 113.50 & 113.48 & 0.15 & 1.7 & 0.72 & 0.61 & 6.2 $\times$ 1.7 & 0.10 & $-$77 \\
DOY: 280 & 113.42 & 113.44 & 0.16 & 2.0 & 0.99 & 0.98 & 10.0 $\times$ 1.6 & 0.04 & $-$81\\
DOY: 300 & 113.44 & 113.38 & 0.14 & 1.3 & 0.99 & 1.00 & 3.0 $\times$ 1.5 & 0.16 & $-$64\\
DOY: 336 & 113.48 & 113.45 & 0.15 &  1.0& 0.95 & 1.00 & 2.7 $\times$ 0.6 & 0.13 & $-$78\\
\hline

\end{tabular}
\end{table*}

\begin{table*}
\centering
\caption{continued}
\begin{tabular}{lccccrrccc}
\hline
No. of cloudlet & $V_{\rm p}$ & $V_{\rm fit}$ & FWHM & $S_{\rm fit}$ & $r_{\rm s}$ & $r_{\rm v}$ & $\Delta$RA$\times\Delta$Dec & $V_{\rm grad}$ & PA\\
Epoch & (km s$^{-1}$) & (km s$^{-1}$) & (km s$^{-1}$) & (Jy beam$^{-1}$) &  &  &  (mas$\times$mas) & (\kms~mas$^{-1}$) & (\degr) \\
\hline
\textbf{Cloudlet 9} \\ 
DOY: 270& 112.81 & - & - & - & 0.71 & 0.90 & 2.3 $\times$ 1.8 & 0.12 & $+$64\\
DOY: 271 & 112.88 & 113.21 & 0.14 & 0.2 & 0.84 & 0.84 & 4.1 $\times$ 3.3 & 0.11 & $+$50 \\
        &        & 112.87 & 0.12 & 1.1 & \\
DOY: 280 & 112.81 & 113.25 & 0.16 & 0.5 & 0.93 & 0.94 & 4.0 $\times$ 3.4 & 0.15 & $+$49 \\
        &        & 112.84 & 0.14 & 1.3 & & & & &\\
DOY: 300& 112.73 & 113.23 & 0.20 & 0.3 & 0.89 & 0.93 & 3.8 $\times$ 3.0 & 0.18 & $+$56\\
   &        & 112.79 & 0.14 & 0.6 & & &\\
DOY: 336& 112.78 & 112.87 & 0.20 & 0.4 & 0.94 & 0.95 & 3.8 $\times$ 4.1 & 0.11 & $+$54 \\
\textbf{Cloudlet 10} \\ 
DOY: 270 & 111.93 & 111.89 & 0.21 & 0.3 & 0.99 & 0.99 & 0.2 $\times$ 1.6 & 0.89 & $+$11 \\
     &        & 111.24 & 0.21 & 0.2 & 0.55 & 1.00 & 0.5 $\times$ 0.4 & 0.58 & $-$64 \\

DOY: 271 & 111.87 & 111.93 & 0.19 & 0.9 & - & - & 1.8 $\times$ 3.3 & - & -\\ 
        &        & 111.28 & 0.17 & 0.5 & - & - & 1.1 $\times$ 1.2 & - & - \\
DOY: 280 & 111.93 & 111.91 & 0.21 & 1.0 & 0.68 & 0.95 & 0.2 $\times$ 1.6 & 0.33 & $-$6 \\
        &        & 111.26 & 0.19 & 0.8 & 0.73 & 0.95 & 1.4 $\times$ 4.3 & 0.33 & $-$42 \\
DOY: 300 & 111.85 & 111.87 & 0.22 & 0.4 & 0.76 & 0.99 & 0.5 $\times$ 1.5 & 0.51 & $+$30\\
   &        & 111.22 & 0.18 & 0.4 & 1.00 & 0.95 & 0.9 $\times$ 1.8 & 0.25 & $-$27 \\
DOY: 336 & 111.37 & 111.95 & 0.22 & 0.3 & - & - & 0.4 $\times$ 1.5 & - & -\\
   &        & 111.30 & 0.21 & 0.4 & 0.62 & 0.71 & 0.9 $\times$ 2.7 & 0.19 & $-$26\\
\textbf{Cloudlet 11} \\ 
DOY: 270 & 110.18 & - & - & - &- & - & - & - & -\\
DOY: 271 & 110.20 & 110.21 & 0.05 & 0.3 & 0.71 & 0.94 & 1.0 $\times$ 0.8 & 0.07 & $-$61\\
DOY: 280 & 110.17 & 110.19 & 0.15 & 0.3 & - & - & 0.2 $\times$ 0.4 & - & - \\ 
DOY: 300 & - & - & - & - &- & - & - & - & -\\
DOY: 336 & - & - & - & - &- & - & - & - & - \\
\textbf{Cloudlet 12} \\ 
DOY: 270 & 110.18 & 110.14 & 0.23 & 0.1 & 0.99 & 0.87 & 2.6 $\times$ 9.6 & 0.07 & $+$16\\
DOY: 271 & 110.16 & 110.50 & 0.07 & 0.3 & 0.99 & 0.90 & 3.9 $\times$ 12.0 & 0.06 & $+$17\\
        &        & 110.15 & 0.18 & 1.9 \\
DOY: 280 & 110.17 & 110.18 & 0.25 & 2.5 & 0.98 & 0.93 & 2.8 $\times$ 10.3 & 0.09 & $+$15\\ 
        &        & 109.17 & 0.15 & 0.2 & - & - & 0.3 $\times$ 1.4 & - & -\\ 
DOY: 300 & 110.10 & 110.11 & 0.22 & 0.9 & 0.99 & 0.78 & 1.5 $\times$ 7.0 & 0.10 & $+$12\\
DOY: 336 & 110.14 & 110.19 & 0.23 & 0.4 & 0.98 & 0.96 & 2.2 $\times$ 7.3 & 0.07 & $+$14 \\
\textbf{Cloudlet 13} \\ 
DOY: 270 & 109.48 & 109.45 & 0.42 & 0.2 & 0.69 & 0.91 & 0.2 $\times$ 2.5 & 0.20 & $+$9\\
DOY: 271 & 109.63 & 109.56 & 0.30 & 0.5 & - & - & 2.7 $\times$ 2.9 & -& - \\
        &        & 109.41 & 0.10 & 0.2 & & &\\
DOY: 280 & 109.56 & 109.52 & 0.32 & 0.6 & 0.75 & 0.98 & 2.7 $\times$ 2.9 & 0.24 & $+$50\\
DOY: 300 & 109.57 & 109.51 & 0.32 & 0.2 & 0.91 & 0.94 & 0.9 $\times$ 1.4 & 0.30 & $+$36\\
DOY: 336 & - & - & - & - &- & - & - & - & -\\
\textbf{Cloudlet 14} \\ 
DOY: 270 & 108.25 & - & - & - &- & - & - & - & -\\
DOY: 271 & 108.22 & 108.25 & 0.11 & 0.6 & 0.74 & 0.86 & 2.0 $\times$ 0.7 & 0.21 & $-$65\\
DOY: 280 & 108.24 & 108.22 & 0.14 & 0.6 & 0.68 & 0.86 & 0.2 $\times$ 0.6 & 0.53 &$-$27\\
DOY: 300 & 108.17 & 108.19 & 0.14 & 0.2 & - & - & 0.3 $\times$ 0.7 & - & -\\
DOY: 336 & - & - & - & - &- & - & - & - & -\\
\textbf{Cloudlet 15} \\ 
DOY: 270 & 107.54 & - & - & - &- & -& - & - & -\\
DOY: 271 & 107.61 & 107.61 & 0.15 & 0.2 & - & - & 1.3 $\times$ 0.7 & - & -\\
DOY: 280 & 108.33 & 108.35 & 0.14 & 0.2 & 0.96 & 0.97 & 0.2 $\times$ 1.0 & 0.18 & $+$14\\
DOY: 300 & 107.64 & - & - & - &- & - & - & - & -\\
DOY: 336 & - & - & - & - &- & - & - & - & - \\

\hline
\end{tabular}
\end{table*}

\begin{table*}
\centering
\caption{Same as Table \ref{tablegaussmeth6} but for the 12.2~GHz methanol maser line.}
\label{tablegaussmeth12}
\begin{tabular}{lccccrrccc}
\hline
No. of cloudlet & $V_{\rm p}$ & $V_{\rm fit}$ & FWHM & $S_{\rm fit}$ & $r_{\rm s}$ & $r_{\rm v}$ & $\Delta$RA$\times\Delta$Dec & $V_{\rm grad}$ & PA\\
Epoch & (km s$^{-1}$) & (km s$^{-1}$) & (km s$^{-1}$) & (Jy beam$^{-1}$) &  &  &  (mas$\times$mas) & (\kms~mas$^{-1}$) & (\degr) \\
\hline
\textbf{Cloudlet 1} \\ 
DOY:270 & 111.80 & - & - & - & - & - & 1.4 $\times$ 2.7 & - & -\\  
DOY:300 & 111.95 & 112.01 & 0.21 & 0.1 & - & - & 0.8 $\times$ 1.3 & - & -\\
DOY:336 & - &- & - & - & - & - & - & - & -\\
\textbf{Cloudlet 2} \\ 
DOY:270 & 111.99 &111.96 & 0.18 & 0.1 & - & - & 0.5 $\times$ 1.4 & - & -\\  
DOY:300 & 111.95 & 111.92 & 0.15 & 0.08 & - & - & 0.1 $\times$ 0.7 & - & -\\
DOY:336 & - &- & - & - & - & - & - & - & -\\
\textbf{Cloudlet 3} \\ 
DOY:270 & 110.26 & 110.31 & 0.18 & 1.0 & 0.93 & 0.99 & 0.3 $\times$ 1.0 & 0.51 & $+$22\\ 
DOY:300 & 110.31 & 110.33 & 0.18 & 1.0 & 0.59 & 0.98  & 0.5 $\times$ 2.5 & 0.40  & $+$13 \\
DOY:336 & 110.35 & 110.36 & 0.17 & 0.6 & 0.60  & 0.94 &  0.4 $\times$ 0.8 & 0.63  & $-$12 \\
\hline
\end{tabular}
\end{table*}

\begin{table*}
\centering
\caption{Same as Table \ref{tablegaussmeth6} but for the 22.2~GHz water vapour maser line.}
\label{tablegausswater}
\begin{tabular}{lccccrrccc}
\hline
No. of cloudlet & $V_{\rm p}$ & $V_{\rm fit}$ & FWHM & $S_{\rm fit}$ & $r_{\rm s}$ & $r_{\rm v}$ & $\Delta$RA$\times\Delta$Dec & $V_{\rm grad}$ & PA \\
Epoch & (km s$^{-1}$) & (km s$^{-1}$) & (km s$^{-1}$) & (Jy beam$^{-1}$) &  &  &  (mas$\times$mas) & (\kms~mas$^{-1}$) & (\degr) \\
\hline
\textbf{Cloudlet 1} \\ 
DOY:270 & 125.12 & 126.30 & 0.45 & 1.2 & 0.90 & 0.91 & 1.4 $\times$ 1.4 & 1.42 & $-$50\\
   & & 125.13 & 0.31 & 7.2 & & & & & \\
DOY:300 & 124.98 & 126.11 & 0.46 & 2.0 & 0.96 & 0.94 & 1.2 $\times$ 1.1 & 1.84 & $-$47\\
   & & 125.00 & 0.41 & 8.2 & & & & &\\
DOY:336 & 124.75 & 125.94 & 0.60 & 3.4 & 0.97 & 0.94 &  1.3 $\times$ 1.4 & 1.65 & $-$44\\
   & & 124.78 & 0.40 & 9.1 & & & & &\\
\textbf{Cloudlet 2} \\ 
DOY:270 & 122.48 & 122.92 & 0.41 & 0.6 & 0.78 & 0.89 & 0.2 $\times$ 0.5 & 1.07 & $-$19\\
               & & 122.50 & 0.23 & 2.4 & 0.57 & 0.79 & 0.1 $\times$ 0.3 & 3.31 & $-$57 \\  
DOY:300 & 122.51 & 122.94 & 0.50 & 1.3 & 0.69 & 0.78 &  0.5 $\times$ 0.8 & 1.45 & $+$41\\
               & & 122.49 & 0.23 & 3.2 & 0.88 & 0.92 & 0.2 $\times$ 0.1 & 3.63 & $-$76\\
DOY:336 & 122.54 & 123.02 & 0.48 & 1.9 & 0.83 & 0.79 & 0.5 $\times$ 0.9 & 0.93 & $+$42\\
               & & 122.53 & 0.24 & 3.3 & 0.18* & 0.90 & 0.3 $\times$ 0.2 & 3.63 & $-$96*\\
               
\hline               
\multicolumn{10}{l}{$*$ - value left to show decline in V$_{grad}$, PA can be unrealistic.}

\end{tabular}
\end{table*}

\begin{table*}
\centering
\caption{Same as Table \ref{tablegaussmeth6} but for the 6.7~GHz 2009 EVN observations*.}
\label{tablegaussmeth2009}
\begin{tabular}{lccccrrccc}
\hline
No. of cloudlet & $V_{\rm p}$ & $V_{\rm fit}$ & FWHM & $S_{\rm fit}$ & $r_{\rm s}$ & $r_{\rm v}$ & $\Delta$RA$\times\Delta$Dec & $V_{\rm grad}$ & PA\\
Epoch & (km s$^{-1}$) & (km s$^{-1}$) & (km s$^{-1}$) & (Jy beam$^{-1}$) &  &  &  (mas$\times$mas) & (\kms~mas$^{-1}$) & (\degr) \\
\hline
1 & 115.41 & 115.45 & 0.15 & 0.2 &  - &  -& 0.6 $\times$ 4.3 &- & -\\
2 & 115.32 & - & - & - & - &- &  0.8 $\times$ 2.2 & - &-\\
3 & 112.86 & 112.86 & 0.09 & 0.4 & 1.00 & 0.98 & 1.5 $\times$ 2.3 & 0.06 & $+$33\\
4 & 112.86 & 112.86 & 0.09 & 0.4 & - & - & 0.6 $\times$ 1.4 & - & - \\
5 & 112.95 & - & - & - & - & - & 0.7 $\times$ 1.5 & - & - \\
6 & 111.99 & 111.93 & 0.18 & 0.4 &  - &- & 1.9 $\times$ 1.7 & - & -\\
7 & 111.99 & 111.96 & 0.17 & 0.5 & 0.83 & 0.90 & 1.4 $\times$ 1.6 & 0.10 & $+$23\\
8  & 110.40 & 110.41 & 0.15 & 0.6 & 0.81 & 0.94 & 2.3 $\times$ 1.7 & 0.11 & $+$54\\
9 & 110.40 & 110.33 & 0.13 & 1.3 & 0.94 & 0.93 & 2.0 $\times$ 3.0 & 0.11 & $+$36\\
10 & 110.23 & 110.18 & 0.15 & 0.6 & 0.92 & 0.86 & 0.9 $\times$ 2.2 & 0.11 & $+$27\\
\hline
\multicolumn{10}{l}{$*$ - data set from \citep{bartkiewicz2016}.}
\end{tabular}
\end{table*}

\begin{figure}[ht!]
\centering
\includegraphics[scale=0.4]{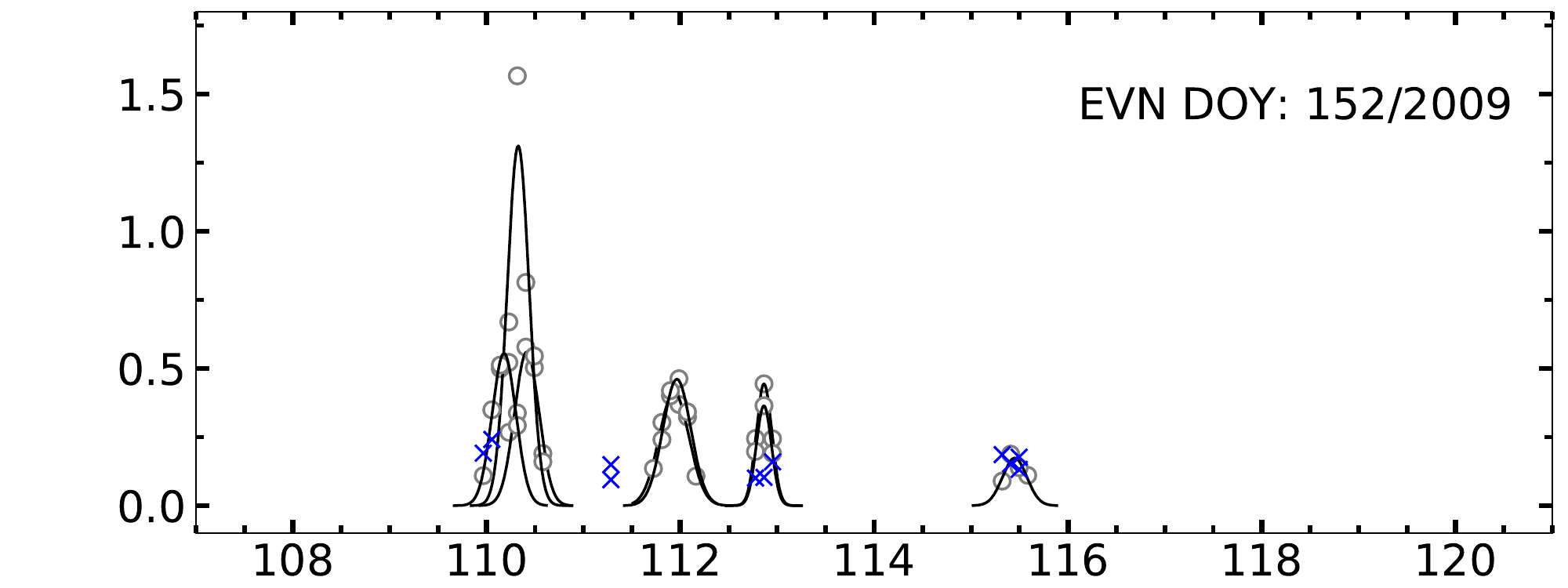}
\includegraphics[scale=0.4]{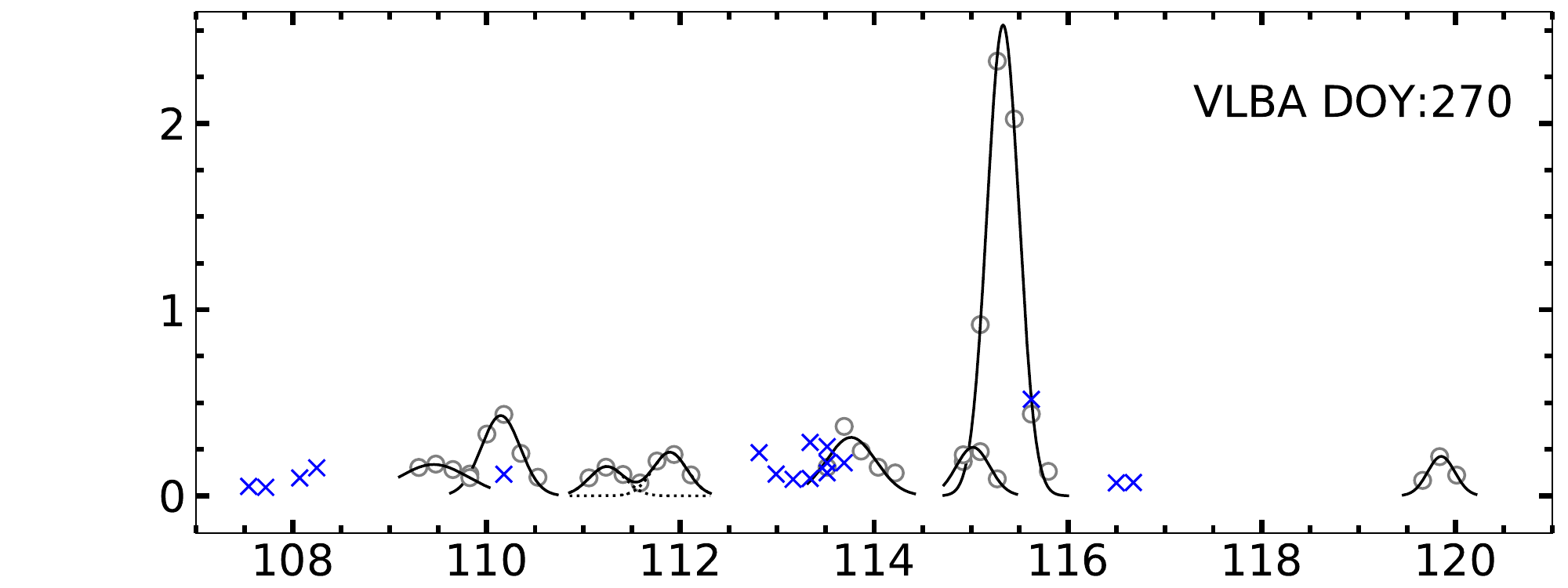}
\includegraphics[scale=0.4]{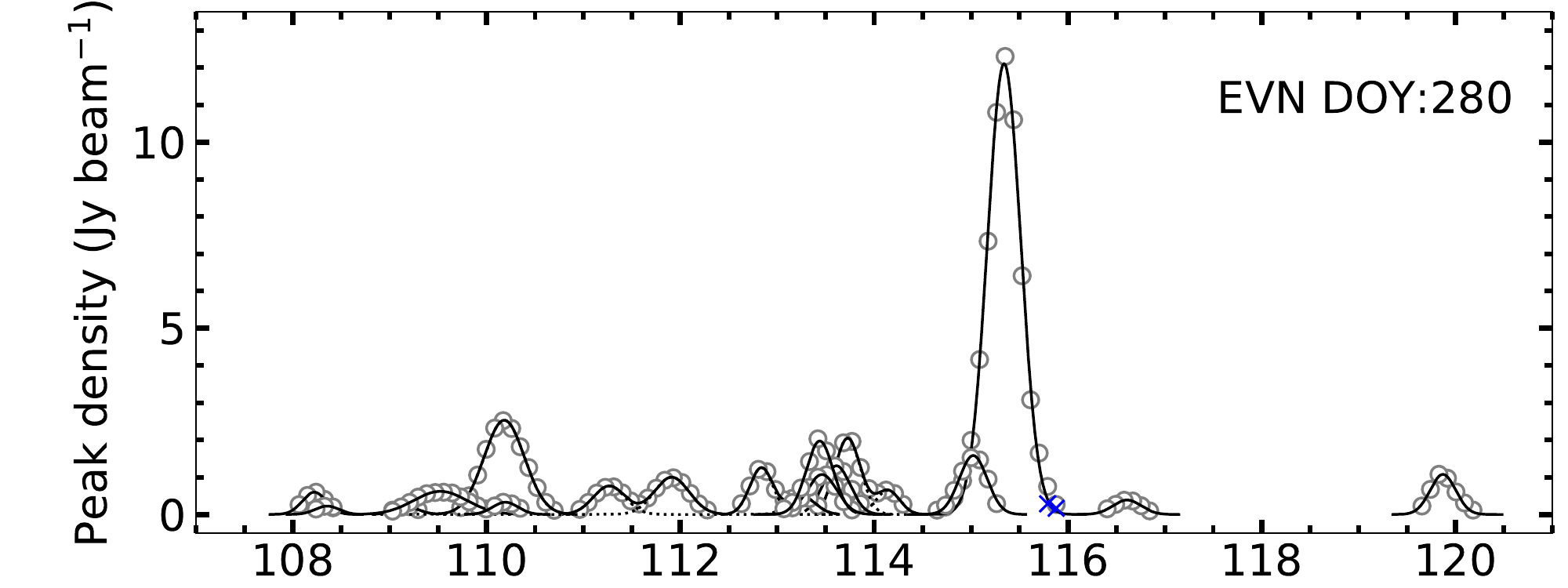}
\includegraphics[scale=0.4]{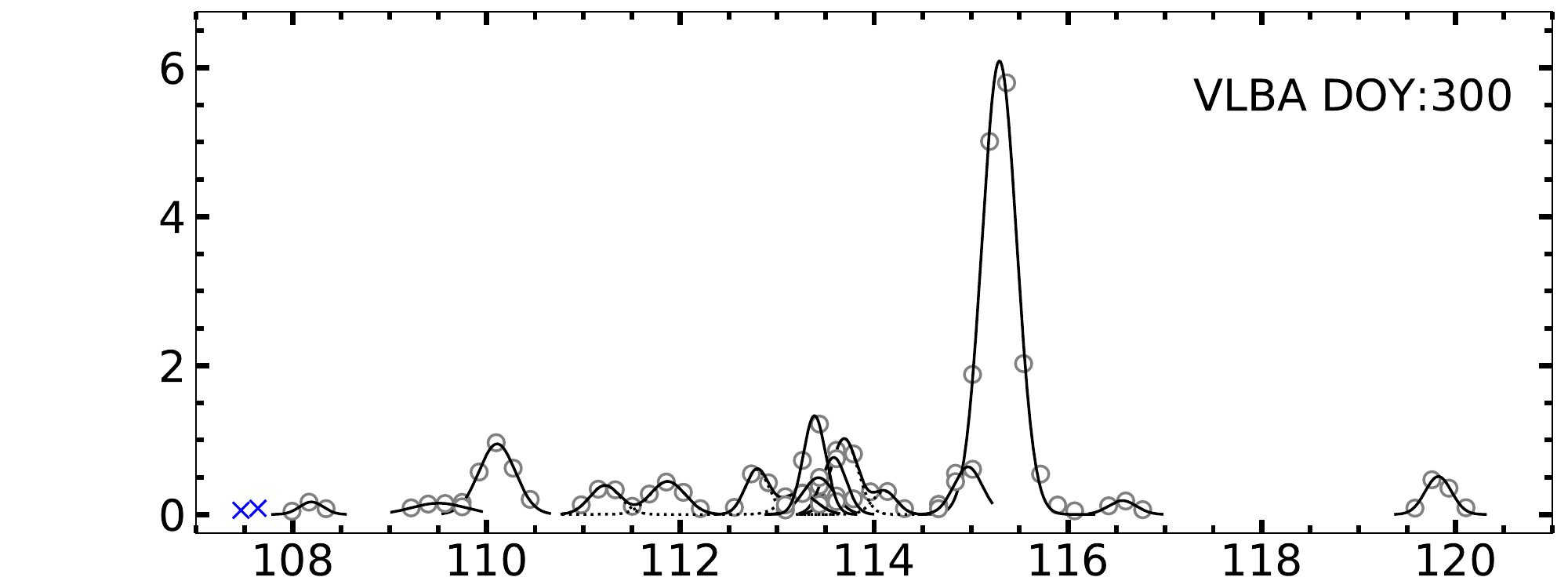}
\includegraphics[scale=0.328]{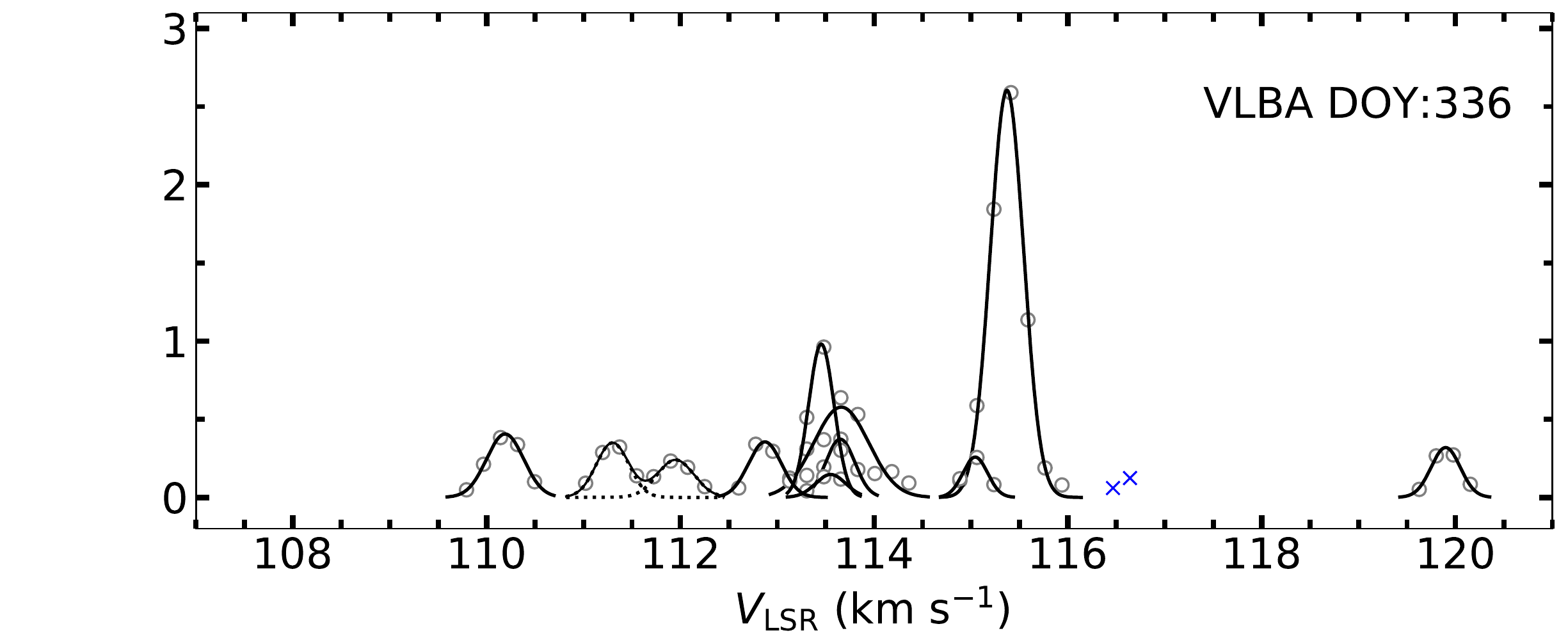}
\caption{Cross-correlation spectra of 6.7~GHz methanol maser line with Gaussian fittings of individual cloudlets. The fits (solid lines) present the Gaussian profiles. Each circle traces the emission level of a single maser spot, as presented in Fig.~\ref{fig:sixepochs}. The blue crosses indicate spots in cloudlets with no Gaussian profile.}
\label{fig:gaussmeth6}
\end{figure}

\begin{figure}[ht!]
\centering
\includegraphics[scale=0.4]{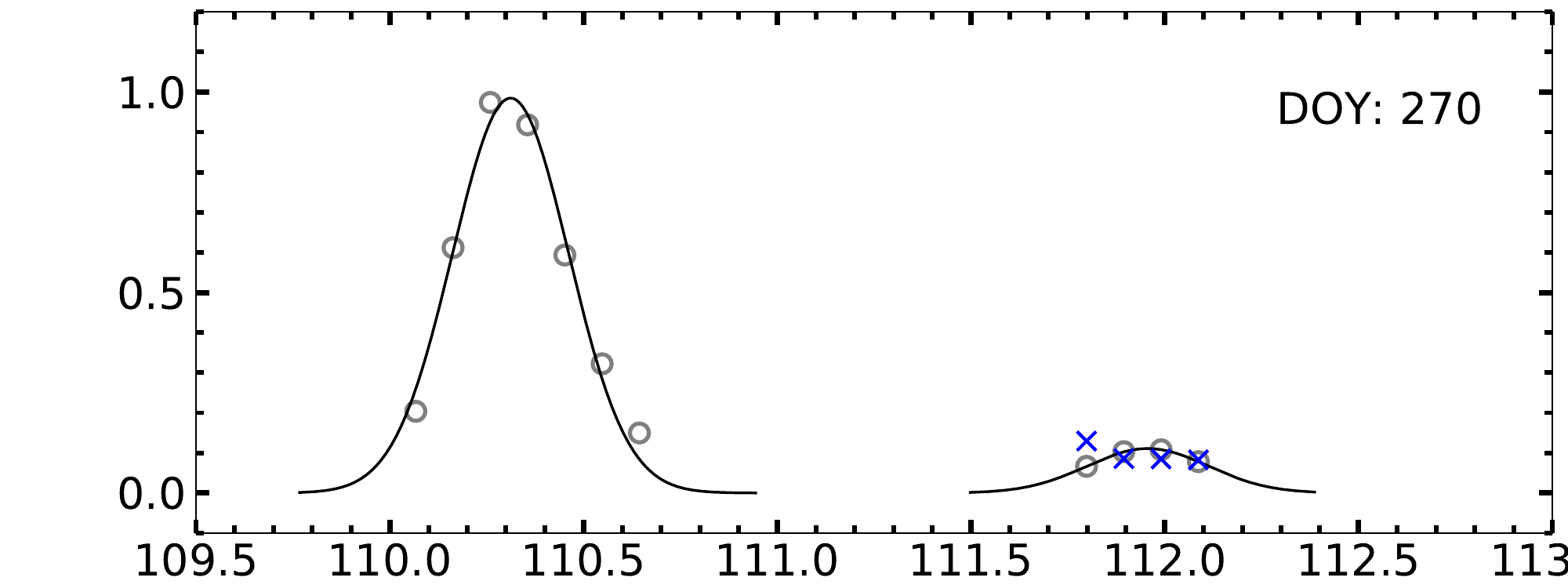}
\includegraphics[scale=0.4]{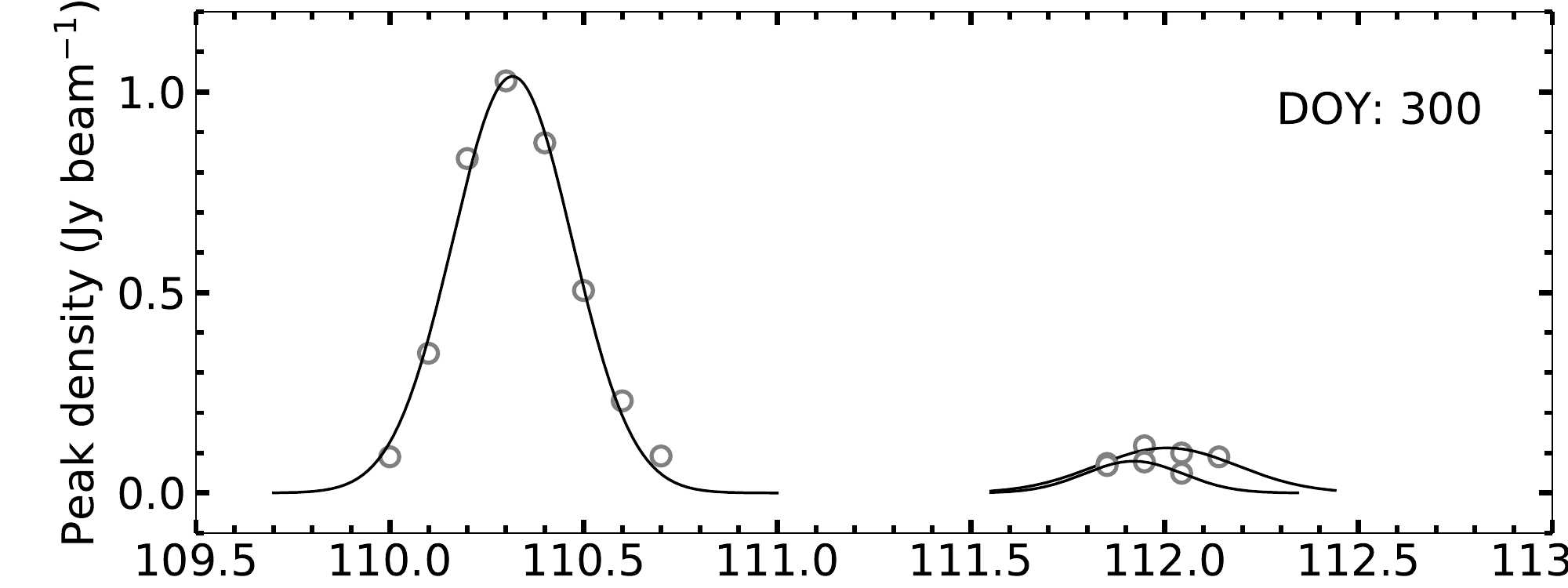}
\includegraphics[scale=0.328]{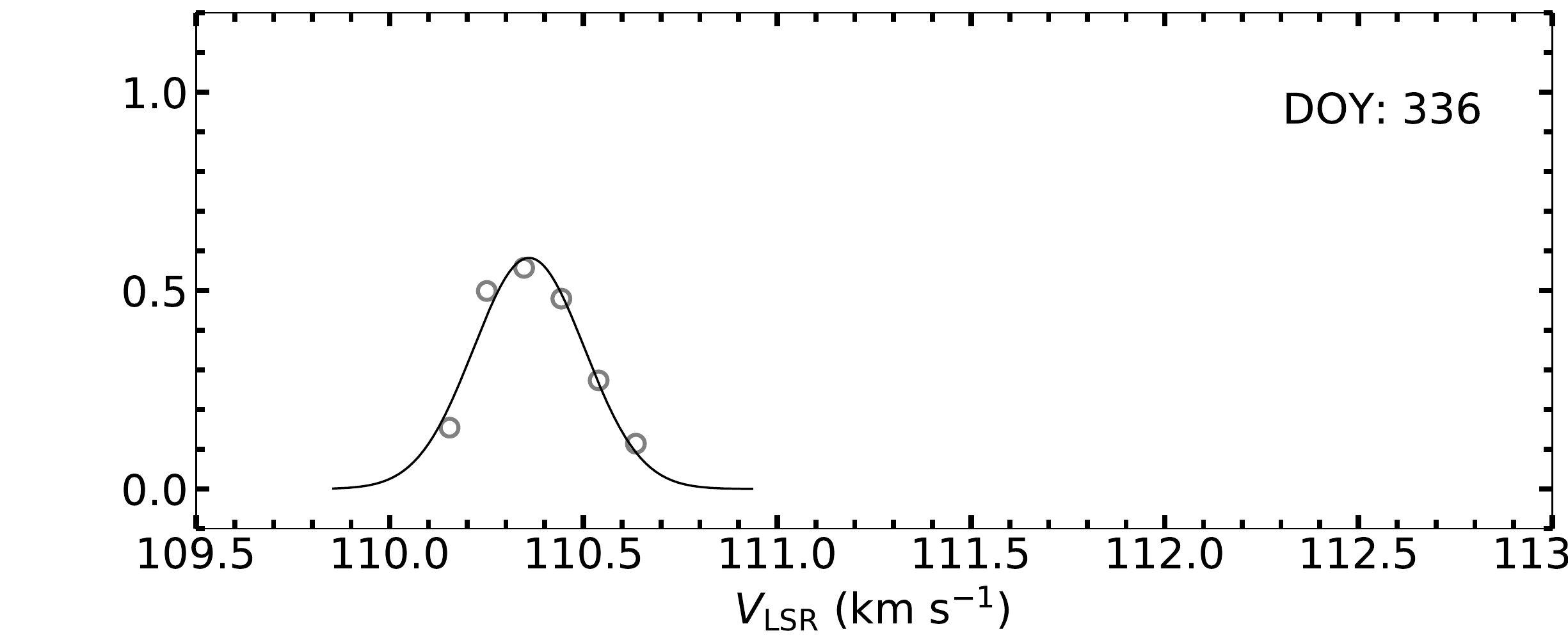}
\caption{Same as Fig.~\ref{fig:gaussmeth6}, but for 12.2~GHz methanol emission.}
\label{fig:gaussmeth12}
\end{figure}

\begin{figure}[ht!]
\centering
\includegraphics[scale=0.4]{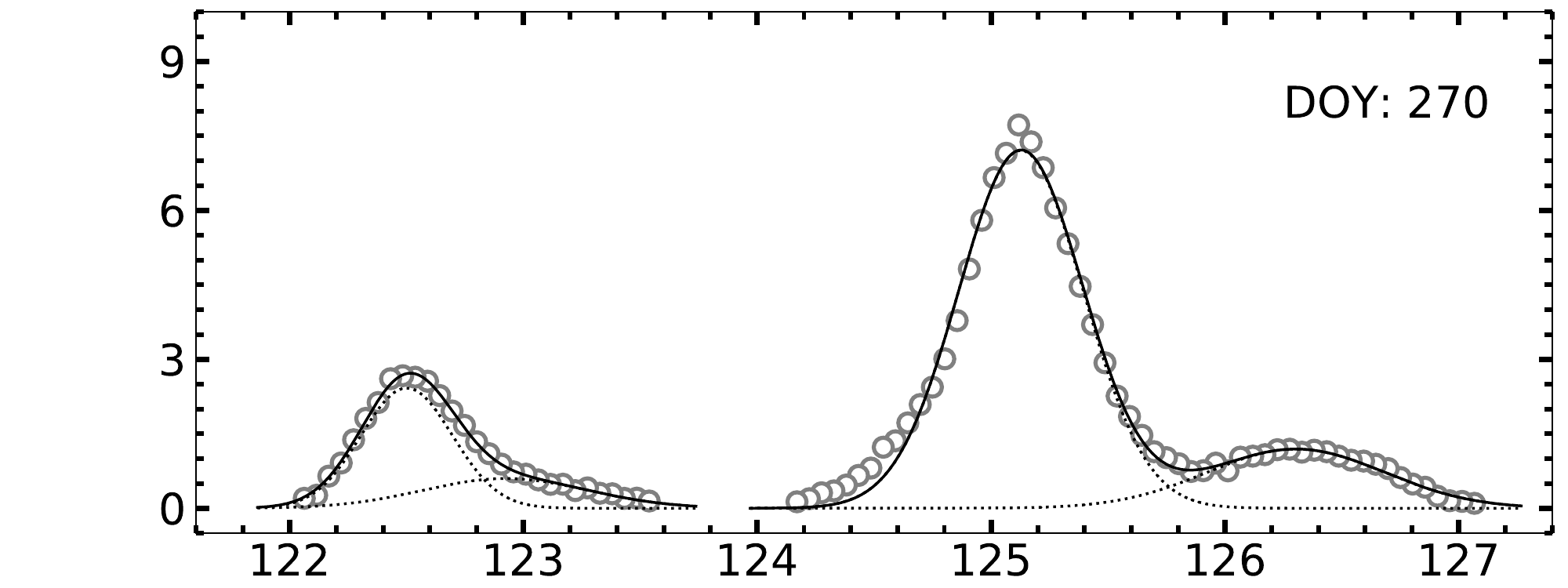}
\includegraphics[scale=0.4]{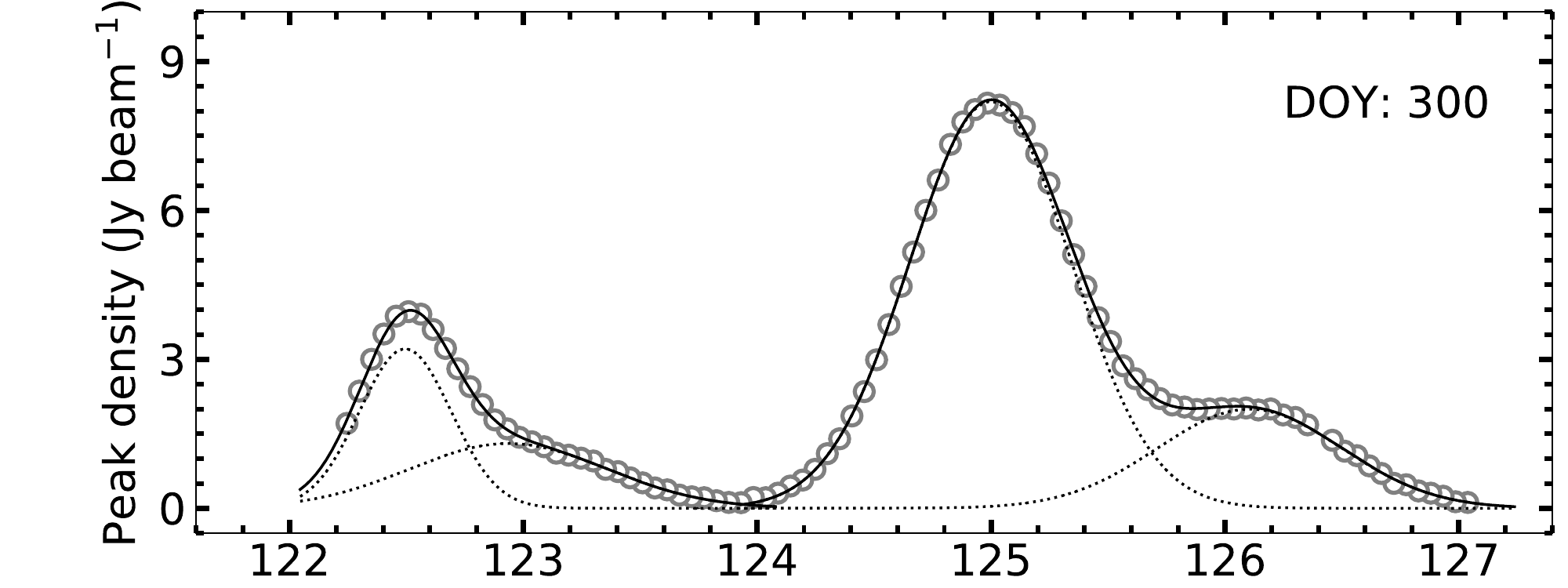}
\includegraphics[scale=0.328]{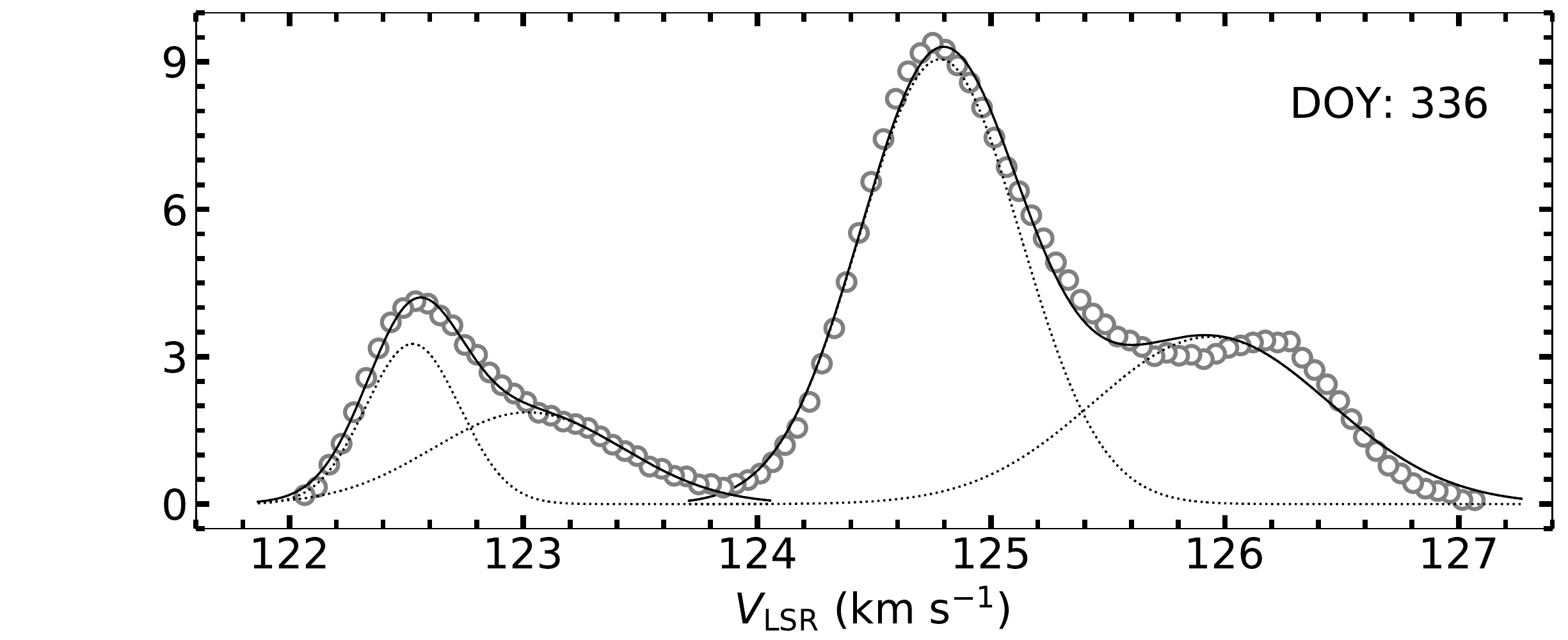}
\caption{Same as Fig.~\ref{fig:gaussmeth6}, but for 22.2~GHz water emission.}
\label{fig:gausswater}
\end{figure}

\begin{figure*}
\centering
\includegraphics[width=0.67\textwidth]{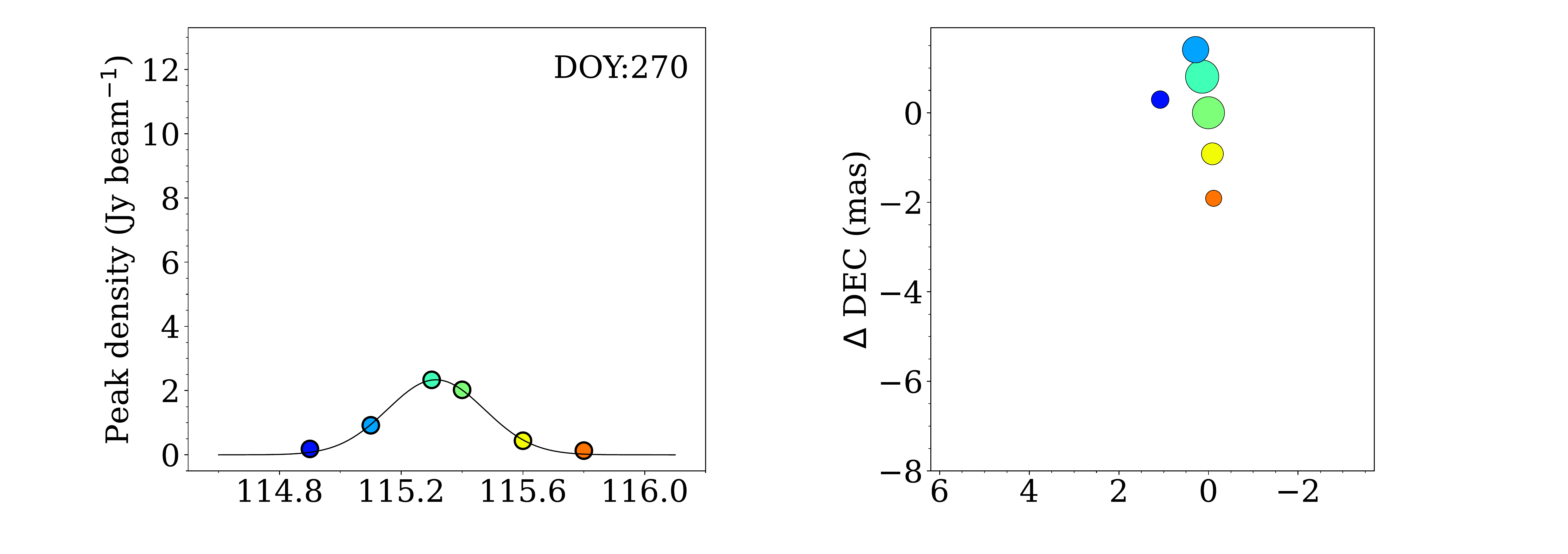}
\includegraphics[width=0.67\textwidth]{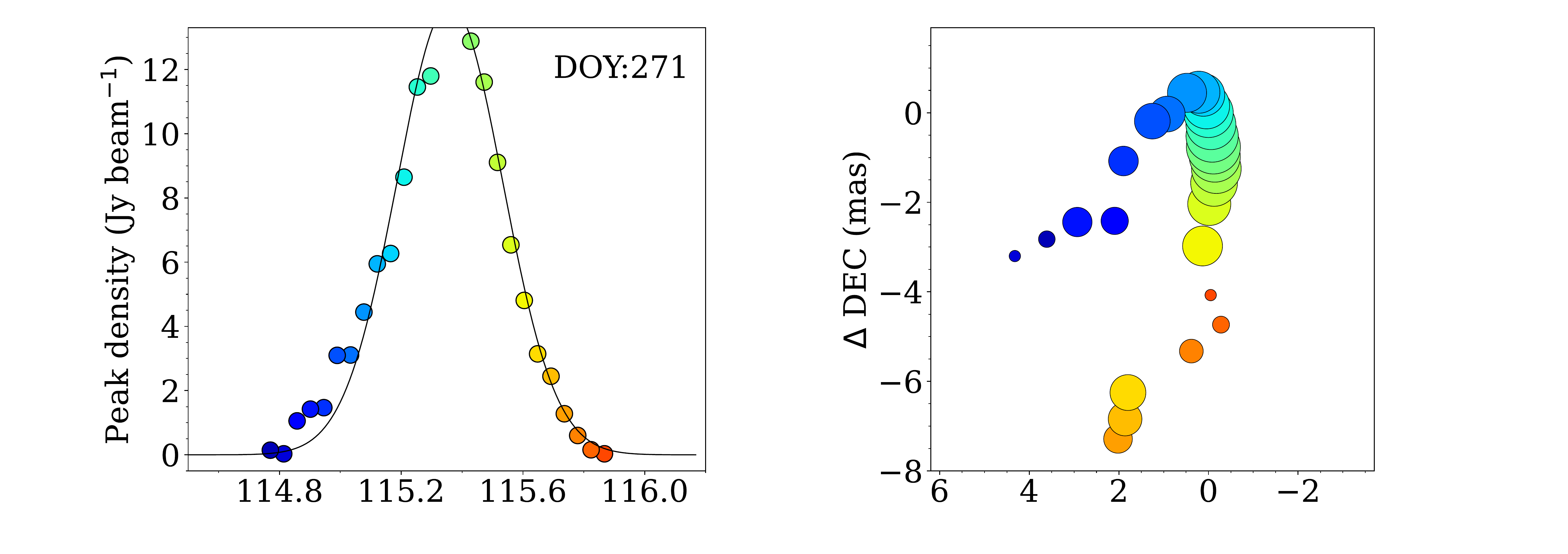}
\includegraphics[width=0.67\textwidth]{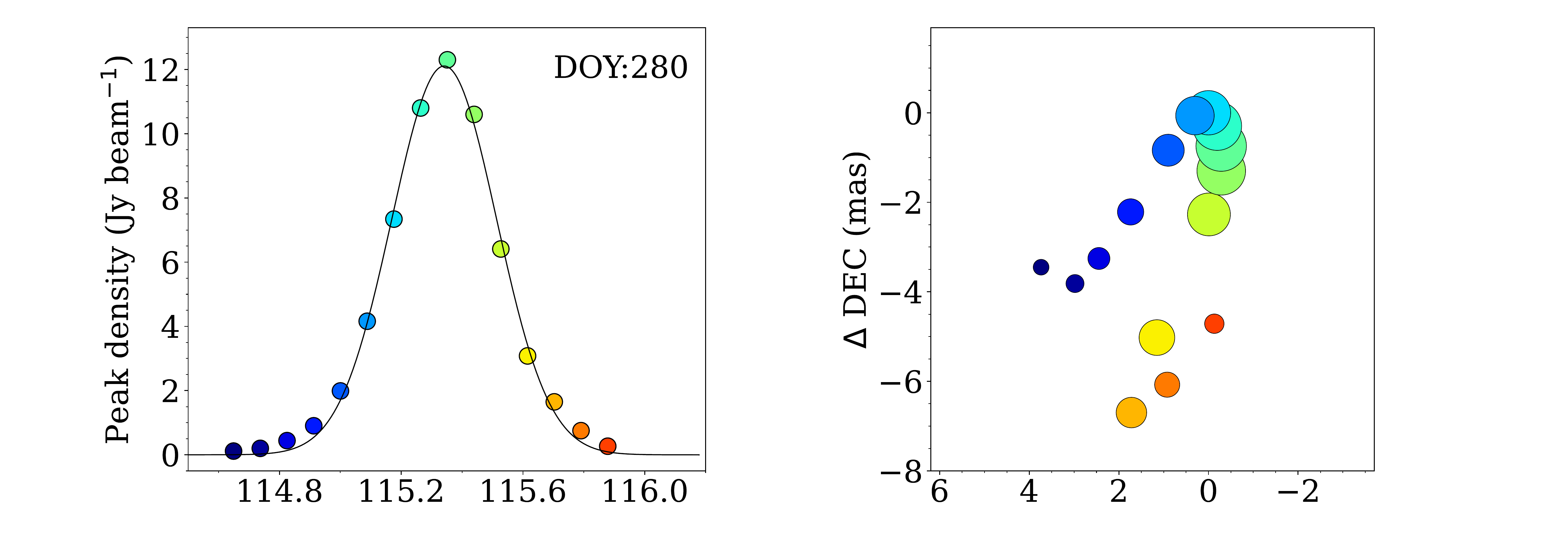}
\includegraphics[width=0.67\textwidth]{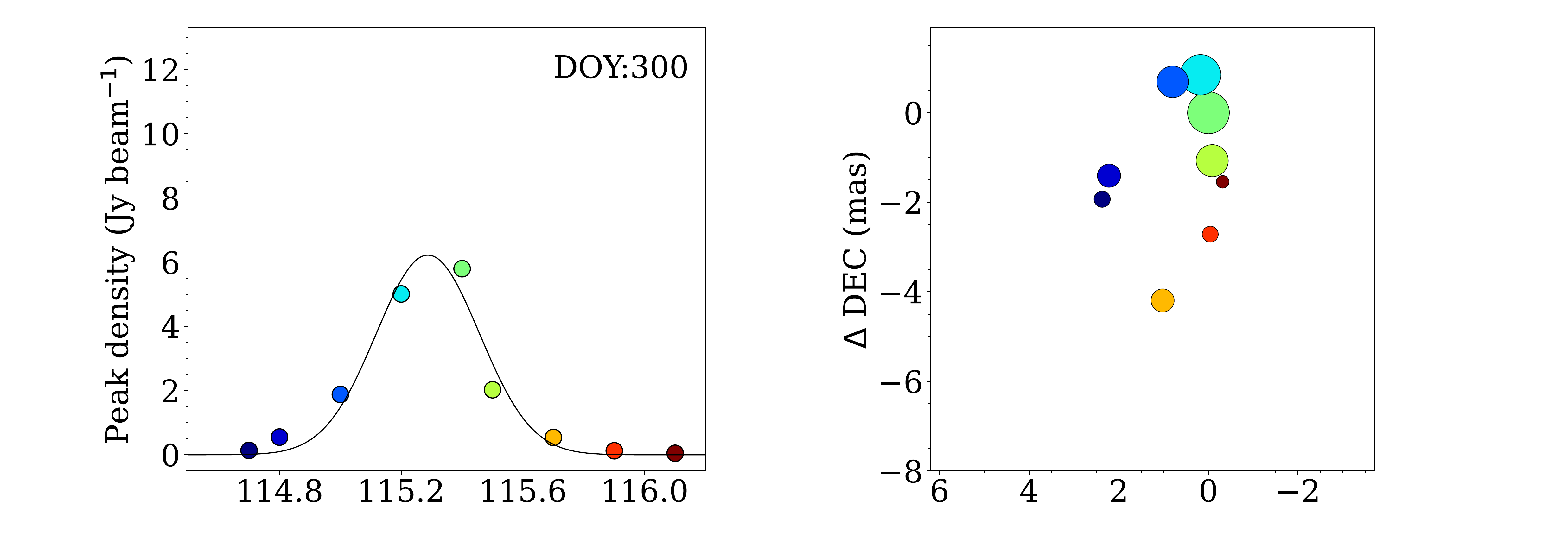}
\includegraphics[width=0.67\textwidth]{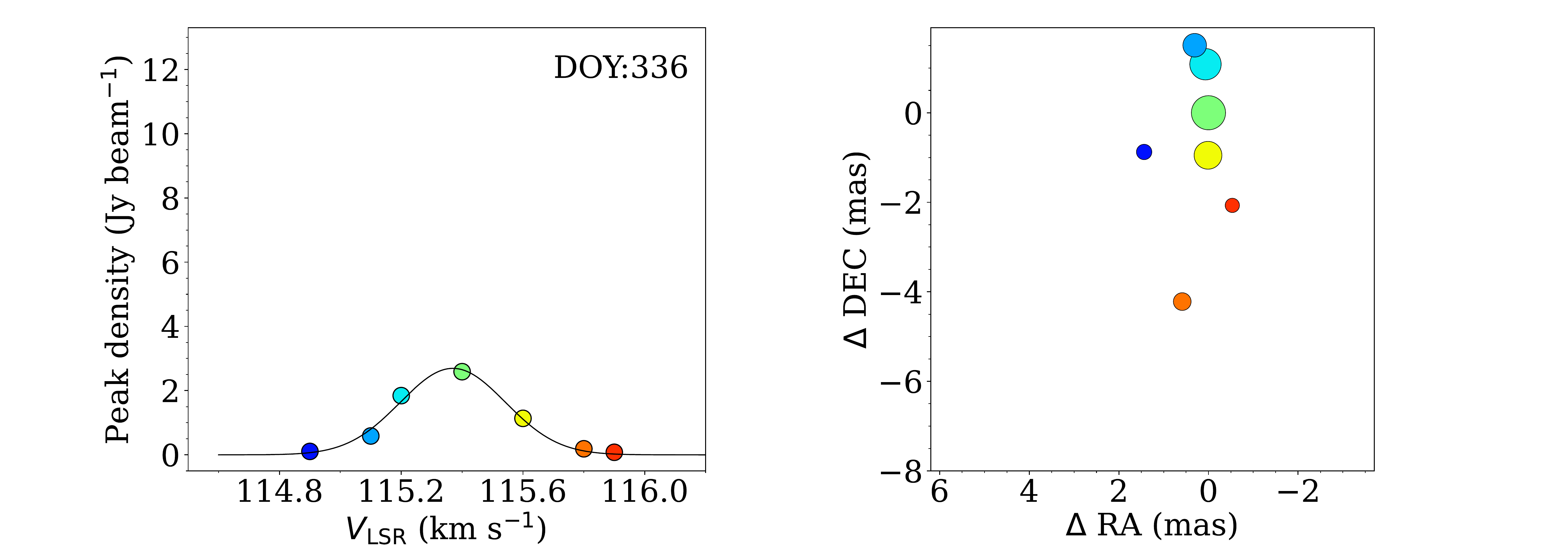}
\caption{Examples of Gaussian characteristics of the individual spectra (left) and the spatial distributions of single spots (right) of the 6.7~GHz methanol maser cloudlets. Here {\bf Cloudlet~3} is presented (Table~\ref{tablegaussmeth6}). The cloudlet shows the arched structure in all epochs of observations, and therefore no linear fits were  done.}
\addtocounter{figure}{-1}
\label{fig:fitmeth67}
\end{figure*}


\begin{figure*}
\centering
\includegraphics[width=\textwidth]{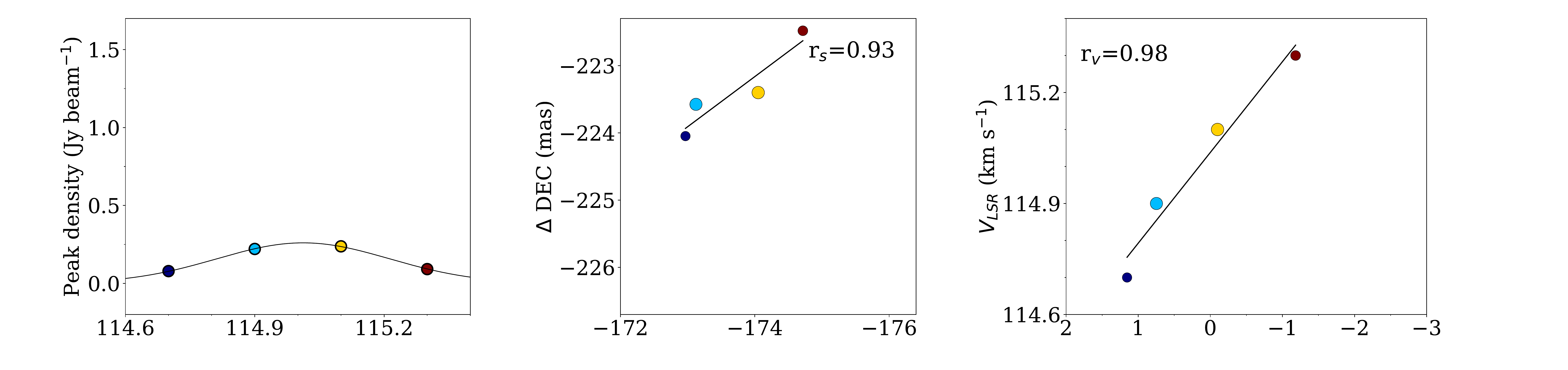}
\includegraphics[width=\textwidth]{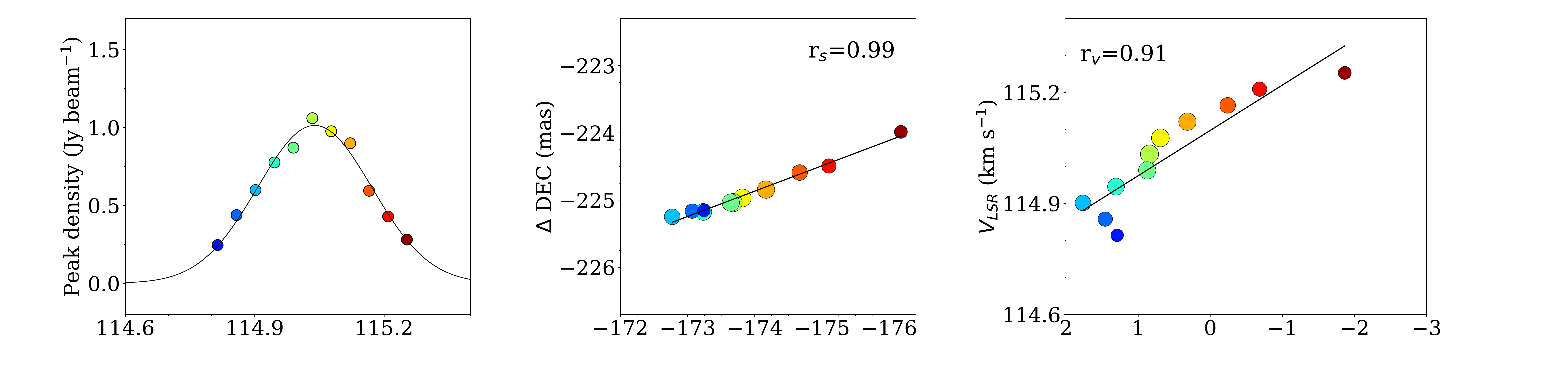}
\includegraphics[width=\textwidth]{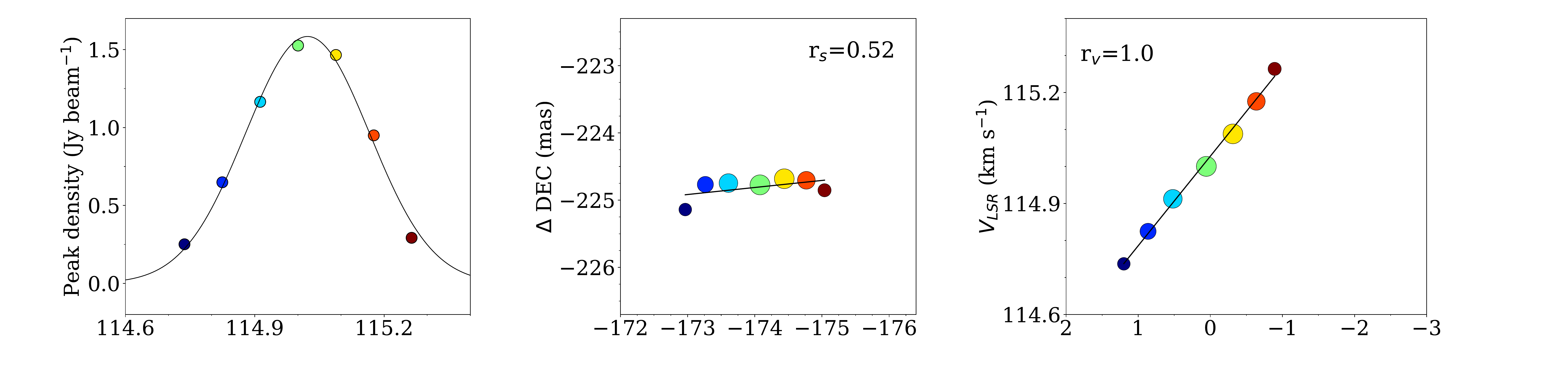}
\includegraphics[width=\textwidth]{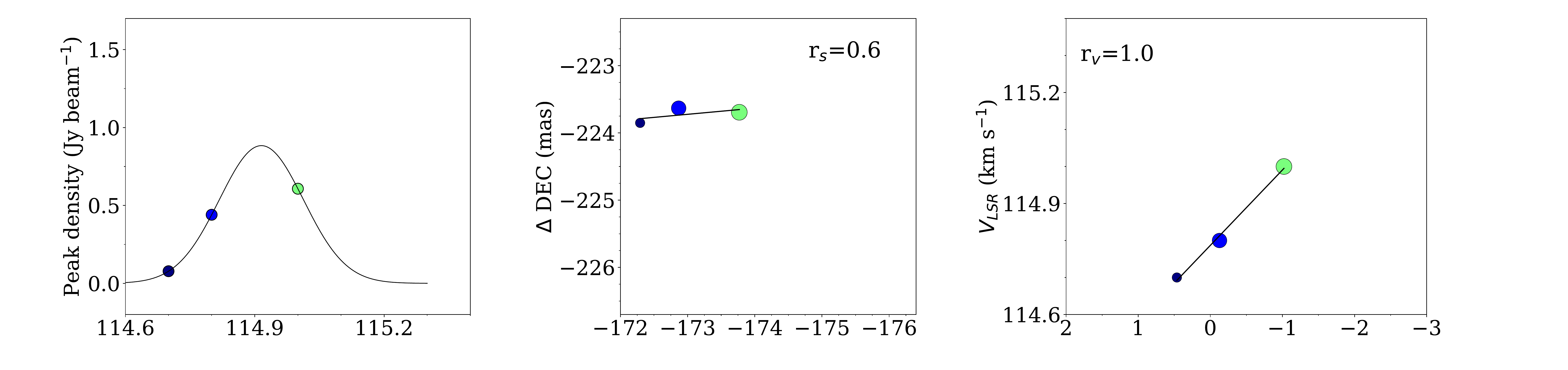}
\includegraphics[width=\textwidth]{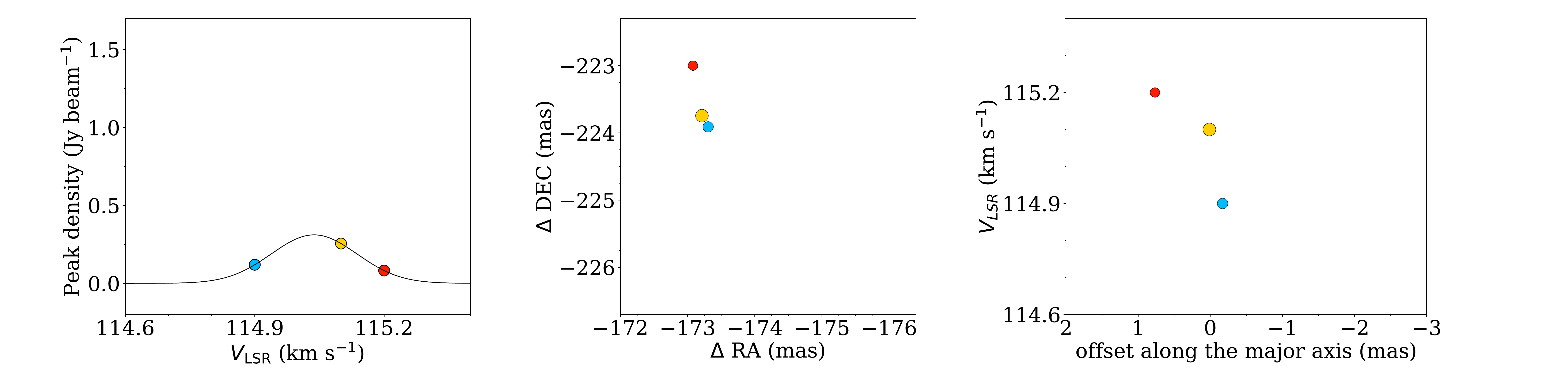}
\caption{continued. {\bf Cloudlet~4} as in Table~\ref{tablegaussmeth6}. The left and middle panels are the same as for Cloudlet 3. The right panel shows the spot V$_{\rm LSR}$--position offset along the major axis of the spot distributions. There is no fit in DOY: 336 due to the weakness of the spots.}
\addtocounter{figure}{-1}
\end{figure*}


\begin{figure*}
\centering
\includegraphics[width=\textwidth]{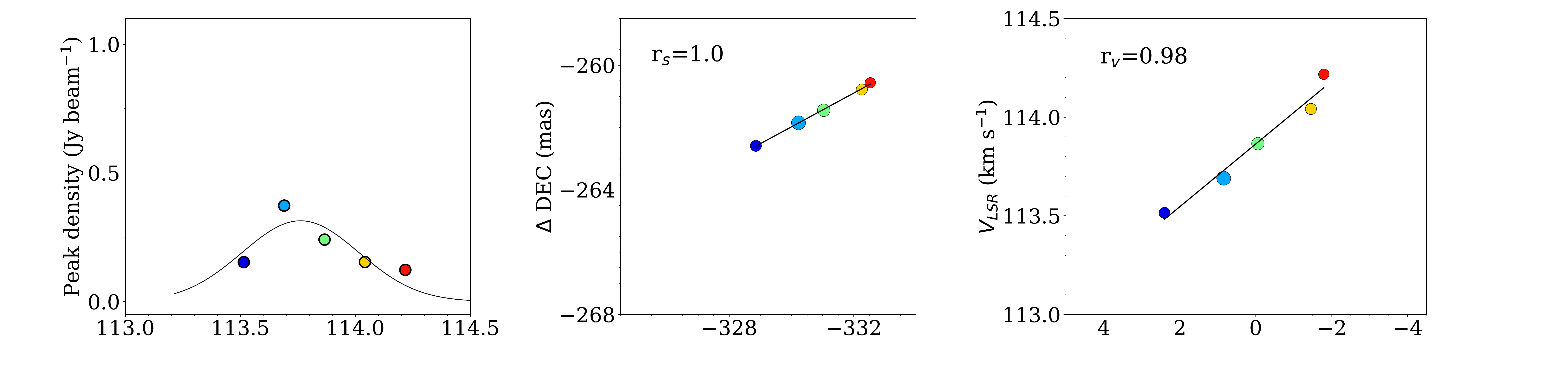}
\includegraphics[width=\textwidth]{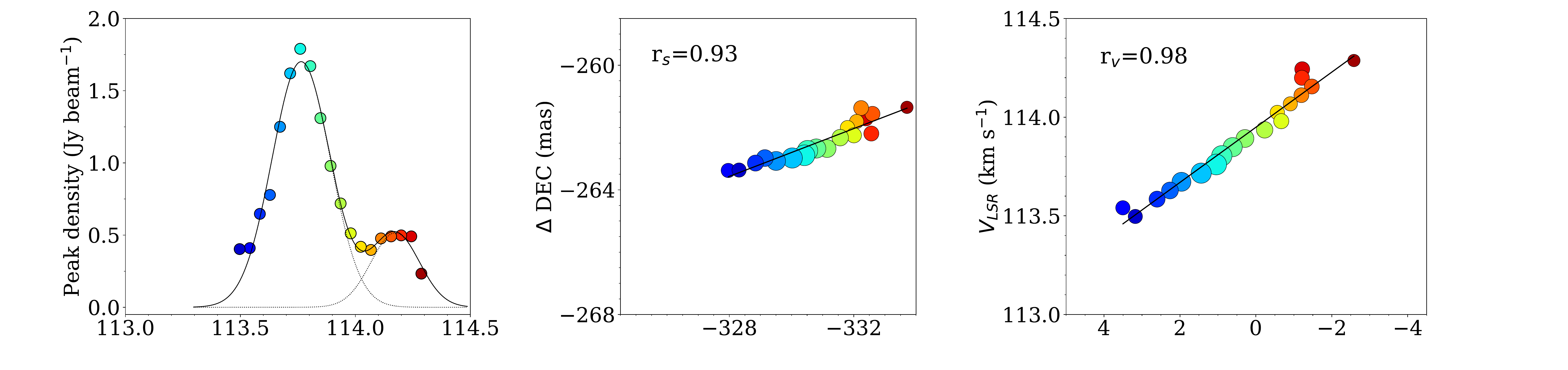}
\includegraphics[width=\textwidth]{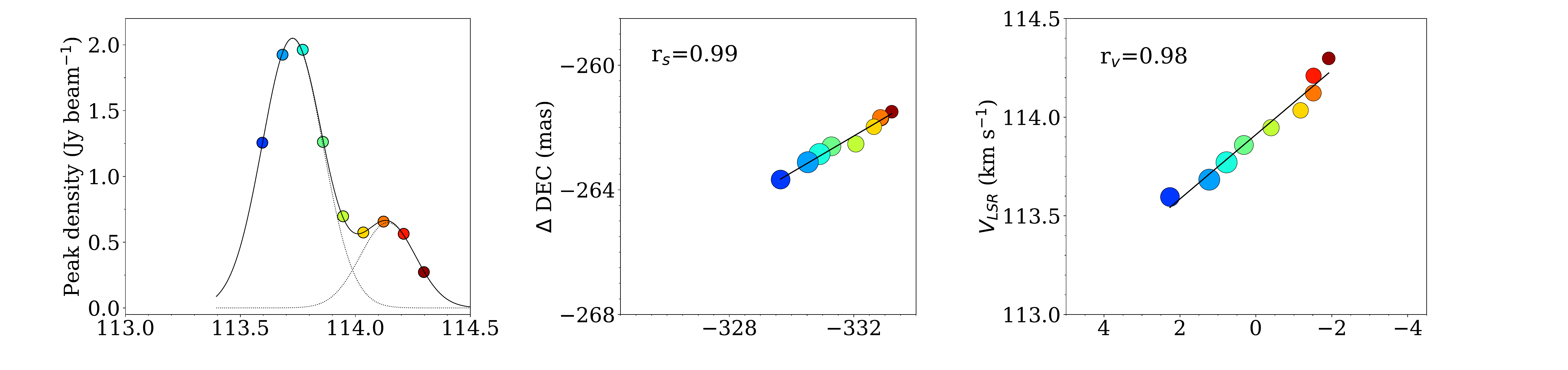}
\includegraphics[width=\textwidth]{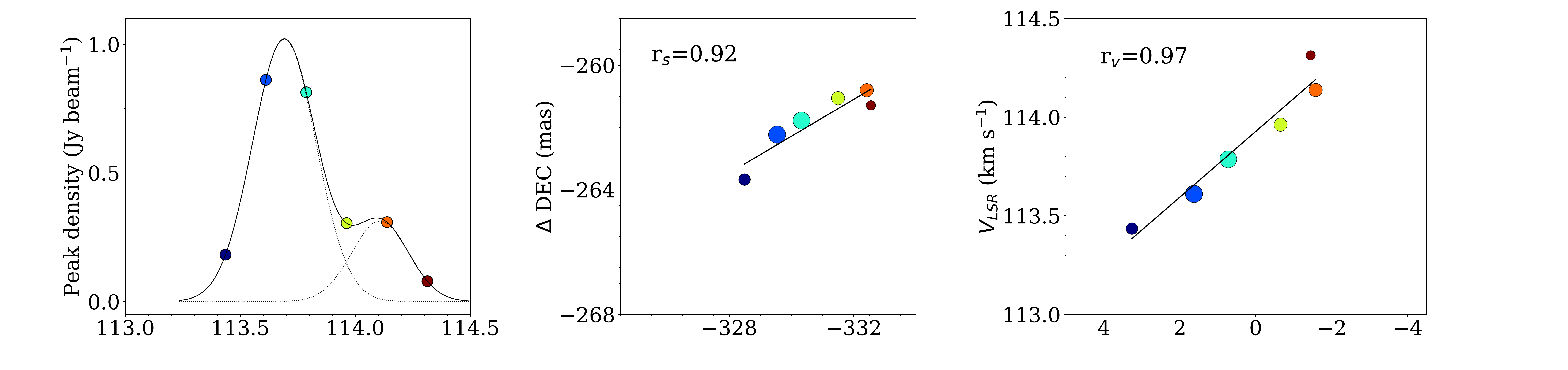}
\includegraphics[width=\textwidth]{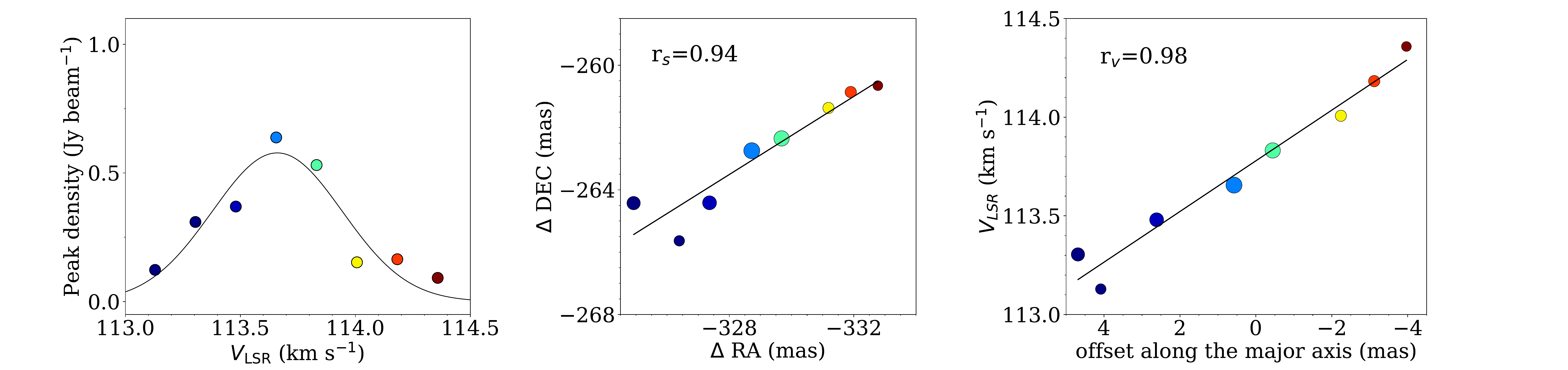}
\caption{continued. {\bf Cloudlet~6}.}
\addtocounter{figure}{-1}
\end{figure*}

\begin{figure*}
\centering
\includegraphics[width=\textwidth]{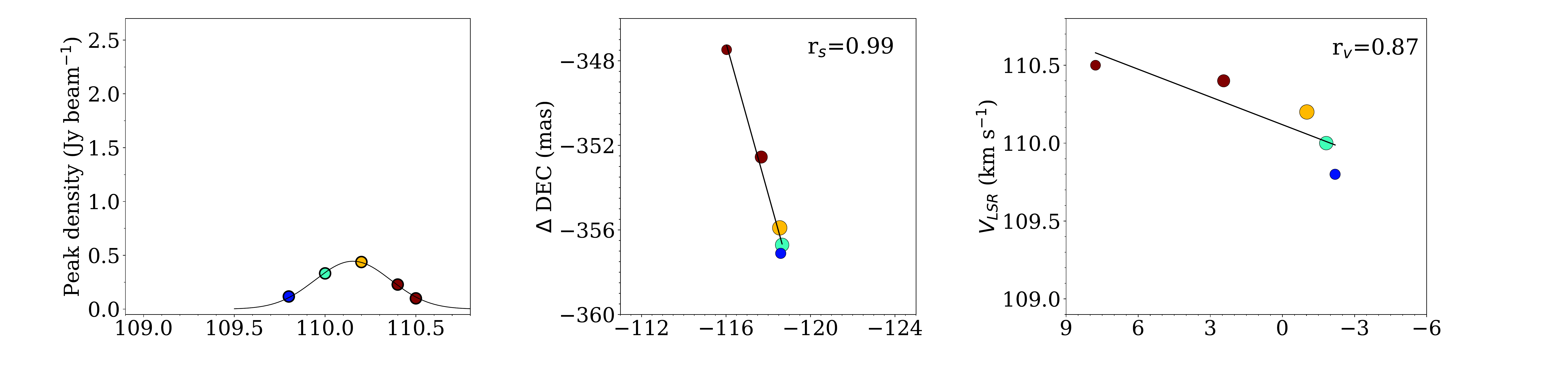}
\includegraphics[width=\textwidth]{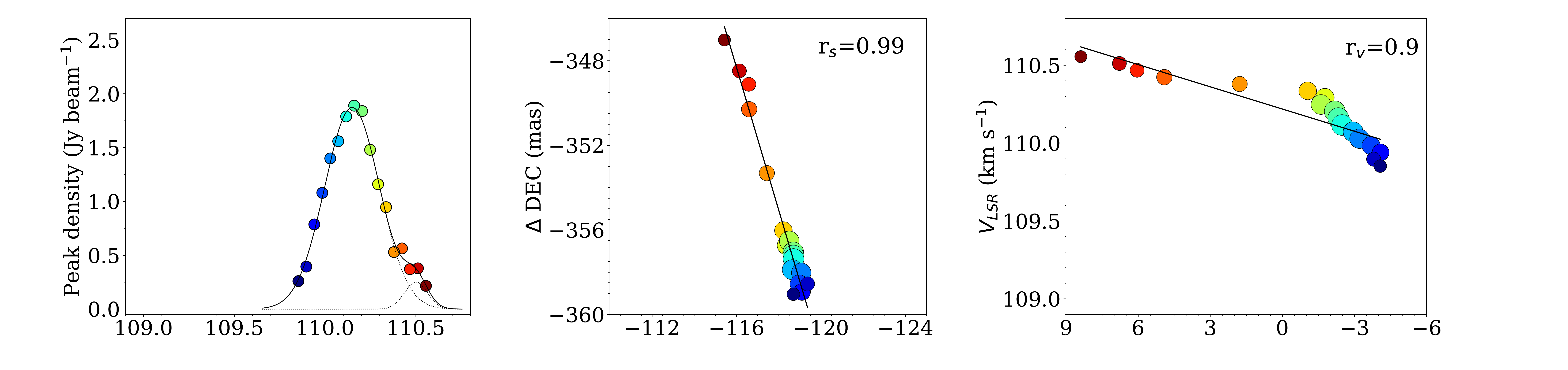}
\includegraphics[width=\textwidth]{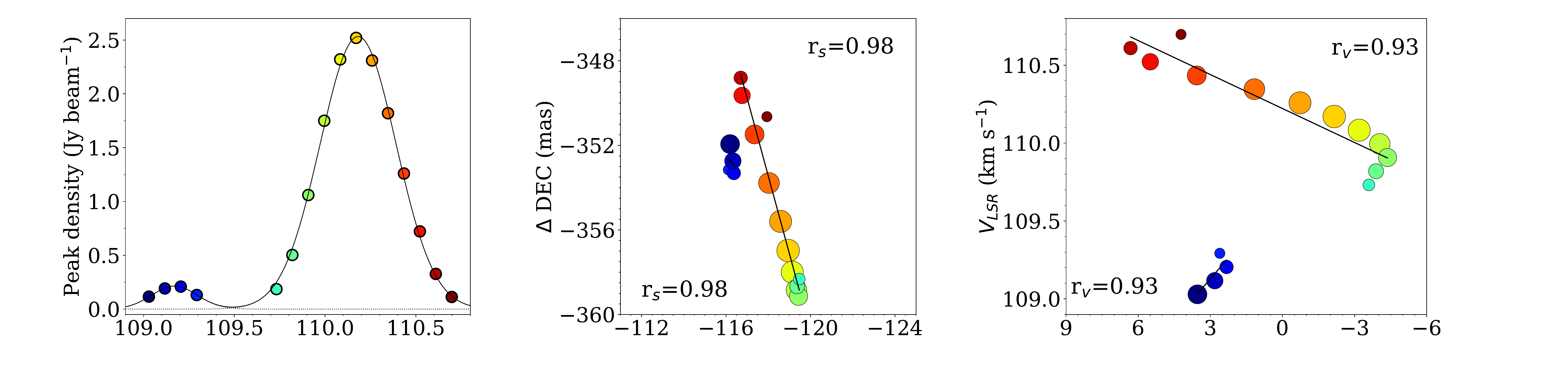}
\includegraphics[width=\textwidth]{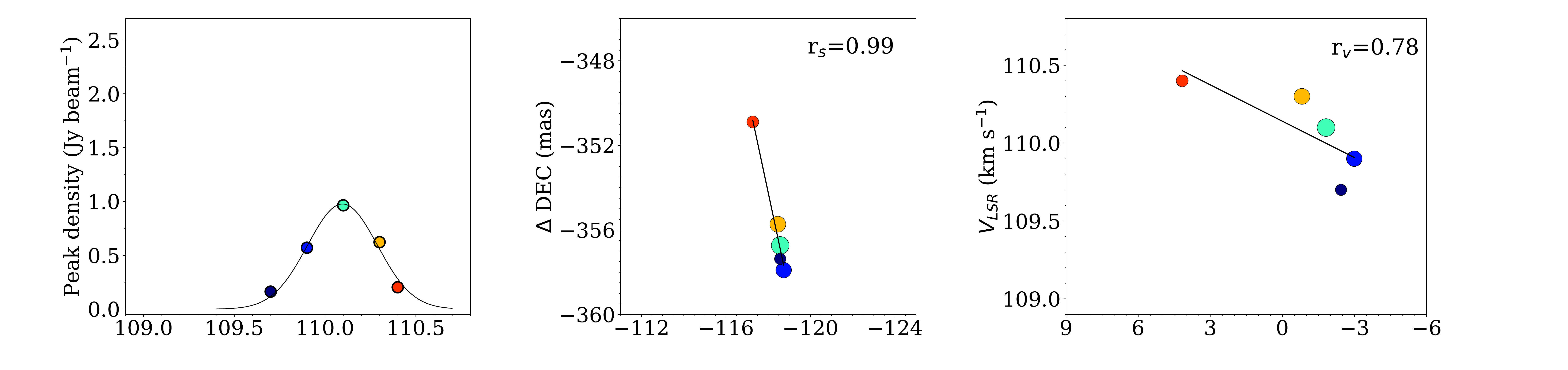}
\includegraphics[width=\textwidth]{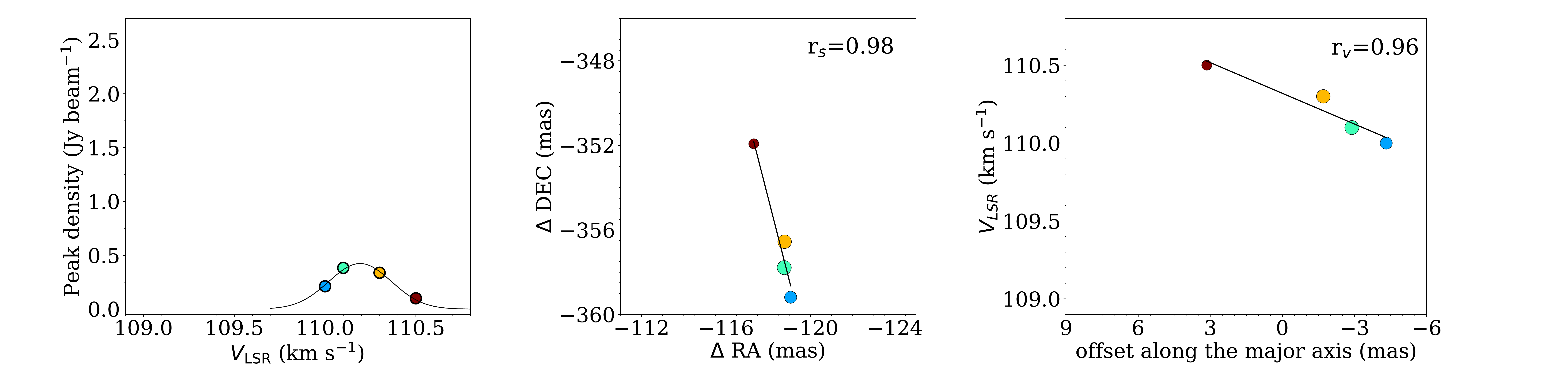}
\caption{continued. {\bf Cloudlet~12}.}
\end{figure*}

\begin{figure*}
\centering
\includegraphics[width=\textwidth]{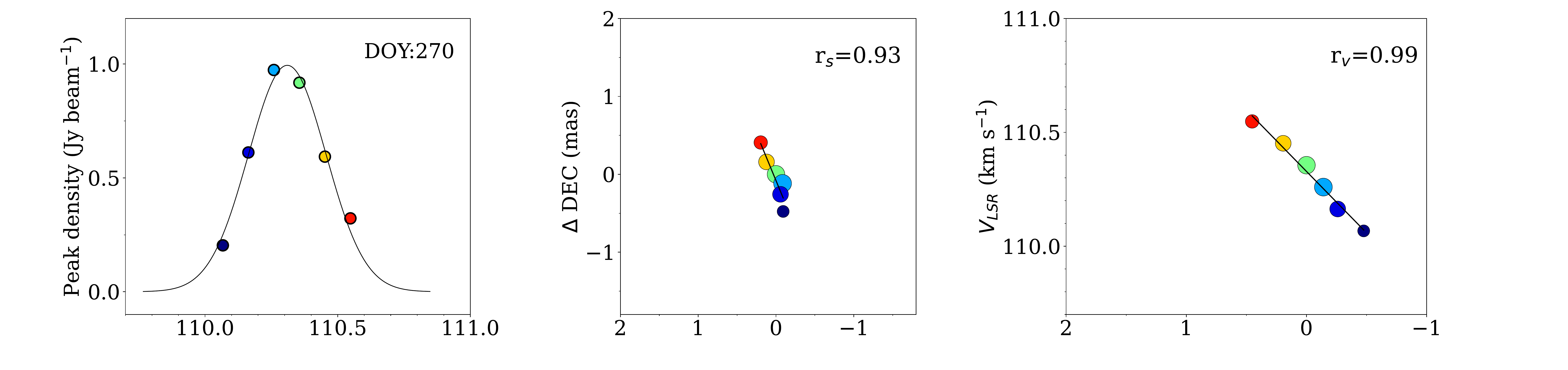}
\includegraphics[width=\textwidth]{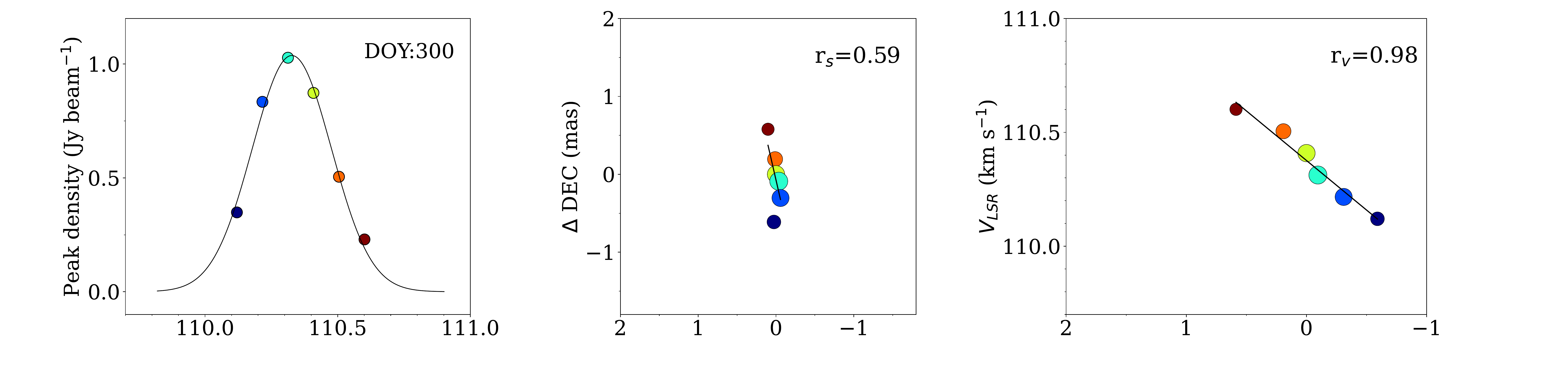}
\includegraphics[width=\textwidth]{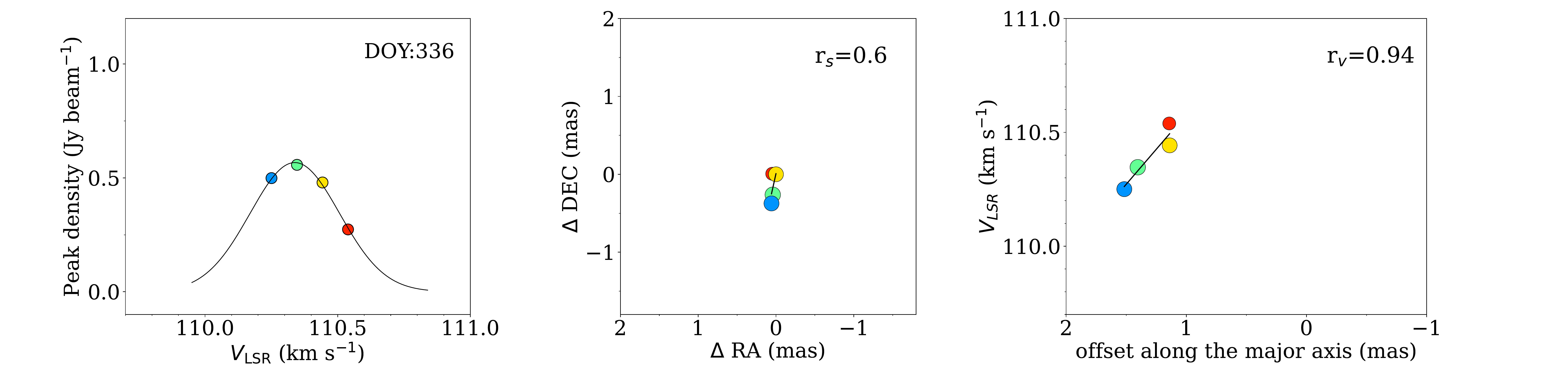}
\caption{Gaussian characteristics of the 12.2~GHz methanol maser Cloudlet 3 (Table~\ref{tablegaussmeth12}). Shown are the individual spectra (left), the spatial distributions of single spots (middle), and the spot V$_{\rm LSR}$--position offset along the major axis of the spot distributions (right).}
\label{fig:fit12group}
\end{figure*}

\begin{figure*}
\centering
\includegraphics[scale=0.2]{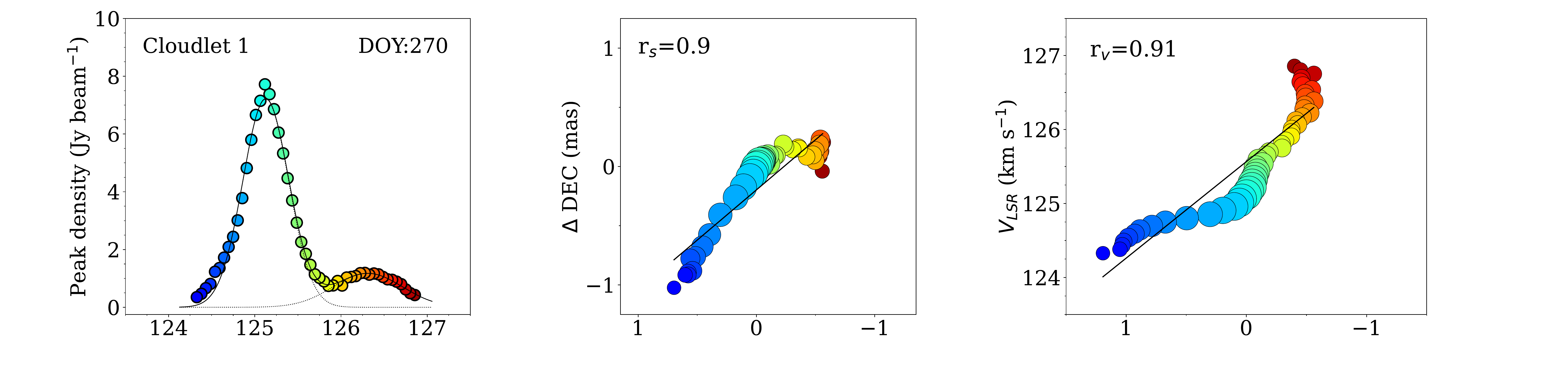}
\includegraphics[scale=0.2]{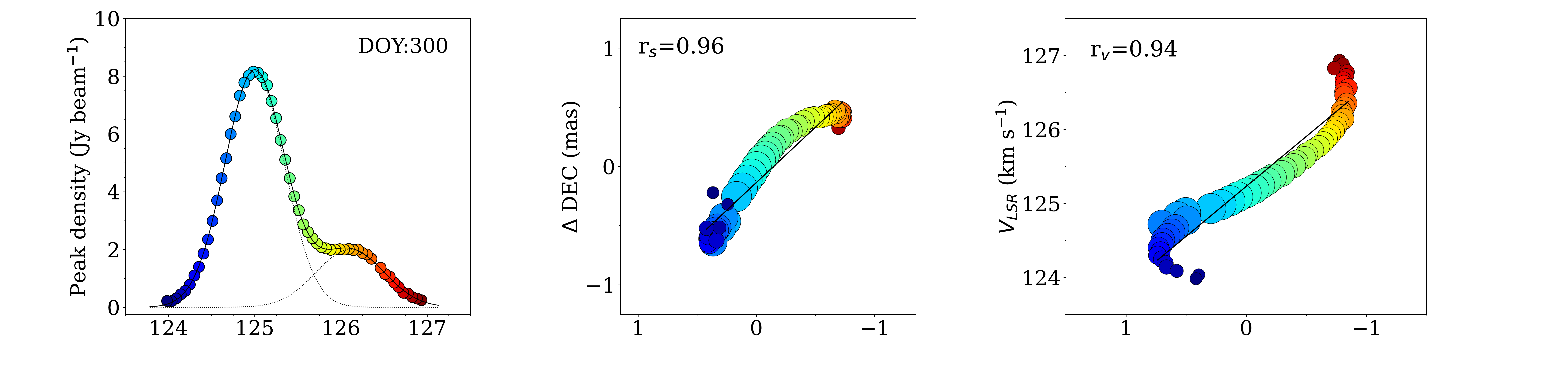}
\includegraphics[scale=0.2]{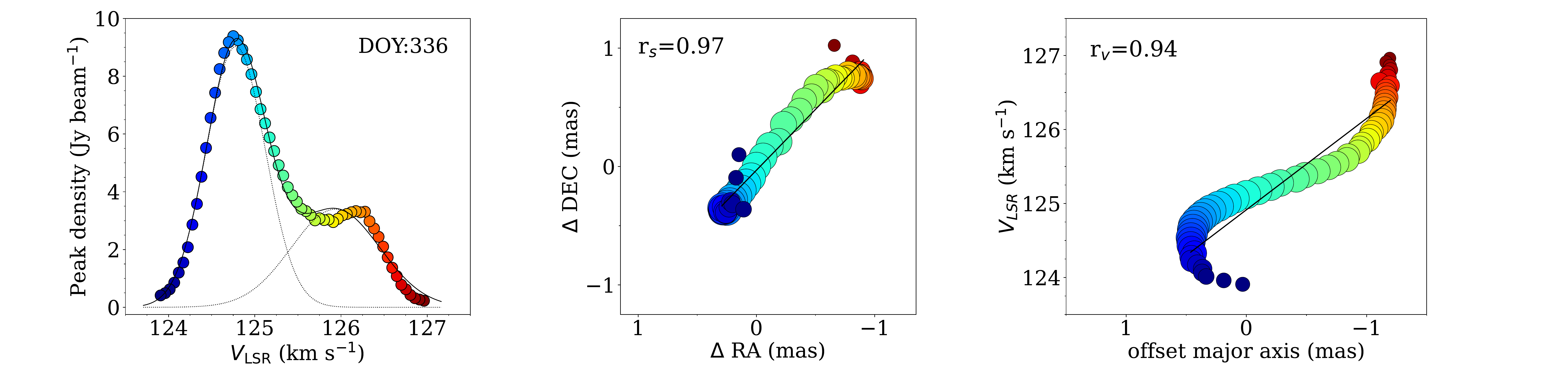}

\includegraphics[scale=0.2]{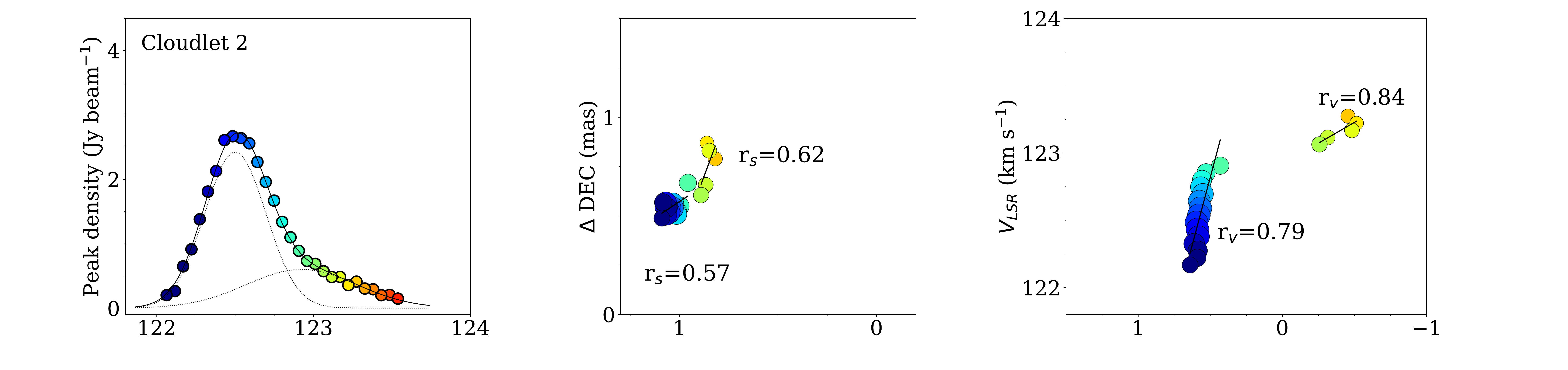}
\includegraphics[scale=0.2]{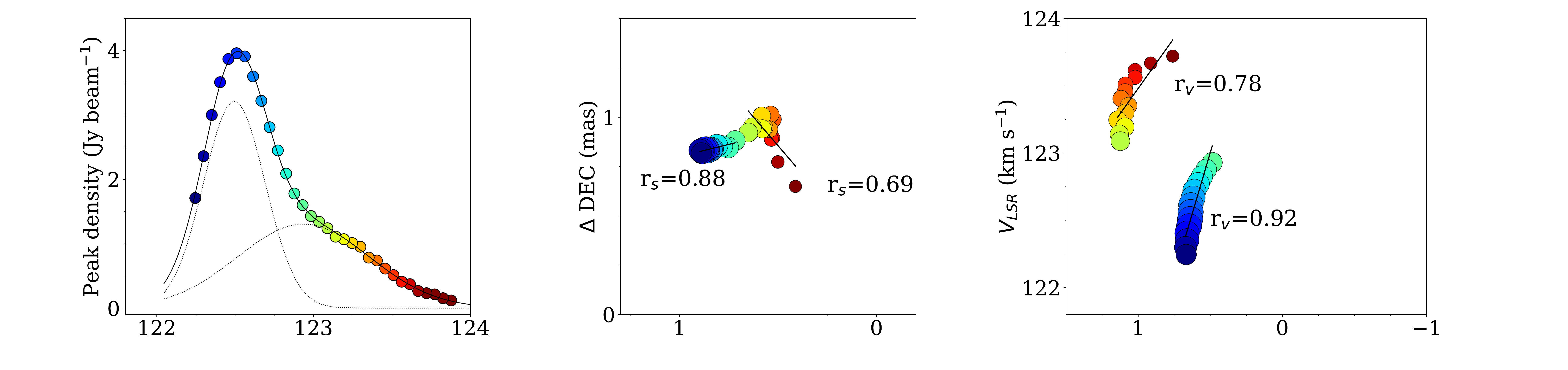}
\includegraphics[scale=0.2]{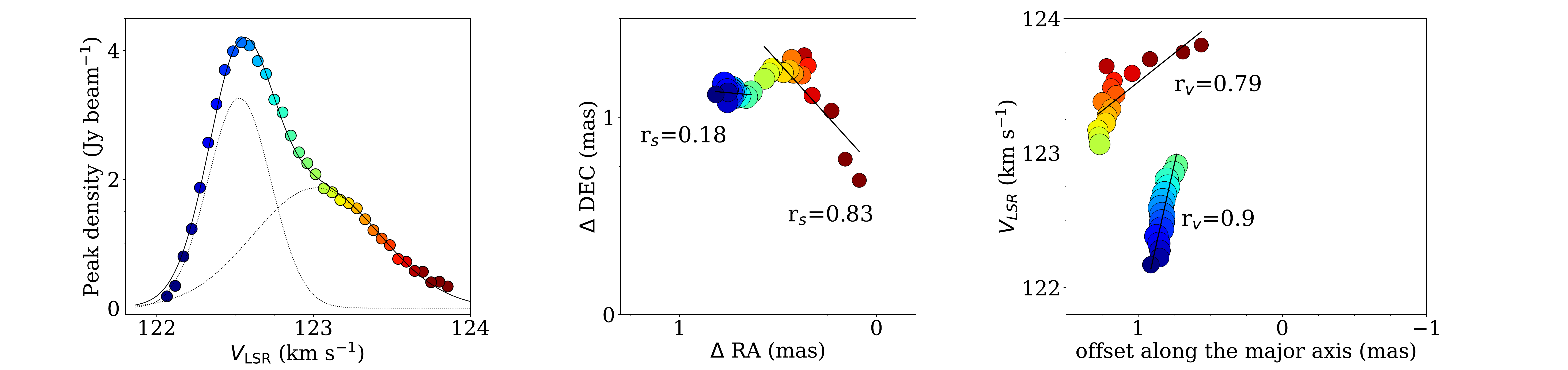}
\caption{Gaussian characteristics of 22.2~GHz water maser Cloudlet~1 (top three  panels) and Cloudlet~2 (bottom three panels) (Table~\ref{tablegausswater}). Shown are the individual spectra (left), the spatial distributions of single spots (middle), and the spot V$_{\rm LSR}$--position offset along the major axis of the spot distributions (right).}
\label{fig:fit22group}
\end{figure*}



\begin{figure*}
\centering
\includegraphics[width=1.0\textwidth]{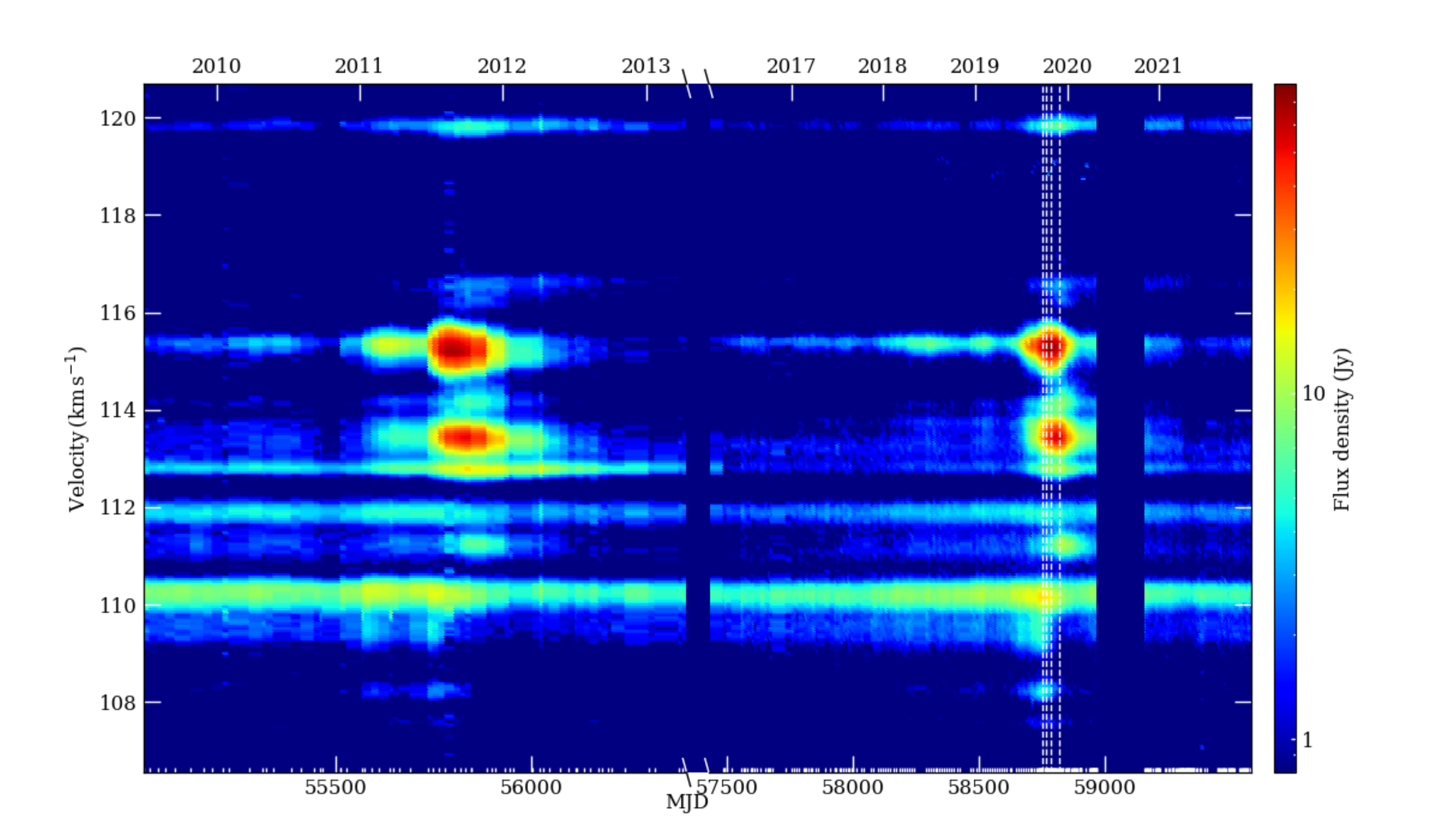}
\caption{Dynamic spectrum of 6.7\,GHz methanol maser emission of G24. The velocity is measured with respect to the local standard of rest. The vertical bars on the bottom ordinate correspond to the dates of the observed spectra. The white dashed vertical lines indicate the epochs of VLBI observations. There were no observations in the periods 56387$<$MJD$<$57482 and 58967$<$MJD$<$59153. 
\label{fig:g24_dynamic_spectrum_by_md} }
\end{figure*}


\begin{figure*}
\centering
\includegraphics[scale=0.5]{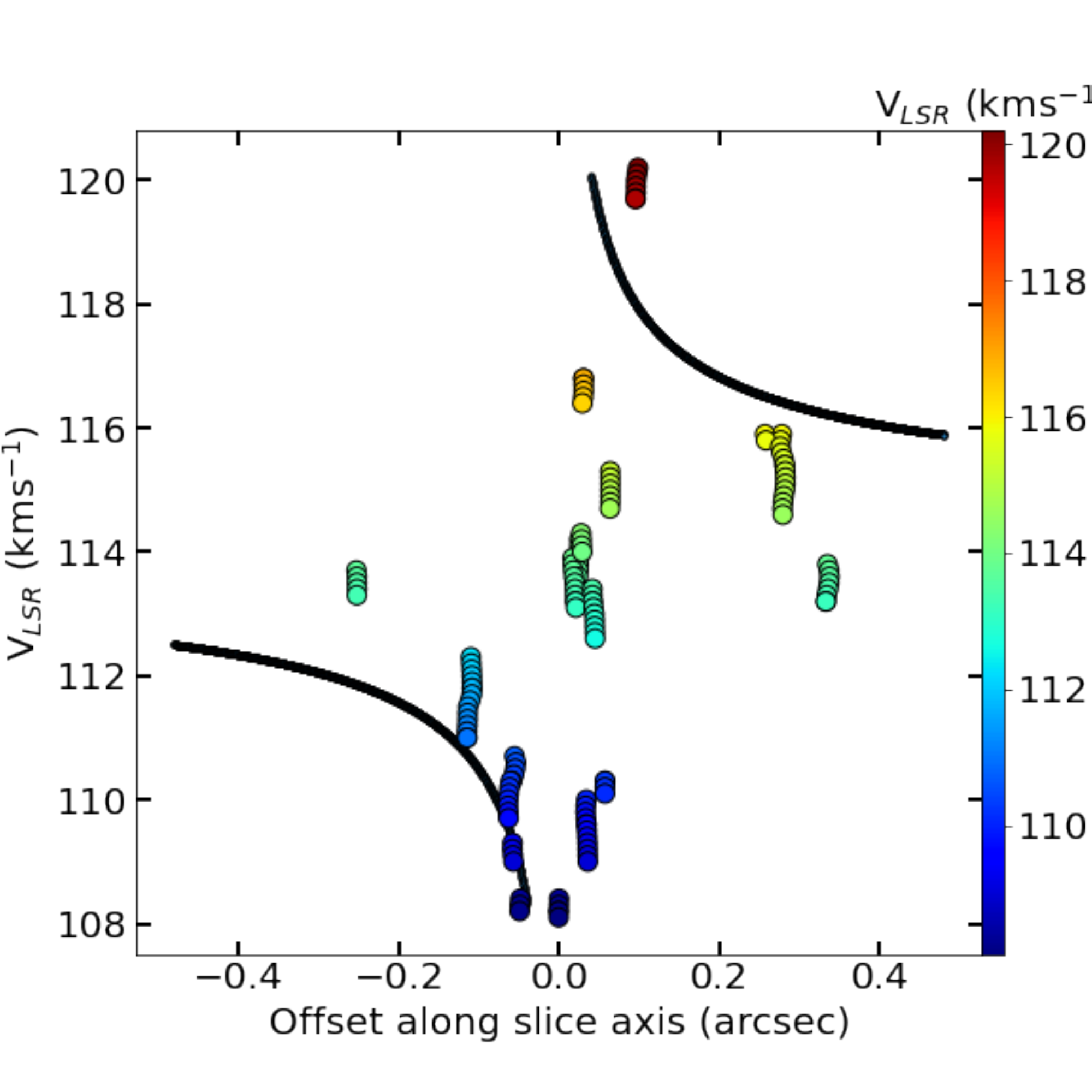}
\caption{Best fit Keplerian disk rotation model (bold dashed  black lines) derived using a central mass of 8.8\,M$_\odot$ \citep{hirota:2022}, inner and outer disk radii of 10~AU and 120~AU, and disk inclination of 85$^{\circ}$, overlaid on the position-velocity diagrams of the observed maser features.
\label{fig:g24_33_pv} }
\end{figure*}

\end{appendix}
\end{document}